\documentclass[eepic,epic,epsfig,12pt,twoside]{article}
\makeatletter
\usepackage{epsfig}
\usepackage{amsfonts}
\usepackage{amsmath}
\usepackage{euscript}
\usepackage{amscd}
\usepackage{graphics}
\usepackage{times}

\usepackage{geometry}
\usepackage[normal,hang]{subfigure}
\usepackage{lscape,graphicx}
\usepackage[normal,bf]{caption2}

\usepackage{pst-plot}
\usepackage{rotating}
\rotdriver{dvips}

\def\eop{\hbox{{\vrule height8pt width4pt depth0pt}}}
\def\epr{\hspace*{\fill}\eop}

\newtheorem{lemma}{Lemma}[section]
\newtheorem{corollary}[lemma]{Corollary}
\newtheorem{theorem}[lemma]{Theorem}

\newtheorem{definition}[lemma]{Definition}
\newtheorem{remark}[lemma]{Remark}

\newtheorem{example}[lemma]{Example}

\newenvironment{proof}{\noindent \begin{text}{\em Proof:} }{\end{text}\hfill{$\epr$}}

\renewcommand{\baselinestretch}{1.2}

\makeatletter
\newcommand\figcaption{\def\@captype{figure}\caption}
\newcommand\tabcaption{\def\@captype{table}\caption}
\makeatother

\newcounter{mySubCounter}
\newcommand{\mySubCaption}{}
\newenvironment{myFirstSubFigure}[1] {
  \addtocounter{figure}{1}
  \begin{mySubFigure}{#1}%
}{%
  \end{mySubFigure}%
}
\newenvironment{mySubFigure}[1] {
 \renewcommand\mySubCaption{#1}
 \refstepcounter{mySubCounter} 
} 
{%
  \vspace{1ex} 
  \addtocontents{lof}{\protect\contentsline %
         {figure}%
         {\numberline{%
            \arabic{section}.\arabic{figure}\alph{mySubCounter}}{\mySubCaption}}%
         {\arabic{page}}%
}\\
 \textbf{Figure \thefigure\alph{mySubCounter} : }\mySubCaption
}
\newenvironment{myLastSubFigure}[1] {
  \begin{mySubFigure}{#1}
}{%
  \end{mySubFigure}%
  \setcounter{mySubCounter}{0}
}

\newcommand{\oneandhalfspacing}{\renewcommand{\baselinestretch}{1.5}}
\oneandhalfspacing

\begin{document}
\title{{\bf Shape preservation behavior of spline curves}}
\author{Ravi Shankar Gautam\\
Department of Mathematics, Indian Institute of Technology Bombay,\\ 
Mumbai-400 076, Maharashtra, India\\
Email: ravishankargautam@gmail.com, \\
gautam@math.iitb.ac.in}
\date{}
\maketitle
\begin{abstract}
Shape preservation behavior of a spline consists of criterial conditions for 
preserving convexity, inflection, collinearity, torsion and coplanarity shapes 
of data polgonal arc. We present our results which acts as an improvement in 
the definitions of and provide geometrical insight into each of the above 
shape preservation criteria. We also investigate the effect of various results 
from the literature on various shape preservation criteria. These results have
not been earlier refered in the context of shape preservation behaviour of 
splines. We point out that each curve segment need to satisfy more than one
shape preservation criteria. We investigate the conflict between different 
shape preservation criteria 1)on each curve segment and 2)of adjacent curve 
segments. We derive simplified formula for shape preservation criteria for 
cubic curve segments. We study the shape preservation behavior of cubic 
Catmull-Rom splines and see that, though being very simple spline curve, it 
indeed satisfy all the shape preservation criteria. 
\end{abstract}
\section{Introduction}
Designers in industries need to create splines which can
interpolate the data points in such a way that they preserve the
shape of polygonal arc formed by data points. Among the properties that 
the spline curves need to satisfy following properties are of common interest
to almost all the designers:

\begin{itemize}
\item Smoothness
\item Preservation of shape of the data polygon
\item Each curve segment to be a low order polynomial curve.
\end{itemize}

\begin{figure}[h]
\centering{\includegraphics{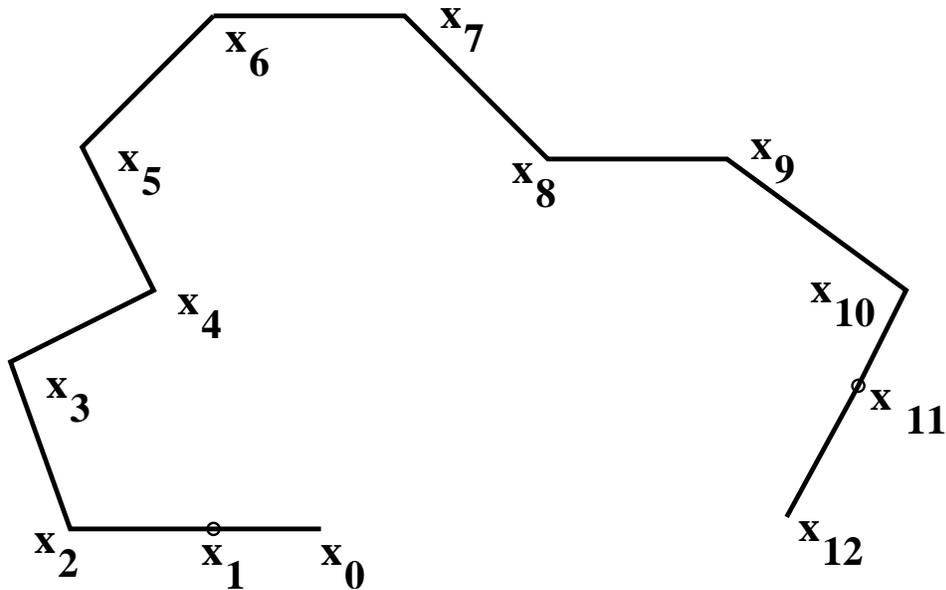}}
\caption{Data points to be interpolated} \label{interpdata}
\end{figure}

\begin{figure}[h]
\centering{\includegraphics{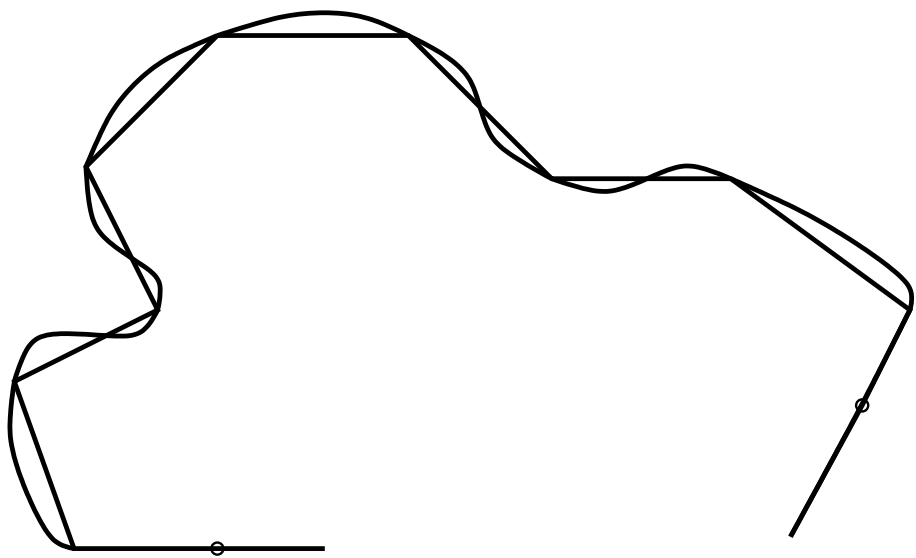}}
\caption{Data polygon with a shape preserving interpolating spline} \label{interpdataspl}
\end{figure}

\begin{figure}[h]
\centering
\includegraphics{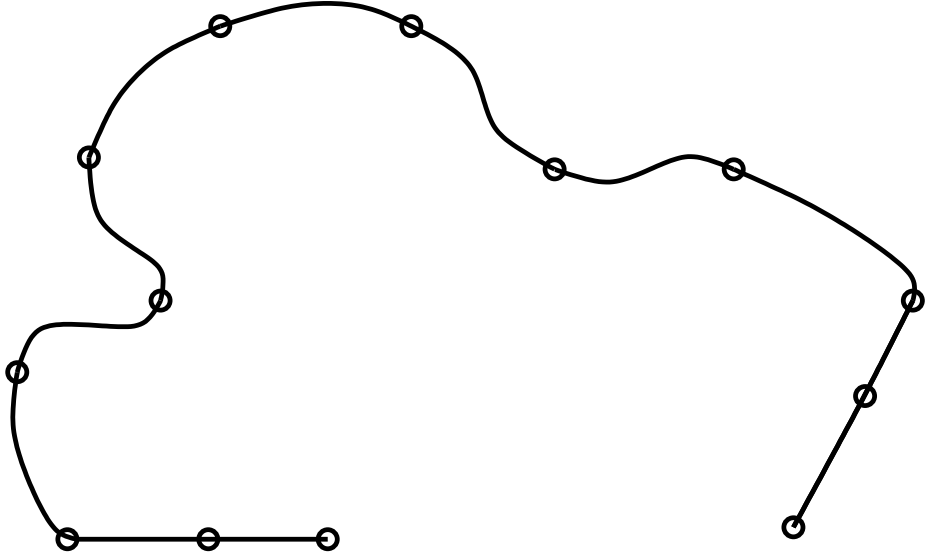}
\caption{Data points with shape preserving interpolating spline} \label{interpspline}
\end{figure}

We first illustrate the shape preservation behaviour of a
spline interpolating planar data points with the help of figures \ref{interpdata},
\ref{interpdataspl} and \ref{interpspline}.

One can observe that 
\begin{itemize}
\item Inflection depicted by data points ${\bf x}_{3}$ ${\bf x}_{4}$ 
${\bf x}_{5}$ and ${\bf x}_{6}$ is preserved by suitable inflection 
of the curve segment between ${\bf x}_{4}$ and ${\bf x}_{5}$
\item Convexity depicted by data points ${\bf x}_{4}$ ${\bf x}_{5}$ 
${\bf x}_{6}$ and ${\bf x}_{7}$ is preserved by the convex shape 
of the curve segment between ${\bf x}_{5}$ and ${\bf x}_{6}$
\item Collinearity depicted by data points ${\bf x}_{0}$ ${\bf x}_{1}$ and 
${\bf x}_{2}$ is preserved by collinearity of the curve segment between
${\bf x}_{0}$ and ${\bf x}_{2}$
\end{itemize}
an so on.
It can also be observed that shape preserving behaviour of a spline makes it 
more close to mimicing free-hand curve drawing.

However, modelling of shape preservation behaviour of interpolating
splines in $R^{3}$ is relatively difficult.
Typically shape preservation criteria that have been studied for the generation 
of interpolating splines consists of conditions for the preservation of
 1)Convexity 
 2)Inflection 
 3)Collinearity 
 4)Torsion and
 5)Coplanarity shapes of data polygonal arc (formed by line joining the 
consecutive points of ordered set of data points).
In this paper we present literature survey and also our analysis
for each of the above criteria. 
In sections \ref{secshapeconvexplan} and \ref{secshapeconvexcrit} 
we present our analysis for convexity preservation criteria.
In literature, basically, two definitions for convexity preservation criteria 
are followed. One which is followed in 
\cite[Kaklis and Karavelas, 1997]{kaklis}, 
\cite[Costantini et. al.]{constantinigoodman, constantiniisabella} etc. 
and another which is followed in 
\cite[Kong and Ong, 2002]{kong},
\cite[Goodman and Ong, 1997]{goodmanong}  etc.
The latter definition includes the conditions of previous definition
and hence we investigate further on the later definition. We  
find that in the later definition, for a large set of interpolating spline 
curve (containing set of rational spline, excluding straight lines, curves as 
small subset) one of the two conditions is redundant (that is, the condition 
is actually taken care by other condition of the definition) for all the points 
of the curve except for set of points on the curve of measure zero (the set is 
finite for rational curves excluding straight lines). Also at these points the 
error (if at all occurs) is very negligible and it is observed that in all
almost all the algorithm staright line segments in the spline curves are
considered for collinearity preservation criteria. We use our 
lemma \ref{lemsuper} and the characterization of convexity of planar curve 
presented in \cite[Liu and Traas, 1997]{liutraas} for our 
analysis of convexity criteria of interpolating splines. In 
\cite[Liu and Traas, 1997]{liutraas} the characterization 
of convexity 
of planar curve on $R^{2}$, that is, $XY$ plane have been derived. 
We use this characterization to get the characterization of planar 
curve on any plane in $R^{3}$. Then the modified characterization has 
been used to improve the definition of convexity preservation criteria of splines. 
We further state simplified characterization of convexity preservation 
criteria for cubic splines in terms of control polygon of individual 
B\'{e}zier segments. 

In sections \ref{planarinflex} to \ref{inflexcrit} we present 
the analysis for inflection preservation criteria of splines. We refer two papers
\cite[Goodman, 1991]{goodman} and 
\cite[Goodman and Ong, 1997]{goodmanong} for our analysis 
for inflection criteria. In \cite[Goodman, 1991]{goodman} 
author has stated definitions and results for inflection counts 
for planar and space curves and polygonal arcs. In 
\cite[Goodman and Ong, 1997]{goodmanong} authors have 
defined the inflection criteria of splines and have constructed a spline
satisfying the criteria. However, in \cite[Goodman and Ong, 1997]{goodmanong}, 
authors haven't indicated
any connection with the analysis of 
\cite[Goodman, 1991]{goodman}. Also the analysis in 
\cite[Goodman, 1991]{goodman} is not used in 
\cite[Goodman and Ong, 1997]{goodmanong}  for the analysis of inflection criteria
of splines. We observe in section \ref{inflexcrit} that our lemma \ref{lemsuper}
acts as a connection between the analysis in sections \ref{planarinflex} and 
\ref{spaceinflex} and conditions stated in definition of convexity criteria of
splines.  

In section \ref{planarinflex} we state definitions and results from \cite[Goodman, 1991]{goodman}
for inflection count for planar curves and polygonal arcs. The relation
between the inflection counts of planar B-spline and B\'{e}zier curves and
inflection counts of their control polygons are stated. In section 
\ref{spaceinflex} we state definitions and results for inflection counts
for space curves and polygonal arcs from \cite[Goodman, 1991]{goodman}. 
Definitions in section \ref{spaceinflex} uses the definitions in section
\ref{planarinflex}. In section \ref{inflexcrit} we state the definition 
and analysis for the convexity criteria for splines. In section 
\ref{negativeinflex} we state the results from 
\cite[Goodman, 1991]{goodman} which states 
(via lemma \ref{lemsuper}) the conditions under which a spline curve does not 
satisfy inflection criteria or convexity criteria.

In sections \ref{seccollin}, \ref{sectorsion} and \ref{seccoplan} we analyze
and give improved conditons for collinearity , torsion and coplanarity 
preservation criteria respectively for splines. 
In section~\ref{conflictsame} and \ref{conflictadjacent} we analyze and give
condtions on a spline curve to resolve conflict between different shape
preservation conditions on a curve segment or a pair of adjacent curve segments.

Almost every geometric modeler necessarily uses cubic splines for generating 
3D models of products. Cubic splines are computationally most
viable solution for various applications requiring complicated geometric 
operations. They are also the splines of least degree that can exhibit 
torsion. Thus we see that it is necessary that cubic splines should preserve
shape of the data points they interpolate. In section \ref{chapshapehermite} we describe our 
analysis and results for shape preserving criteria for cubic splines.
The conditions for shape preserving criteria for cubic splines derived in this 
section are expressed in terms data points and slopes at data points.
During our analysis we obtain simplified formula for discrete shape 
measures and a new property for B\'{e}zier curves.

In subsection \ref{secdervbezcubic} we state aome additional notations 
required in section \ref{chapshapehermite}. 
In subsections \ref{secconvexcubic}, \ref{secinflexcubic}, \ref{sectorsioncubic}, 
\ref{seccollincubic} and \ref{seccoplancubic} we derive the conditions for 
convexity, inflection, torsion, collinearity and coplanarity preservation 
criteria respectively,
for the cubic curve segment in terms of the two data points at their ends and 
slope vectors at them. 
In subsection \ref{sectorsioncubic} we simplify the expression for torsion of cubic
B\'{e}zier curves. 
In subsection \ref{seccollincubic} we derive an expression for sine of the angle
between a point vector on a B\'{e}zier curve with a given vector and state it in theorem
\ref{thmnewbezproperty} and we use it to get simplified formula for collinearity 
preservation criteria for cubic curve segments. Further we use this analysis in   
subsection \ref{seccoplancubic} to get simplified formula for coplanrity preservation 
criteria for cubic curve segments. In section \ref{secshapenetwork} we investigate 
shape preservation behavior of cubic Catmull-Rom splines.
Finally in section \ref{conclusion} we state our conclusions about the analysis in this 
paper.

\section{Notations and prelimnaries}
Let ${\bf x}_{i} \in R^{3}$, $i=0,...,n$ be $n+1$ data points and $\mathcal{D}$ be 
the polyline or polygonal arc formed by joining the points with each side being 
$L_{i}= {\bf x}_{i}-{\bf x}_{i-1}$, $i=1, ..., n$. 
Let $N_{i} = L_{i-1} \times L_{i}$. In discrete differential geometry 
\cite[Sauer, 1970]{sauer} discrete binormal is defined as 
$\displaystyle{ \frac{N_{i}}{|N_{i}|} }$. 
For a curve $\gamma(t)=[x(t),y(t),z(t)]$,
$t \in [0,1]$ in $R^{3}$ let $\omega(t)= \gamma'(t) \times \gamma''(t)$.

\section{Convexity preservation criteria for interpolating splines}\label{secshapeconvexcrit}
\setcounter{equation}{0}
In \cite[Karvelas and Kaklis, 2000]{karavelas}, 
\cite[Kaklis and Karavelas, 1997]{kaklis} 
\cite[Costantini, Goodman, Manni, 2000]{constantinigoodman} 
\cite[Costantini, Cravero, Manni, 2002]{constantiniisabella}, 
\cite[Manni, Pelosi, 2004]{mannipelosi} we have the 
following definition
\begin{definition}\label{deffirstconvex}
Convexity preservation criteria for a curve $\gamma(t)$ interpolating data points 
consists of following condition:
\begin{enumerate}
\item if $N_{i-1} \cdot N_{i} > 0$, then $\omega(t) \cdot N_{m} > 0$, 
$t \in [t_{i-1},t_{i}]$, $m = i-1 \mbox{, }i$.
\end{enumerate}
\end{definition}

\noindent The above definition \ref{deffirstconvex} takes into account the ability of 
a spline to appear as convex curve along only two viewpoints $N_{i-1}$ and $N_{i}$.

\begin{figure}[h]
\centerline{ 
\includegraphics*[width=14cm,height=4cm]{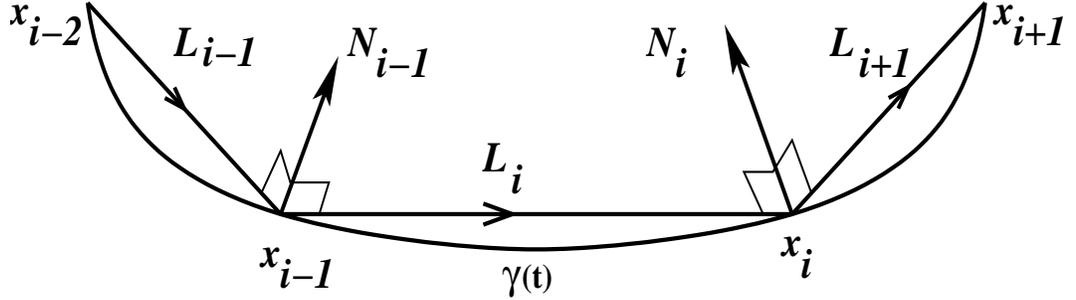} 
}
\caption{Data point with $N_{i-1} \cdot N_{i} >0$ requiring convexity 
preservation by $i^{th}$ curve segment} \label{shapespline}
\end{figure}

\noindent 
The following definition is from \cite[Goodman and Ong, 1997]{goodmanong}, 
\cite[Kong and Ong, 2002]{kong}.
\begin{definition}\cite[Goodman and Ong, 1997]{goodmanong}\label{defsecondconvex}
Convexity preservation criteria for a curve $\gamma(t)$ interpolating data points 
consists of following condition:
\begin{enumerate}
\item If $N_{i-1} \cdot N_{i} > 0$, then for all $N = \lambda N_{i-1} + \mu N_{i}$,
where $\lambda, \mu \geq 0$, ($\lambda, \mu$)$\neq$($0,0$), the projection
$P_{N^{\perp}} \gamma(t)$, $t$ $\in [t_{i-1},t_{i}]$, is globally convex and 
\item $ \omega (t) \cdot N>0$.  
\end{enumerate}
\end{definition}

Now we state a theorem from 
\cite[Goodman and Ong, 1997]{goodmanong} which makes the 
convexity condition of projection curves simpler.
\begin{theorem}\cite[Goodman and Ong, 1997]{goodmanong}\label{projection}
Suppose that $R$ is a curve in $R^{3}$ and $V_{1}$ and
$V_{2}$ are vectors in $R^{3}$ so that $P_{V^{\perp}_{1}}R$ and $P_{V^{\perp}_{2}}R$ are convex
with the same orientation with respect to $V_{1}$ and $V_{2}$ respectively.
Then $P_{V^{\perp}}R$ is convex with the same orientation with respect to $V$ where 
$V= \lambda V_{1} + \mu V_{2}$ for any $\lambda, \mu \geq 0$, 
($\lambda,\mu$)$\neq$($0,0$).
\end{theorem}

Thus from the theorem (\ref{projection}) we can observe that 
condition requiring $P_{N^{\perp}_{i-1}} \gamma_{i}(t)$  and 
$P_{N^{\perp}_{i}} \gamma_{i}(t)$ to be convex is 
equivalent to the condition requiring curves $P_{N^{\perp}}\gamma_{i}(t)$ for
$N= \lambda N_{i-1} + \mu N_{i}$ for any $\lambda, \mu \geq 0$, 
($\lambda,\mu$)$\neq$($0,0$), to be convex.

\subsection{Convexity of planar curves}\label{secshapeconvexplan}
\setcounter{equation}{0}
\subsubsection{Convexity of planar curves on $XY$ plane}
In \cite[Liu, Traas, 1997]{liutraas} authors have defined local and global
convexity of a planar curve as follows. The author distiguishes convexity and concavity of
planar curves in terms of the orientation we assign to the curve. Consider a curve
$\gamma(t)=(x(t), y(t))$, $t\in [0,1]$ in $R^{2}$.
An oriented planar curve is an ordered set in $R^{2}$, given by
$\displaystyle{
\gamma(t) = [x(t), y(t)] \mbox{, } t \in [0,1]
}$
with direction from $t=0$ to $t=1$.
A global supporting line of an oriented curve $\gamma(t)$ at a point $\gamma(t_{0})$ 
is an oriented line, $L$, having consistent direction with $\gamma(t)$ in $t_{0}$,
and satisfying
(a) $\gamma(t_{0})$ is a point of $L$;
(b) the entire curve $\gamma(t)$, $t \in [0,1]$, lies in one closed half-plane
with respect to $L$.

A global supporting line of an oriented curve $\gamma(t)$ at a point $\gamma(t_{0})$ 
is an oriented line, $L$, having consistent direction with $\gamma(t)$ in $t_{0}$,
and satisfying
\begin{enumerate}
\item $\gamma(t_{0})$ is a point of $L$;
\item The entire curve $\gamma(t)$, $t \in [0,1]$, lies in one closed half-plane
with respect to $L$. 
\end{enumerate} 

A local supporting line of an oriented curve $\gamma(t)$ at a point $\gamma(t_{0})$ 
is an oriented line, $L$, having consistent direction with $\gamma(t)$ in $t_{0}$,
and satisfying
\begin{enumerate}
\item $\gamma(t_{0})$ is a point of $L$;
\item A local neighborhood $\gamma(t)$, $t \in [t_{1},t_{2}]$, of $\gamma(t_{0})$, lies
in one closed half-plane with respect to $L$, where $t_{1}$ and $t_{2}$ satisfy
\begin{eqnarray}
0 \leq t_{1} < t_{0} < t_{2} \leq 1 \mbox{ or } 
0 = t_{1} = t_{0} < t_{2} \leq 1 \mbox{ or } 
0 \leq t_{1} < t_{0} = t_{2}=1 \mbox{.}
\end{eqnarray} 
\end{enumerate}

\begin{figure}[h]
\begin{center}
\includegraphics*{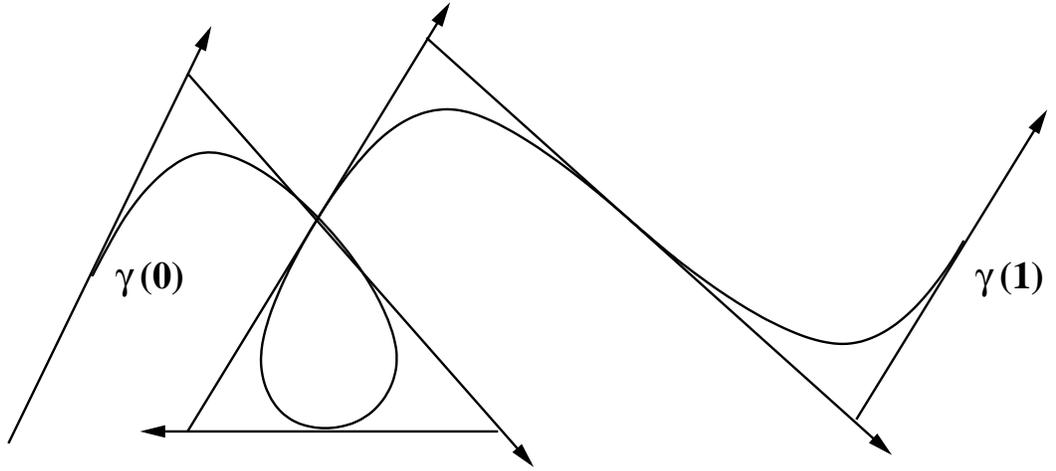}
\caption{Supporting and non-supporting lines [Liu and Traas '97]} \label{support}
\end{center}
\end{figure}

\begin{definition}\cite[Liu, Traas, 1997]{liutraas}\label{d3}
$\gamma(t)$, $t \in [0,1]$, is a globally convex curve if it satisfy:
\begin{enumerate}
\item There is at-least one global supporting line at every point of 
$\gamma(t)$;
\item The entire curve lies in the right closed half-plane with the 
supporting line as its left boundary.
\end{enumerate}
\end{definition}

\begin{figure}[h]
  \centerline{
    \subfigure[closed]{
\epsfig{file=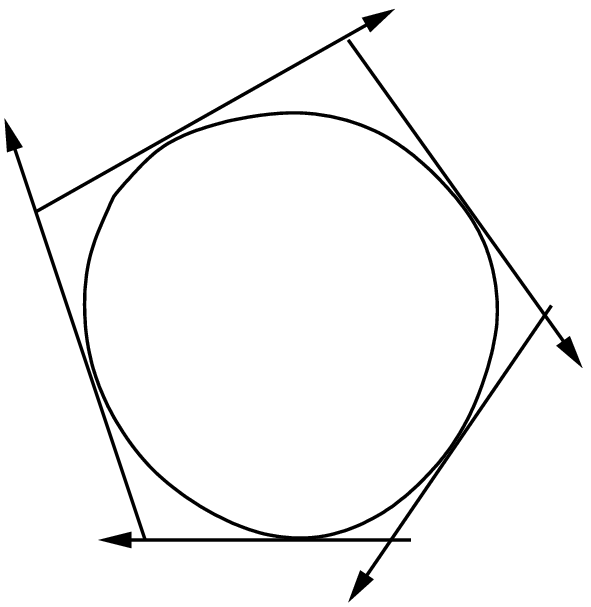}}
\hspace{1cm}
    \subfigure[open]{
\epsfig{file=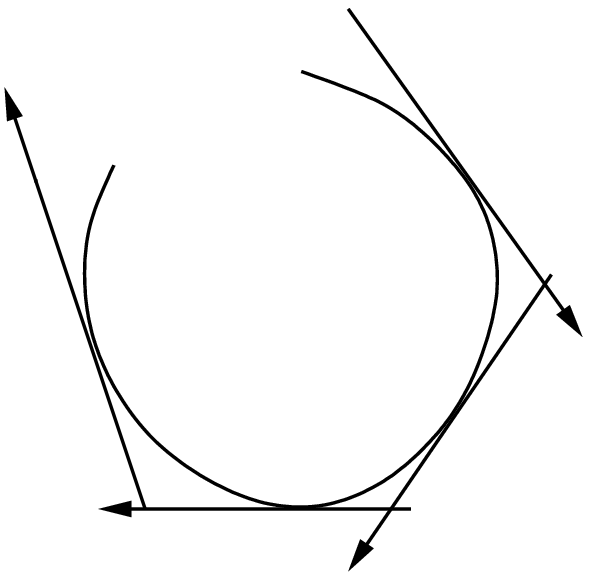}}
}
\caption{Globally convex curve [Liu and Traas '97]}
\end{figure}

\begin{definition}\cite[Liu, Traas, 1997]{liutraas}\label{d4}
$\gamma(t)$, $t \in [0,1]$, in definition (\ref{d3}) is a locally convex curve 
if in (1) the global supporting line is replaced by a local one and (2) is 
true only for a related local neighborhood.
\end{definition}

A globally and locally concave curve has the same definition as for globally and locally 
convex curve respectively with left and right interchanged.

\begin{figure}[h]
  \centerline{
    \subfigure[]{
\epsfig{file=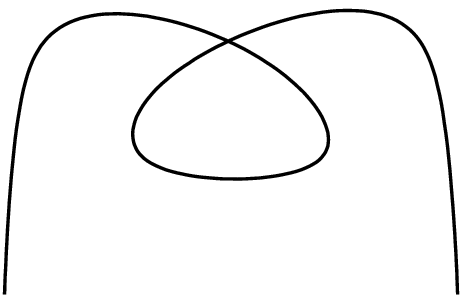}}
\hspace{1cm}
    \subfigure[]{
\epsfig{file=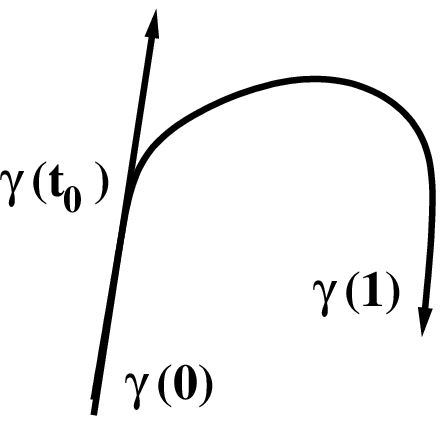}}
}
\caption{(a) Locally convex curve [Liu and Traas '97] 
(b) Condition \ref{globconvexcondn2} for global convexity [Liu and Traas '97]}
\end{figure}

We observe that the above definitions 
\footnote{There are other definitions of a convex curve. For example, in
\cite[Farin]{Farin} it is defined as a part of boundary of a convex set.
In \cite[Liu and Traas, 1997]{liutraas} is has been proved, 
using Hahn-Banach theorem, that 
this definition is included in definition (\ref{d3}).} hold for any planar 
curve in $R^{3}$. However, the two main theorems of 
\cite[Liu and Traas, 1997]{liutraas} that
we state below need some modifications.

\begin{theorem}\cite[Liu and Traas, 1997]{liutraas}\label{planloc}
$\gamma(t)$, $t \in [0,1]$, is locally convex if and only if
\begin{eqnarray}
\gamma'(t) \times \gamma''(t) \leq 0 \mbox{, } t \in [0,1] \label{loconvexcondn}
\end{eqnarray}
where $\gamma(t)$ is $C^{2}$-continuous, $\gamma'(t)$, $\gamma''(t)$ are first and second
derivatives of $\gamma(t)$, and 
\begin{eqnarray}
\gamma'(t) \times \gamma''(t) = \left|
\begin{array}{cc}
x'(t) & x''(t)\\
y'(t) & y''(t)
\end{array}
\right|
= x'(t)y''(t)-x''(t)y'(t) \mbox{.}\label{plancross}
\end{eqnarray}\hfill{$\epr$}
\end{theorem}

\begin{theorem}\cite[Liu and Traas, 1997]{liutraas}\label{planglob}
A curve $\gamma(t)$ satisfying the condition
\begin{eqnarray}
\gamma(t) \neq \gamma(0) \mbox{ for } t \in (0,1)
\end{eqnarray}
is globally convex if and only if
\begin{eqnarray}
\gamma'(t) \times \gamma''(t) & \leq & 0 \mbox{,} \label{globconvexcondn1}\\
(\gamma(t)-\gamma(0)) \times \gamma'(t) & \leq & 0 \mbox{, } t \in [0,1] \mbox{,}\label{globconvexcondn2}\\
\gamma'(0) \times (\gamma(t)-\gamma(0)) & \leq & 0 \mbox{, } t \in [0,1] \mbox{.}\label{globconvexcondn3}
\end{eqnarray}  \hfill{$\epr$}
\end{theorem}

\begin{remark}\label{remarkequality}
For a curve $\gamma (t) \in R^{2}$, $\gamma'(t) \times \gamma''(t)$ is its 
curvature. Therefore, in \ref{loconvexcondn} and \ref{globconvexcondn1} the 
equality holds at $t=t_{1}$ if and only if $\gamma (t)$ behaves locally as 
straight line (turns with angle $0^{o}$) at $t=t_{1}$. One can replace 
\ref{loconvexcondn} and 
\ref{globconvexcondn1} by $\gamma'(t) \times \gamma''(t) <  0$ by requiring 
strict convexity of $\gamma (t)$. But we did not find this to be significant
requirement as set of points for which equality holds is of measure zero
for almost all curves (except straight lines) used for interpolation (and 
therefore their local behaviour as straight line have negligible effect). 
The conditions \ref{globconvexcondn2} and \ref{globconvexcondn3} ensures
that the curve doesn't intersect itself and equality in these conditions
may cause the curve to become straight line in some cases.
\end{remark}

\begin{figure}[h]
  \centerline{
    \subfigure[]{\label{cnxitcondn2_neg}
\epsfig{file=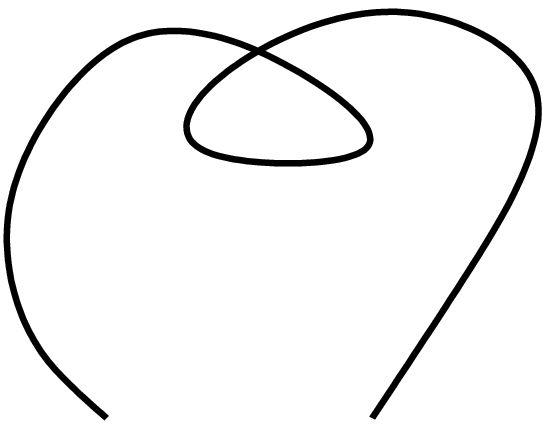}}
\hspace{1cm}
    \subfigure[]{\label{cnxitcondn3_neg}
\epsfig{file=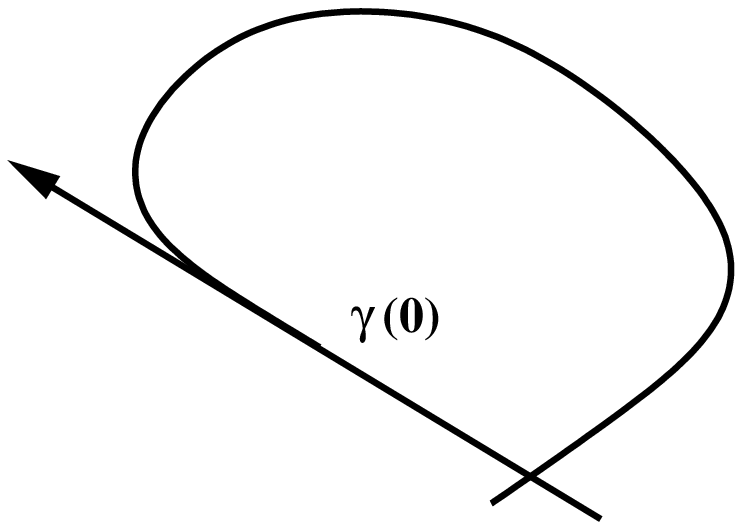}}
}
\caption{Non-convex curve (a) without \ref{globconvexcondn2} 
(b) without \ref{globconvexcondn3} [Liu and Traas '97]}
\end{figure}


\subsubsection{Convexity of planar curves in $R^{3}$}
In this subsection we extend the characterization of globally convexity of 
curve in $R^{2}$ to the characterization of global convexity of planar curve 
$\gamma(t)=[x(t),y(t),z(t)]$, $t\in [0,1]$ in $R^{3}$. 
In the remaining part of this chapter, except in section \ref{planarinflex},
we  use the operator $\times$ to denote the cross product of two
vectors, that is,
\begin{eqnarray}
A \times B &=&\left| 
\begin{array}{ccc}
i & j & k\\
A_{x} & A_{y} & A_{z}\\
B_{x} & B_{y} & B_{z}
\end{array}
\right|
\end{eqnarray}
where $A=[A_{x},A_{y},A_{z}]$, $B=[B_{x},B_{y},B_{z}] \in R^{3}$,
$i$, $j$ and $k$ are unit vectors along $x$, $y$ and $z$ axis
respectively. 
We observe that \cite[Liu and Traas, 1997]{liutraas} the 
operator $\times$ (in $R^{2}$) is mainly used to understand the direction in 
which the curve bends with respect to the orientation induced by $z-$axis.
We now explain our characterization for global convexity 
of planar curve in $R^{3}$ which solves our above purpose as follows.

It has been observed in \cite[Liu and Traas, 1997]{liutraas}, 
the convexity and concavity depends on its orientation, that is, direction 
in which is traversed. By inverting the orientation, a
convex curve turns into a concave curve and vice versa. 
Definition based on direction helps to distinguish between two
curves as convex and concave curves in most of the practical
situations. For example, in the case of two curves $\gamma(t)$ and
$Q(t)$ ($t \in [0,1]$) such that $\gamma(1)$ and $Q(1)$ lie on the
right side of $\gamma(0)$ and $Q(0)$ respectively.

\begin{figure}[h]
\centerline{ 
\includegraphics*{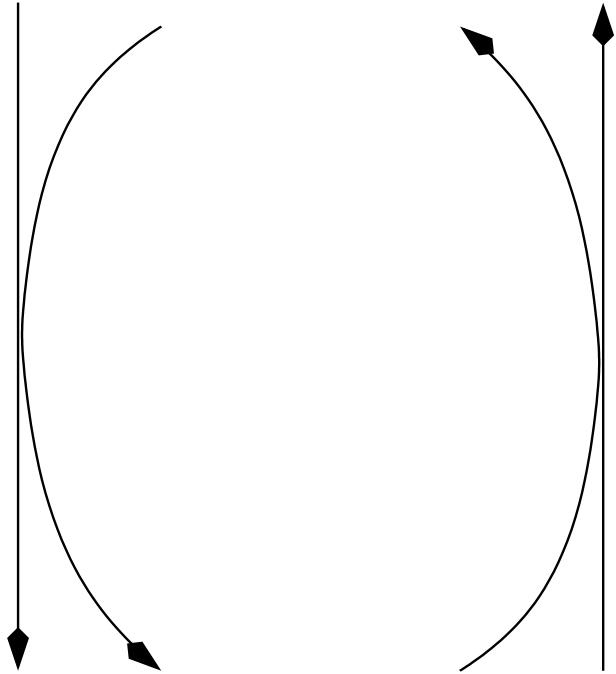} 
}
\caption{Convex curves with different orientations [Liu and Traas '97]} \label{orientconvex}
\end{figure}

But the direction of orientation itself gets inverted if the normal 
to the plane is taken to be the normal with its direction opposite
to the given normal. 
(Note that in (\ref{plancross}) "$\times$" can be interpreted as to denote
the dot product of the cross product between two vectors in $x-y$ plane with 
unit vector along $z$-axis (which is along normal dierction to the 
$x-y$ plane)).
In our case we need the convexity of the curve according to 
orientation induced by a specified normal vector as it requires that 
the spline curve to be convex in the same direction as the (data)
polygonal arc. The normal vector is specified as the normal to the
plane containing a pair of adjacent line segments in the data polygonal
arc.
(Therefore we do not consider the case of concavity.)

Thus we have the following definition for local and global convexity 
of a planar curve lying on a plane $\Pi$ in $R^{3}$ with respect to 
orientation of normal vector $N$ of the plane $\Pi$.
\begin{definition}\label{d9}
A $C^{2}$-continuous planar curve $\gamma(t)$, $t \in [0,1]$, is locally convex if and 
only if $(\gamma'(t) \times \gamma''(t)) \cdot N \geq 0 \mbox{, } t \in [0,1]$.
\end{definition}

\begin{definition}\label{d10}
A $C^{2}$-continuous planar curve $\gamma(t)$, $t \in [0,1]$, satisfying the condition
\begin{eqnarray}
\gamma(t) \neq \gamma(0) \mbox{ for } t \in (0,1)
\end{eqnarray}
is globally convex if and only if it satisfies following conditions.
\begin{enumerate}
\item $(\gamma'(t) \times \gamma''(t)) \cdot N  \geq  0$
\item $((\gamma(t)-\gamma(0)) \times \gamma'(t)) \cdot N  \geq  0 \mbox{, } t \in [0,1]$
\item $(\gamma'(0) \times (\gamma(t)-\gamma(0))) \cdot N  \geq  0 \mbox{, } t \in [0,1]$
\end{enumerate}
\end{definition}

Following remarks about defintions \ref{d9} and \ref{d10} are very significant.
\begin{remark}
Since $\gamma ' (t)$, $\gamma '' (t)$ and $(\gamma(t)-\gamma(0))$, 
$t \in [0,1]$ are parallel to the plane $\Pi$. Therefore sign of 
dot product of each cross product with $N$ actually represents the bending 
of the curve according to the orientation induced by $N$.
\end{remark}

\begin{remark}
Remark \ref{remarkequality} about equality holds for
the inequalities appearing in the definitions \ref{d9} and \ref{d10} holds.
\end{remark}
\subsection{Improvement in the condition of convexity preservation criteria}
Now using our lemma, stated below, we find that for almost all interpolating 
splines the condition 
$\omega (t) \cdot N >0$ in definition \ref{defsecondconvex}
is implied by global convexity of $P_{N^{\perp}} (\gamma (t))$ 
(and need not be stated separately) for almost all values of $t$. 
Thus we further simplify definition of convexity 
criteria. This lemma also helps to modify the definition for
inflection criteria to get a simpler definition in section \ref{inflexcrit}.

\begin{lemma}\label{lemsuper} 
Let $P_{N^{\perp}} (\gamma (t))$ be denoted as $\gamma_{N} (t)$,
$\omega (t) = \gamma ' (t) \times \gamma '' (t)$ and 
$\omega_{N} (t)= \gamma_{N} ' (t) \times \gamma_{N} '' (t)$. Then
$\omega_{N} (t) \cdot N = \omega (t) \cdot N$.
\end{lemma}

\begin{proof} 
A plane with normal vector $N$ is given by
$\displaystyle{\frac{(x,y,z) \cdot N +d}{\Vert N \Vert} =0}$, $d \in R$ .
We know that 
$\displaystyle{\gamma_{N}(t) = \gamma (t) + \frac{\gamma (t) \cdot N +d}{\Vert N \Vert^{2}} N}$. 
Therefore 
$\displaystyle{\gamma_{N}'(t) = \gamma'(t) + \frac{\gamma'(t) \cdot N +d}{\Vert N \Vert^{2}} N}$,\\  
$\displaystyle{\gamma_{N}''(t) = \gamma''(t) + \frac{\gamma''(t) \cdot N +d}{\Vert N \Vert^{2}} N}$. 
From the above we get\\
$\displaystyle{
\omega_{N} (t) = (\gamma ' (t) \times \gamma '' (t)) + 
(\gamma ' (t) \times \frac{\gamma '' (t) \cdot N +d}{\Vert N \Vert^{2}} N)+
(\frac{\gamma ' (t) \cdot N +d}{\Vert N \Vert^{2}} N \times \gamma '' (t))
}$.
Thus\\ 
$\displaystyle{
\omega_{N} (t) \cdot N = }$\\
$\displaystyle{
(\gamma ' (t) \times \gamma '' (t)) \cdot N+ 
(\gamma ' (t) \times \frac{\gamma '' (t) \cdot N +d}{\Vert N \Vert^{2}} N) \cdot N+
(\frac{\gamma ' (t) \cdot N +d}{\Vert N \Vert^{2}} N \times \gamma '' (t)) \cdot N 
= \omega (t) \cdot N
}$.
Hence proved.
\end{proof}

\noindent Now we first analyse the condition $\omega (t) \cdot N \geq 0$.
\begin{itemize}
\item For a large class of curves, including rational curves (except for 
straight lines) as a small subset, $\omega (t) \cdot N = 0$ holds for a 
set of values of $t \in [0,1]$ whose measure is zero.

\item $\omega (t) \cdot N = 0$ at $t=t_{1}$ if and only if 
$|\omega_{N} (t)| =0$ that is, $\gamma_{N} (t)$ behaves as straight line at 
$t=t_{1}$. Such behavior at $t=t_{1}$ and not in its neighbbourhood, has 
negligible effect on shape of the projected curve (along a viewpoint).

\item For some cases $\omega (t) \cdot N = 0$ at $t=t_{1}$ may also imply that
$\omega (t)$, which is binormal of the curve $\gamma (t)$, is perpendicular
to $N$ at $t=t_{1}$. That is, $N$ is parallel to the osculating plane
(plane on which curve lies locally) of $\gamma (t)$ at $t=t_{1}$. This might 
not be good unless such a torsion is required. Also, this may occur to great 
extent even for the case where $\epsilon > \omega (t) \cdot N >0$, for 
sufficiently small values of $\epsilon$.
However, possibility of such torsion, if undesired, can be controlled 
by other shape preservation criteria like torsion, coplanarity and collinearity
preservation criteria.
\end{itemize}
 
From the above analysis and lemma \ref{lemsuper} we get following two 
observations.
First, from definition \ref{d9} (for local convexity of a curve), we see that 
the definition \ref{deffirstconvex} requires the projection of curve on the 
plane perpendicular to $N_{i-1}$ and $N_{i}$ be only locally convex which one 
can observe from figure \ref{cnxitcondn2_neg} that the spline may not 
always serve the purpose of shape preservation. 

Second, according to the definition \ref{d10} for convexity of a planar curve in $R^{3}$
one of the condition that $\gamma_{N} (t)$ need to satisfy, to be globally convex,
is $\omega_{N} (t) \cdot N \geq 0$. 
From our lemma \ref{lemsuper}, global convexity of $P_{N^{\perp}} \gamma(t)$ 
implies $\omega (t) \cdot N>0$. Thus we see that the condition $\omega(t) \cdot N>0$ is redundant 
in the definition \ref{defsecondconvex}. 

We now state our results which will further simplify the conditions in the 
definition \ref{defsecondconvex}, as lemmas. The proofs of these lemmas are similar to that
of lemma \ref{lemsuper}

\begin{lemma}\label{lemsuperglob1} With the notation same as that in lemma \ref{lemsuper}
we have
\begin{eqnarray*}
((\gamma_{N^{\perp}}(t)-\gamma_{N^{\perp}}(0)) \times \gamma_{N^{\perp}}'(t)) \cdot N &=&
((\gamma (t)-\gamma (0)) \times \gamma '(t)) \cdot N 
\end{eqnarray*}
\end{lemma}

\begin{lemma}\label{lemsuperglob2} With the notation same as that in lemma \ref{lemsuper}
we have
\begin{eqnarray*}
(\gamma_{N^{\perp}}'(0) \times (\gamma_{N^{\perp}}(t)-\gamma_{N^{\perp}}(0))) \cdot N &=&
(\gamma (0) \times (\gamma (t)-\gamma(0))) \cdot N
\end{eqnarray*}
\end{lemma}

Using theorem \ref{projection} and lemmas \ref{lemsuper}, \ref{lemsuperglob1} and 
\ref{lemsuperglob2}, we have the following improved definition for convexity criteria for 
splines.
\begin{definition}\label{impconvex}
A spline curve $\gamma(t)$ interpolating data points satisfies convexity criteria if,
for $j=i-1, i$,
\begin{enumerate}
\item $\omega (t) \cdot N_{j}  \geq  0$,
\item $((\gamma (t)-\gamma (0)) \times \gamma '(t)) \cdot N_{j} \geq  0 $,
\item $(\gamma '(0) \times (\gamma (t)-\gamma (0))) \cdot N_{j} \geq  0 $,
\end{enumerate}
$t \in [t_{i-1},t_{i}]$, whenever $N_{i-1} \cdot N_{i} > 0$.
\end{definition}

\section{Inflection of a planar curves and polygonal arcs}\label{planarinflex}
\setcounter{equation}{0}
In \cite[Goodman, 1991]{goodman} authors have given definitions and 
conditions for the existence of inflections in a planar curve as follows. 
In this section for $A=(A_{1},A_{2})$, $B=(B_{1}, B_{2})$ in $R^{2}$ we write
\begin{eqnarray}
A \times B &=& A_{1} B_{2} - A_{2}B_{1}
\end{eqnarray}
For any sequence $a=(a_{1}, \hdots, a_{n})$ in $R^{n}$, we define 
$S(a)= S(a_{1}, \hdots, a_{n})$ to be the number of strict sign changes in the 
sequence. 

\begin{definition}\cite[Goodman, 1991]{goodman}\label{d11}
We say a polygonal arc $P_{0}P_{1} \hdots P_{n}$ for points
$P_{0}$, $P_{1}$, $\hdots$, $P_{n}$ in $R^{2}$ is regular
if the following hold
\begin{enumerate}
\item It turns through a total angle of magnitude at most $\pi$, that is, for 
some $V$ in $R^{2}$, $V \cdot (P_{i}-P_{i-1}) \geq 0$, $i=1, \hdots, n$. 
\item It does not turn through an angle of $\pi$ at any vertex, that is, for 
any $0 < i \leq j <n$ with $P_{i-1} \neq P_{i} = P_{i+1} \hdots =P_{j} \neq 
P_{j+1}$, $P_{i} - P_{i-1} \neq \lambda (P_{j+1}-P_{j})$ for any $\lambda < 0$.
\end{enumerate}
\end{definition}

\begin{definition}\cite[Goodman, 1991]{goodman}\label{d100}
For a regular polygonal arc
$P_{0} P_{1} \cdots P_{n}$ in $R^{2}$, with the condition $P_{i-1} 
\neq P_{i}$, $i=1, \hdots, n $ denoted by $\Gamma$ we define inflection count 
as 
\begin{eqnarray}
i(\Gamma) &=& S(V_{1}, \hdots, V_{n-1})
\end{eqnarray}
where
\begin{eqnarray}
V_{i} &=& (P_{i} - P_{i-1}) \times (P_{i+1}-P_{i}) \mbox{ }i=1, \hdots, n-1 \mbox{.}
\end{eqnarray}
\end{definition}


For any function $f:(a,b) \rightarrow R$ we define $S(f)$ to be the 
number of strict sign changes in $f(t)$, $a < t < b$, that is 
$S(f)=sup S(f(t_{1}), \hdots, f(t_{n}))$, where the supremum is taken over all 
sequences $a < t_{1} < \hdots < t_{n}=b$, for all $n$.
For a curve $\gamma (t) \in R^{2}$, $t \in [a, b]$ which is constant for $t$
in $[\alpha, \beta] \subset (a,b)$ but not on any larger interval, we define
\begin{eqnarray}
K(t)=
\left\{
\begin{array}{ll}
\frac{1}{2}\{ u(\alpha^{-}) \times u'(\alpha^{-}) + u(\beta^{-}) \times u'(\beta^{-})\}, & \mbox{ if } u(\alpha^{-}) = u(\beta^{+}), \\
u(\alpha^{-}) \times u(\beta^{+}), & \mbox{ if } u(\alpha^{-}) \neq u(\beta^{+}) \mbox{.}
\end{array} 
\right.
\end{eqnarray} 

\begin{definition}
Inflection count for a curve $\gamma(t) \in R^{2}$, $t \in [a, b]$, is defined 
as $i(\gamma):= S(K)$. 
It is actually the number of times the curve changes from turning in a 
clockwise direction to turning in an anti-clockwise direction, or vice-versa. 
\end{definition}

With the above definitions regarding inflection count we state the following relation 
between inflection count of B-spline curve and its control polygon from \cite[Goodman, 1991]{goodman}.
\begin{theorem}\cite[Goodman, 1991]{goodman}\label{planbspl}
Suppose
\begin{eqnarray}
\gamma (t) &=& \sum_{i=-n}^{m-1} A_{i} N_{i} (t) 
\mbox{, } t_{0} \leq t \leq t_{m}\mbox{, } 
\end{eqnarray}
and $\Gamma$ denotes the polygonal arc $A_{-n}A_{-n+1} \cdots A_{m-1}$. If 
$A_{i-n} \cdots A_{i}$ is regular for $i=0, \hdots, m-1$, then
\begin{eqnarray}
i(\gamma) & \leq & i(\Gamma) \mbox{.}
\end{eqnarray}
where $N_{i}$ denote B-spline basis function. 
\end{theorem}

We now state the relation from \cite[Goodman 1991]{goodman} between inflection count of B\'{e}zier
curve and its control polygon which follows as a corollary to the above theorem.

\begin{corollary}\cite[Goodman, 1991]{goodman}
If
\begin{eqnarray}
\gamma (t) &=& \sum_{i=1}^{n} A_{i} (^{n}_{i}) t^{i} (1-t)^{n-i} \mbox{, }
0 \leq t \leq 1 \mbox{, }
\end{eqnarray}
and the polygonal arc $\Gamma = A_{0} A_{1} \cdots A_{n}$ is regular, then
\begin{eqnarray}
i(\gamma) \leq i(\Gamma) \mbox{.}
\end{eqnarray} 
\end{corollary}

Now we state a theorem from \cite[Goodman, 1991]{goodman} 
which says that the above result does not hold if the control polygon is 
not regular. 

\begin{theorem}\cite[Goodman 1991]{goodman}\label{contraplancubic}
Let $\gamma (t)$ be cubic B\'{e}zier curve given by
\begin{eqnarray}
\gamma (t) &=& A (1-t)^{3} +3 B t(1-t)^{2} + 3 C t^{2} (1-t) + D t^{3} 
\label{plancubic}
\end{eqnarray}
where $A$, $B$, $C$, $D \in R^{2}$ are control points and $\Gamma$ denote
the polygonal arc ABCD.
Suppose $\Gamma$ turns through an angle of magnitude $> \pi$ and let $P$
be the point of intersection of the line through $A$ and $B$ and the line 
through $C$ and $D$. Then
\begin{eqnarray}
i(\gamma) &=& 
\left\{
\begin{array}{l}
0 \mbox{, if} |B-A||C-D|/|B-P||C-P| \leq 4,\\
2 \mbox{, if} |B-A||C-D|/|B-P||C-P| > 4.
\end{array}
\right.
\end{eqnarray} 
\end{theorem}

\begin{figure}[h]
\centerline{ 
\includegraphics*[width=340pt,height=230pt]{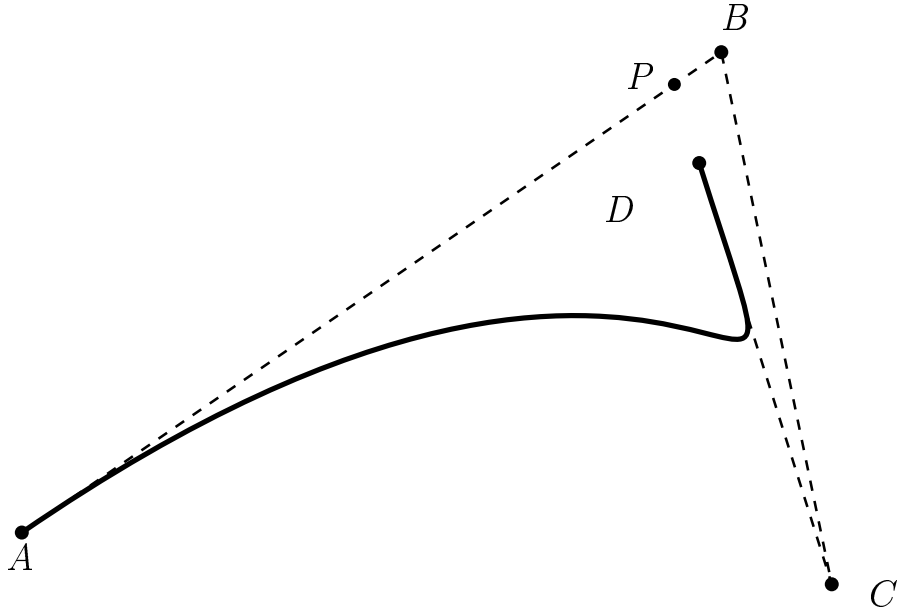} 
}
\caption{Cubic B\'{e}zier curve with two inflection points whose control polygon has no inflection} \label{cubicinflec_curve}
\end{figure}

\begin{figure}[h]
\centerline{ 
\includegraphics*[width=287pt,height=182pt]{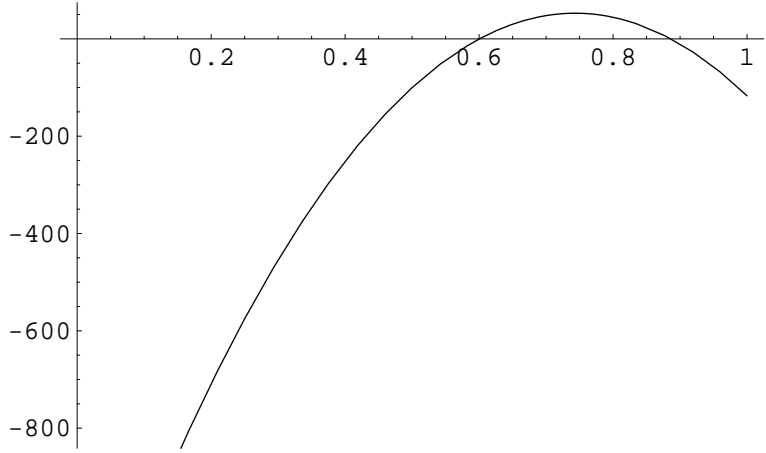} 
}
\caption{Curvature plot of the curve in figure \ref{cubicinflec_curve}}\label{cubicinflec_cture}
\end{figure}

\section{Inflections of curves and polygonal arcs in $R^{3}$}\label{spaceinflex}
\setcounter{equation}{0}
Let $\gamma : [a,b] \rightarrow R^{3}$ be a curve in $R^{3}$.
We also resume the meaning of operator $\times$ as cross 
product of two vectors in $R^{3}$ as defined in subsection 4.1.3.

For any $w$ in $S_{1}=\{ v \in R^{3} : |v| = 1 \}$ we shall denote by $P_{w}$
the orthogonal projection from $R^{3}$ onto the 2-dimensional subspace 
orthogonal to $w$, that is, $P_{w} x = x - (x \cdot w) w$.

\begin{definition}\cite[Goodman, 1991]{goodman}\label{d23}
Inflection count $I(\gamma)$ of the (spatial) curve $\gamma$ to be the maximum
number of inflections that be seen in $\gamma$ by observing from any direction,
that is,
\begin{eqnarray}
I(\gamma) &=& \mbox{ess sup}\{i(P_{w}\gamma) : w \in S_{1}\}\mbox{.}
\end{eqnarray}
\end{definition}

We suppose that $\gamma$ is continuous with piecewise $C^{1}$ unit tangent 
vector $u(t) = \gamma'(t)/|\gamma'(t)|$. As before suppose $\gamma (t)$ is 
constant for $t$ in $[\alpha, \beta] \subset (a,b)$ but not on any larger
interval and we define for $\alpha \leq t \leq \beta$,
\begin{eqnarray}
K(t) &=& 
\left\{
\begin{array}{l}
\frac{1}{2}\{u(\alpha^{-}) \times u'(\alpha^{-}) + u(\beta^{+} \times u'(\beta^{+}))\mbox{, if } u(\alpha^{-}) = u(\beta^{+})\mbox{,}\\
u(\alpha^{-}) \times u(\beta^{+}) \mbox{, if } u(\alpha^{-}) \neq u(\beta^{+})\mbox{.}
\end{array}
\right.
\end{eqnarray}

\begin{theorem}\cite[Goodman, 1991]{goodman}\label{spaceinflexdefn}
Suppose $\gamma : [a,b] \rightarrow R^{3}$ is continuous with piecewise $C^{1}$
unit tangent vector $u(t) = \frac{\gamma' (t)}{|\gamma'(t)|}$. Then 
\begin{eqnarray}
I(\gamma) &=& \mbox{sup } \{ S(w.K) : w \in S_{1}\}
\end{eqnarray}
\end{theorem}

\begin{corollary}\label{spaceplaneinflex}
If the curve $\gamma$, as in theorem (\ref{spaceinflexdefn}), lies in a plane
with a normal $n$, then
\begin{eqnarray}
I(\gamma) &=& S(n \cdot K) \mbox{.}
\end{eqnarray}
\end{corollary}

Theorem (\ref{spaceinflexdefn}), also implies the following definition for
polygonal arc.

\begin{definition}\cite[Goodman, 1991]{goodman}
For a polygonal arc $P_{0} P_{1} \cdots P_{n}$ and $P_{i-1} \neq P_{i}$,
$i=1,...,n$, denoted by $\Gamma$ then its inflection count is
\begin{eqnarray}
I(\Gamma) &=& \mbox{sup } \{ S(w \cdot V_{1}, \hdots, w \cdot V_{n}) : w \in S_{1} \}
\mbox{.}
\end{eqnarray}
\end{definition}

\section{Inflection preservation criteria for interpolating splines}\label{inflexcrit}
\setcounter{equation}{0}
Let ${\bf x}_{i} \in R^{3}$, $i=0,...,n$ be $n+1$ data points and $\mathcal{D}$ be 
the polyline joining the points with each side being 
${\bf L}_{i}= {\bf x}_{i}-{\bf x}_{i-1}$, $i=1, ..., n$. 
Let $N_{i} = L_{i-1} \times L_{i}$. In discrete differential geometry 
[Sauer, 1970] discrete binormal is defined as 
$\displaystyle{ \frac{N_{i}}{|N_{i}|} }$. For a curve $P(t)=[x(t),y(t),z(t)]$,
$t \in [0,1]$ in $R^{3}$ let $\omega(t)= P'(t) \times P''(t)$. 

In \cite[Costantini, Goodman, Manni, 2000]{constantinigoodman}, 
\cite[Costantini Cravero Manni, 2002]{constantiniisabella} 
\cite[Manni, Pelosi, 2004]{mannipelosi} we have the following definition
\begin{definition}\cite[Costantini, Goodman, Manni, 2000]{constantinigoodman}
A curve $\gamma (t)$ interpolating data points satisfies inflection criteria if is 
satisfy the condition that, if $N_{i-1} \cdot N_{i} < 0$, then 
$(\omega(t_{l}) \cdot N_{m})(N_{l} \cdot N_{m}) > 0$, $l,m=i-1,i$, and 
$ \omega (t) \cdot N_{j}$, has precisely one sign change in 
$t \in [t_{i-1},t_{i}]$, $m = i-1 \mbox{, }i$.
\end{definition}

The above definition requires the projection of curve on the plane perpendicular
to $N_{i-1}$ and $N_{i}$ have only one inflection point. 
Thus it takes care about inflection preservation along two viewpoints only.

In \cite[Goodman  and Ong, 1997]{goodmanong}, 
\cite[Kong and Ong, 2002]{kong}
we have the following definition
\begin{definition}\cite[Goodman  and Ong, 1997]{goodmanong}\label{defninflexcrit}
Inflection preservation criteria is defined by the condition that
if $N_{i-1} \cdot N_{i} < 0$, 
\begin{enumerate}
\item $\omega(t_{i-1}) \cdot N_{i-1} > 0$, 
$\omega(t_{i}) \cdot N_{i} > 0$ and 

\item for all $N = \lambda N_{i-1} + \mu N_{i}$, where $\lambda \mu \leq 0$, 
($\lambda, \mu$)$\neq$($0,0$), $\omega (t) \cdot N$ has precisely one sign 
change in $[t_{i-1}, t_{i}]$.
\end{enumerate}
\end{definition}
One can see that the above definition takes care about inflection preservation 
along all the viewpoints between $N_{i-1}$ and $N_{i}$ along the plane 
containing the two normals.

\begin{figure}[h]
\centerline{ 
\includegraphics{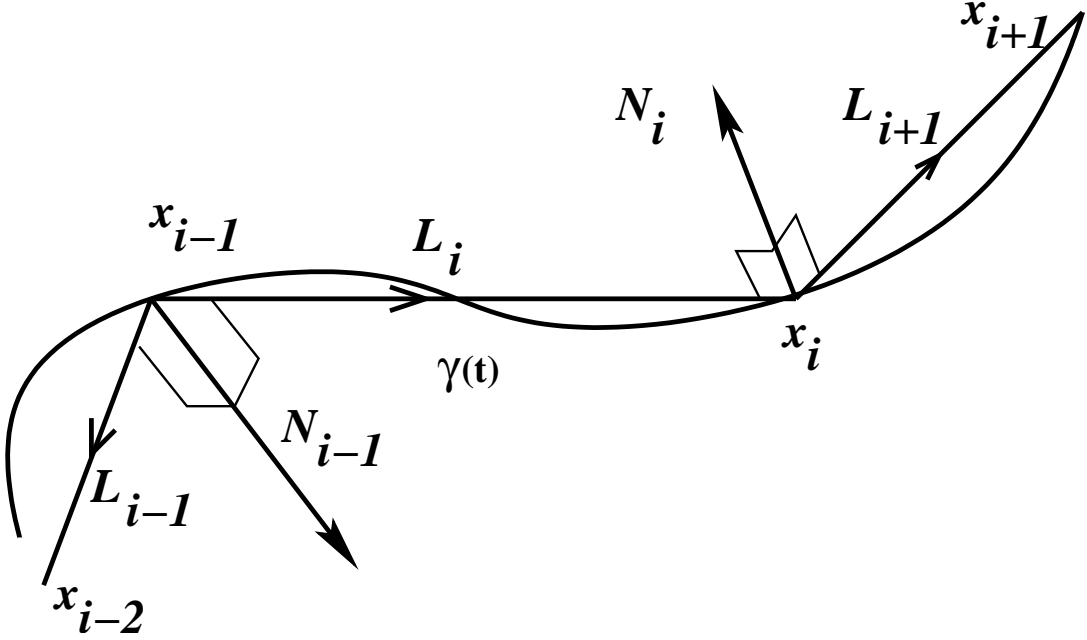} 
}
\caption{Data point with $N_{i-1} \cdot N_{i} <0$ requiring inflection
preservation by $i^{th}$ curve segment}
\end{figure}

We now observe that our lemma \ref{lemsuper} acts as the connection between 
the analysis of inflection of curves and polygonal arcs and the definition
\ref{defninflexcrit}. The two conditions basically states that the projection
of the curve $\gamma(t)$ on the plane with normal vector $N$, 
$P_{N^{\perp}} (\gamma (t))$, $t \in [t_{i-1}, t_{i}]$ should have only
one inflection point. 

We now state the condition under which a curve segment of $\gamma (t)$ with 
B\'{e}zier representation satisfies the inflection criteria.
Let a B\'{e}zier curve $\gamma (t)$ be 
\begin{eqnarray}
\gamma (t) &=& \sum_{i=1}^{n} A_{i} (^{n}_{i}) t^{i} (1-t)^{n-i} \mbox{, }
0 \leq t \leq 1 \mbox{, }
\end{eqnarray}
$A_{i} \in R^{3}$, the polygonal arc $\Gamma = A_{0} A_{1} \cdots A_{n}$, 
$V_{i}=(A_{i}-A_{i-1}) \times (A_{i+1}-A_{i})$. 

If $A_{i}$, $i=0,...,n$ are such that
\begin{enumerate}
\item $N_{1} \cdot V_{1} <0$, $N_{1} \cdot V_{n-1} >0$ and $P_{N_{1}^{\perp}}$ 
have only one inflection 
\item $N_{2} \cdot V_{1} >0$, $N_{2} \cdot V_{n-1} <0$ and $P_{N_{1}^{\perp}}$ 
have only one inflection,
\item $(N_{1} \cdot V_{i})(N_{2} \cdot V_{i}) < 0$, $i=1,...,n-1$. 
\end{enumerate}
and we have two scalars $\lambda$, $\mu$ such that $\lambda \mu \leq 0$, 
$(\lambda, \mu) \neq (0,0)$, $N = \lambda N_{i} + \mu N_{i+1}$ then we have 
$V_{i} \cdot N =\lambda V_{i} \cdot N_{i} +\mu V_{i} \cdot N_{i+1}$ and 
thus $P_{N^{\perp}} \Gamma$ has only one inflection point.

Thus we see that if the $i^{th}$ curve segment of interpolating spline 
$\gamma_{i} (t)$ is a cubic curve and satisfies the first condition of the 
definition (\ref{defninflexcrit}) then it also satisfies the second condition.


\section{Difficulties in the construction of convexity and inflection preserving 
splines}\label{negativeinflex}
\setcounter{equation}{0}
We observe that results on inflection counts, apart from affecting the analysis
of inflection criteria of splines, have significant effect on the analysis for 
convexity preservation criteria of splines. Convexity preservation criteria requires that under certain
conditions projection of a curve segment $\gamma (t)$ on planes with a specified 
normal $N_{c}$ should be convex. Also the condition that inflection count of curve 
$\gamma (t)$ is greater than 1, that is, $I(\gamma) \geq 1$ says that there exist 
a vector $N_{s}$, such that, projection of curve on planes with normal vector $N_{s}$ has
inflection points greater than 1 and hence is not convex. Thus we see that if 
$N_{c}=N_{s}$, then $\gamma (t)$ fails to satisfy the convexity criteria.

\noindent 
Among the results stated below some of them are stated in 
\cite[Goodman, 1991]{goodman} as corollaries
we state them as theorems because of their relevance to us.
 
\begin{theorem}\cite[Goodman, 1991]{goodman}\label{spaceplaneinflex_neg}
If $\gamma$ is a curve, as in theorem (\ref{spaceinflexdefn}), which is not 
planar, then $I(\gamma) \geq 1$. 
\end{theorem}

\noindent Using the theorem (\ref{contraplancubic}) we have following results for 
cubic B\'{e}zier curves.
\begin{theorem}\cite[Goodman, 1991]{goodman}
If $\gamma$ is a cubic polynomial curve which is not planar, $I(\gamma)=2$
\end{theorem}

We have seen in previous sections that convexity and inflection counts
B\'{e}zier and B-spline curve are related to the convexity and inflection
counts of their control polygons. We state few results from 
\cite[Goodman, 1991]{goodman} for inflection count of polygonal arcs. 

\begin{theorem}\cite[Goodman, 1991]{goodman}
If $P_{0}, ..., P_{n}$ are not coplanar, then $I(\Gamma)=1$ if and only if
$V_{1}, ..., V_{n-1}$ lie in order in a plane sector sub-tending an angle 
$\leq \pi$.
\end{theorem}

\begin{theorem}\cite[Goodman, 1991]{goodman}\label{contraconvexcubiccontrol}
If $n=3$ and $P_{0}, ..., P_{3}$ are not coplanar, then $I(\Gamma)=1$.
\end{theorem}

\begin{remark}
The above theorems negates the general perception about the inflection 
counts and convexity of curves and polygonal arcs.
\end{remark}

\begin{remark}
The proofs provided in \cite[Goodman, 1991]{goodman} for 
the theorems stated above are constructive, that is, plane on which 
projection of the curves have inflections points are explicitly constructed. 
\end{remark}

We now illustrate the difficulties in constructing a convexity and inflection 
preserving curve (as indicated in the theorems above) in the following examples. 

\begin{example}
Let 
${\bf x}_{i-2}=(-3,-3,-0.5)$, ${\bf x}_{i-1}=(0,0,0)$, 
${\bf x}_{i}=(0,0,5)$, ${\bf x}_{i+1}=(2,-4,5.5)$. The 
normals at ${\bf x}_{i-1}$ and ${\bf x}_{i}$ are 
${\bf N}_{i-1}=(1.5, -1.5, 0)$ and ${\bf N}_{i} = (2., 1., 0)$ respectively.
Since $N_{i-1} \cdot N_{i} =1.5 >0$ we require that the curve between 
${\bf x}_{i-1}$ and ${\bf x}_{i}$ satisfy convexity preservation criteria.

Data polygonal arc $D$ formed by 
${\bf x}_{i-2}$, ${\bf x}_{i-1}$, ${\bf x}_{i}$ and ${\bf x}_{i+1}$ along with
normals ${\bf N}_{i-1}$ and ${\bf N}_{i}$ as thick lines is shown in figure 
\ref{view0}. The figures \ref{view1}, \ref{view2} and \ref{view3} 
show the curve between ${\bf x}_{i-1}$ and ${\bf x}_{i}$ with data polygonal arc $D$ 
along the viewpoints ${\bf N}_{i-1}$ and ${\bf N}_{i}$ and $V_{1}= (-10,0,1)$ 
(that is projection in the plane with normals ${\bf N}_{i-1}$, ${\bf N}_{i}$
and ${\bf V}_{1}$). We observe that though the curve satisfies the conditions of 
convexity preservation criteria along the viewpoint ${\bf V}_{1}$, it fails to do 
the same for both the viewpoints ${\bf N}_{i-1}$ and ${\bf N}_{i}$. Thus we see that 
it is relatively difficult to construct a curve satisfying the convexity preservation 
criteria using a graphical interface.

\begin{figure}
  \centering
    \subfigure[Data polygon arc with normals as thick lines]{\label{view0}
\includegraphics{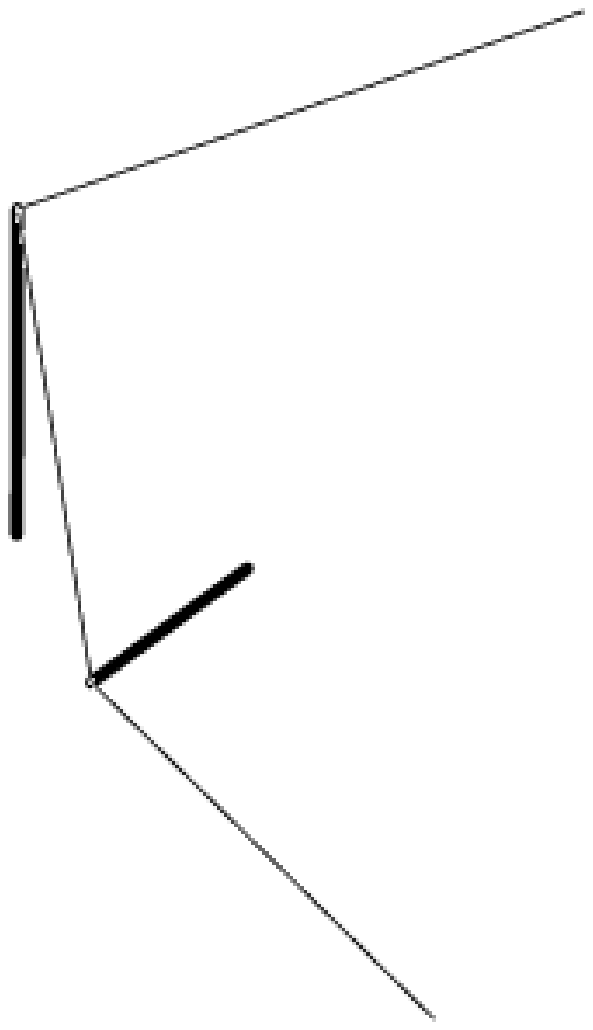}
}
\hspace{1cm}
    \subfigure[Viewpoint ${\bf N}_{i-1}$]{\label{view1}
\includegraphics{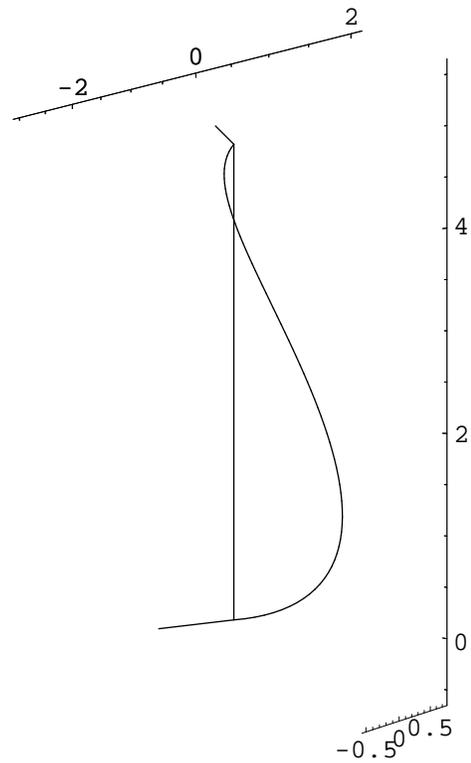}
}\\
    \subfigure[Viewpoint ${\bf N}_{i-1}$]{\label{view2}
\includegraphics{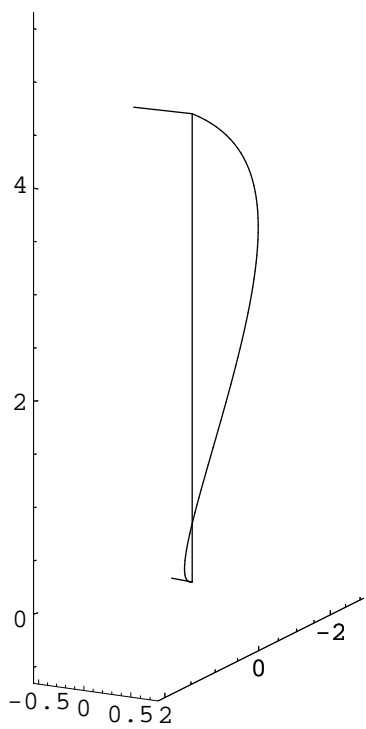}
}
\hspace{1cm}
    \subfigure[Viewpoint ${\bf V}_{1}= (-10,0,1)$]{\label{view3}
\includegraphics{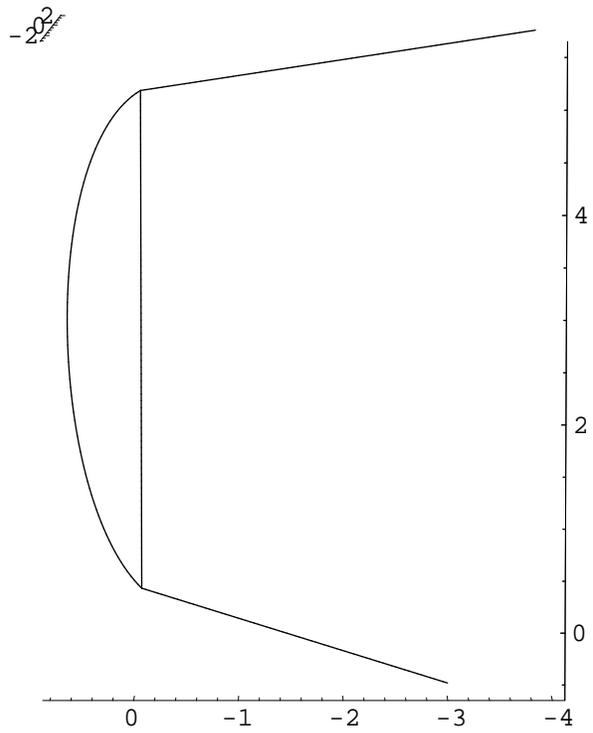}
}
\caption{Data polygonal arc with curve between ${\bf x}_{i-1}$ 
and ${\bf x}_{i}$ along different viewpoints}\label{viewpointex1}
\end{figure}

The mathematica code for the generation of figures 
\ref{view0}, \ref{view1}, \ref{view2} and \ref{view2}    
is as below:\\
{\rm \small (***************************************************************)\\
m1=\{3,1,0.5\};m2=\{2,-1,0.5\};\\
A=\{0,0,0\}; B=\{0,0,5\}; Am=A+m1; Bm=B-m2;\\
P=((1-t)\^{ }3)*A + (3*(1-t)\^{ }2*t)*Am + (3*(1-t)*t\^{ }2)*Bm + (t\^{ }3)*B\\
L0=\{-3,-3,-0.5\}; L1=B-A; L2=\{2,-4,0.5\}; F=B+L2;\\
N1=(1/10)*Cross[(-1)*L0,B]
N2=(1/10)*Cross[B,L2]
DT=N1.N2\\
P0=A+t*L0;
P1=A+t*B;
P2=B+t*L2;\\
(* L0, A, B,F are data points *)\\
Poly=Show[Graphics3D[\{Line[\{L0,A,B,F\}], \{Thickness[0.010],
            Line[\{A,N1\}]\}, \{Thickness[0.010], Line[\{B,N2\}]\}\}],
      ViewPoint$->$\{-1,0,1\}, Boxed$->$False];\\
Poly0=Show[Graphics3D[Line[\{L0,A,B,F\}]], ViewPoint$->$\{-10,0,1\},
      Boxed$->$False];\\
ln0=ParametricPlot3D[P0,\{t,0,1\}, Boxed$->$False];\\
ln1=ParametricPlot3D[P1,\{t,0,1\}, Boxed$->$False];\\
ln2=ParametricPlot3D[P2,\{t,0,1\}, Boxed$->$False];\\
Crv0=ParametricPlot3D[P,\{t,0,1\}, ViewPoint$->$N1, Boxed$->$False];\\
CrvPoly0=Show[\{Crv1,Poly0\}, ViewPoint$->$N1, Boxed$->$False];\\
Crv1=ParametricPlot3D[P, \{t,0,1\}, ViewPoint$->$N2, Boxed$->$False];\\
CrvPoly1=Show[\{Crv1,Poly0\}, ViewPoint$->$N2, Boxed$->$False];\\
Crv2=ParametricPlot3D[P,\{t,0,1\}, ViewPoint$->$\{-10,0,1\}, Boxed$->$False];\\
CrvPoly2=Show[ln0,ln1,ln2,Crv2];\\
Display["d:$\backslash$Gautam\_Viewpoint$\backslash$view0.png", Poly, "PNG"];\\
Display["d:$\backslash$Gautam\_Viewpoint$\backslash$view1.png", CrvPoly0, "PNG"];\\
Display["d:$\backslash$Gautam\_Viewpoint$\backslash$view2.png", CrvPoly1, "PNG"];\\
Display["d:$\backslash$Gautam\_Viewpoint$\backslash$view3.png", CrvPoly2, "PNG"];\\
(***********************************************************)
}


\end{example}

\begin{example}
Here we have ${\bf x}_{i-2}=(-3,-3,-0.5)$, ${\bf x}_{i-1}=(0,0,0)$, 
${\bf x}_{i}=(0,0,10)$, ${\bf x}_{i+1}=(2,-4,10.5)$. The 
normals at ${\bf x}_{i-1}$ and ${\bf x}_{i}$ are ${\bf N}_{i-1}=(30.,-30.,0)$ and
${\bf N}_{i} = (40., 20., 0)$ respectively. 
Since $N_{i-1} \cdot N_{i} =600 >0$ we require that the curve between 
${\bf x}_{i-1}$ and ${\bf x}_{i}$ satisfy convexity preservation criteria.

The figures \ref{newview1}, \ref{newview2}, \ref{newview3}, \ref{newview4}
\ref{newview5} and \ref{newview6} 
show the curve between ${\bf x}_{i-1}$ and ${\bf x}_{i}$ with data polygonal arc 
along the viewpoints ${\bf N}_{i-1}$, ${\bf N}_{i}$, 
${\bf V}_{1} = (7.458, -1.863, -3.506)$,  
${\bf V}_{2} = (-17.458, 1.863, 23.506)$, 
${\bf V}_{3} = (-7.458, 6.863, 33.506)$ and 
${\bf V}_{4} = (-7.458, 30.863, 33.506)$ respectively. 
Observe that the curve along viewpoints ${\bf N}_{i-1}$, ${\bf N}_{i}$,
${\bf V}_{1}$, ${\bf V}_{2}$ is convex where as along viewpoints
${\bf V}_{3}$ and ${\bf V}_{4}$ is not convex. 
This means that if values of ${\bf x}_{i-2}$ or ${\bf x}_{i+1}$ are
altered such that ${\bf N}_{i-1}$ or ${\bf N}_{i}$ changes to 
${\bf V}_{3}$ or ${\bf V}_{4}$ with ${\bf N}_{i-1} \cdot {\bf N}_{i}>0$, 
then the curve doesn't satisfy convexity 
preservation criteria with respect to changed data polygonal arc.

In addition to the above observation in example 
we also note that (figure \ref{newview5})
along the viewpoint ${\bf V}_{3} = (-7.458, 6.863, 33.506)$ 
the curve has two inflections as shown in figure \ref{curvature5}
(the curvature of the projection of curve along the viewpoint 
${\bf V}_{5}$ changes its sign twice). Thus if ${\bf x}_{i-2}$ 
or ${\bf x}_{i+1}$ are such that ${\bf N}_{i-1}$ or ${\bf N}_{i}$ 
respectively are equal to ${\bf V}_{3}$, with 
${\bf N}_{i-1} \cdot {\bf N}_{i} < 0$ then 
the curve would not satisfy inflection preservation criteria.

The mathematica code for the generation of figures 
\ref{newview1}, \ref{newview2}, \ref{newview3}, \ref{newview4}, 
\ref{newview5}, \ref{newview6} is as below:\\
{\rm \small (******************************************************)\\
m1x=1;m1y=1;m1z=0.5;\\
m2x=2;m2y=-3;m2z=0.5;\\
Ax=0;Ay=0;Az=0;\\
Dx=0;Dy=0;Dz=10;\\
Bx=Ax+m1x; By=Ay+m1y;Bz=Az+m1z;(*(1,1,0.5)*)\\
Cx=Dx-m2x;Cy=Dy-m2y; Cz=Dz-m2z;(*(-2,3,9.5)*)\\
Px=Ax*(1-t)\^{ } 3 + Bx*3*(1-t)\^{ } 2*t + Cx*3*(1-t)*t\^{ } 2 + Dx*t\^{ } 3;\\
Py=Ay*(1-t)\^{ } 3 + By*3*(1-t)\^{ } 2*t + Cy*3*(1-t)*t\^{ } 2 + Dy*t\^{ } 3;\\
Pz=Az*(1-t)\^{ } 3 + Bz*3*(1-t)\^{ } 2*t + Cz*3*(1-t)*t\^{ } 2 + Dz*t\^{ } 3;\\
L0x=-3;L0y=-3;L0z=-0.5;(*xi-2=(-3,-3,-0.5)*)\\
L1x=Dx-Ax;L1y=Dy-Ay;L1z=Dz-Az;\\
L2x=2;L2y=-4;L2z=0.5;\\
P0x=Ax+L0x*t;P0y=Ay+L0y*t;P0z=Az+L0z*t;\\
P1x=Ax+Dx*t;P1y=Ay+Dy*t;P1z=Az+Dz*t;\\
P2x=Dx+L2x*t;P2y=Dy+L2y*t;P2z=Dz+L2z*t;(*xi+1=(2,-4,10.5)*)\\
N1=Cross[\{-L0x,-L0y,-L0z\},\{0,0,10\}]
N2=Cross[\{0,0,10\},\{L2x,L2y,10+L2z\}]
A=N1.N2\\
CrvPoly1=ParametricPlot3D[\{\{Px,Py,Pz\}, \{P0x,P0y,P0z\}, \{P1x,P1y,P1z\},
\{P2x,P2y,P2z\}\}, \{t,0,1\}, ViewPoint$->$N1, Boxed$->$False];\\
CrvPoly2=ParametricPlot3D[\{\{Px,Py,Pz\}, \{P0x,P0y,P0z\}, \{P1x,P1y,P1z\},
\{P2x,P2y,P2z\}\}, \{t,0,1\}, ViewPoint$->$N2, Boxed$->$False];\\
CrvPoly3=ParametricPlot3D[\{\{Px,Py,Pz\}, \{P0x,P0y,P0z\}, \{P1x,P1y,P1z\},
\{P2x,P2y,P2z\}\}, \{t,0,1\}, ViewPoint$->$\{7.458, -1.863, -3.506\},
      Boxed$->$False];\\
CrvPoly4=ParametricPlot3D[\{\{Px,Py,Pz\}, \{P0x,P0y,P0z\}, \{P1x,P1y,P1z\},
\{P2x,P2y,P2z\}\}, \{t,0,1\}, ViewPoint$->$\{-17.458, 1.863, 23.506\},
      Boxed$->$False];\\
CrvPoly5=ParametricPlot3D[\{\{Px,Py,Pz\}, \{P0x,P0y,P0z\}, \{P1x,P1y,P1z\},
\{P2x,P2y,P2z\}\}, \{t,0,1\}, ViewPoint$->$\{-7.458, 6.863, 33.506\},
      Boxed$->$False];\\
CrvPoly6=ParametricPlot3D[\{\{Px,Py,Pz\}, \{P0x,P0y,P0z\}, \{P1x,P1y,P1z\},
\{P2x,P2y,P2z\}\}, \{t,0,1\}, ViewPoin$t->$\{-7.458, 30.863, 33.506\},
      Boxed$->$False];\\
Display["newview1.png",CrvPoly1,"PNG"]\\
Display["newview2.png",CrvPoly2,"PNG"]\\
Display["newview3.png",CrvPoly3,"PNG"]\\
Display["newview4.png",CrvPoly4,"PNG"]\\
Display["newview5.png",CrvPoly5,"PNG"]\\
Display["newview6.png",CrvPoly6,"PNG"]\\
%
%
(*******************************************************)
} 
\begin{figure}
  \centering
    \subfigure[Viewpoint ${\bf N}_{i-1}$]{\label{newview1}
\includegraphics{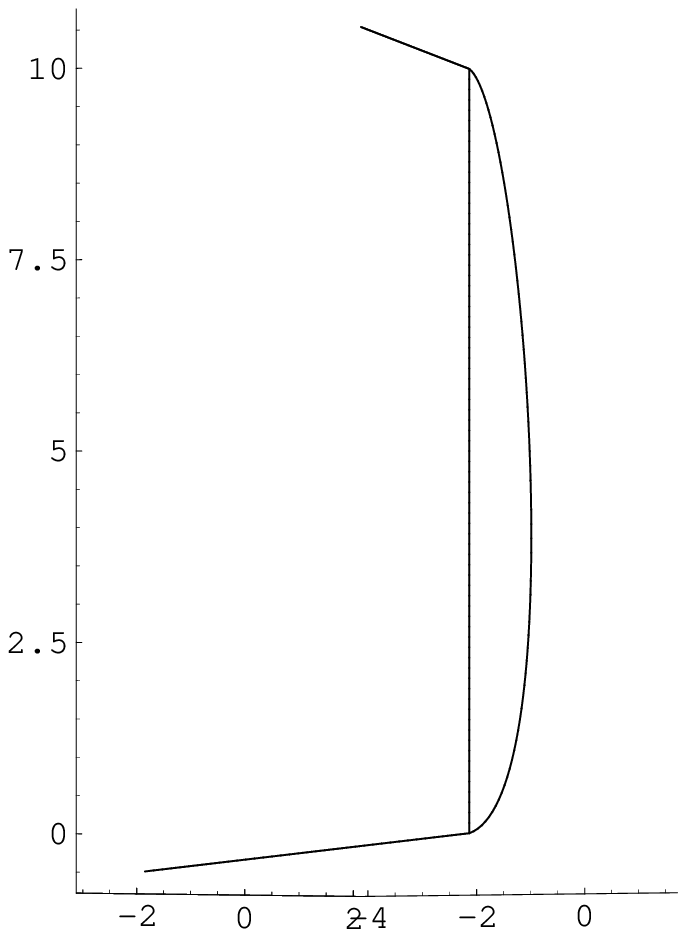}
}
\hspace{1cm}
    \subfigure[Viewpoint ${\bf N}_{i}$]{\label{newview2}
\includegraphics{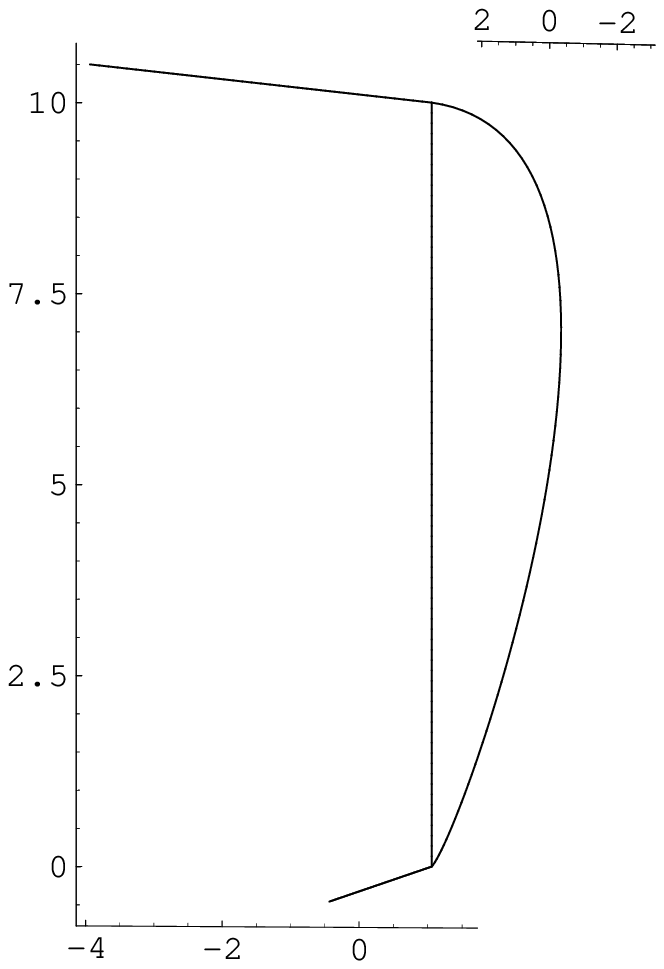}
}\\
    \subfigure[Viewpoint ${\bf V}_{1}=(7.458, -1.863, -3.506)$]{\label{newview3} 
\includegraphics{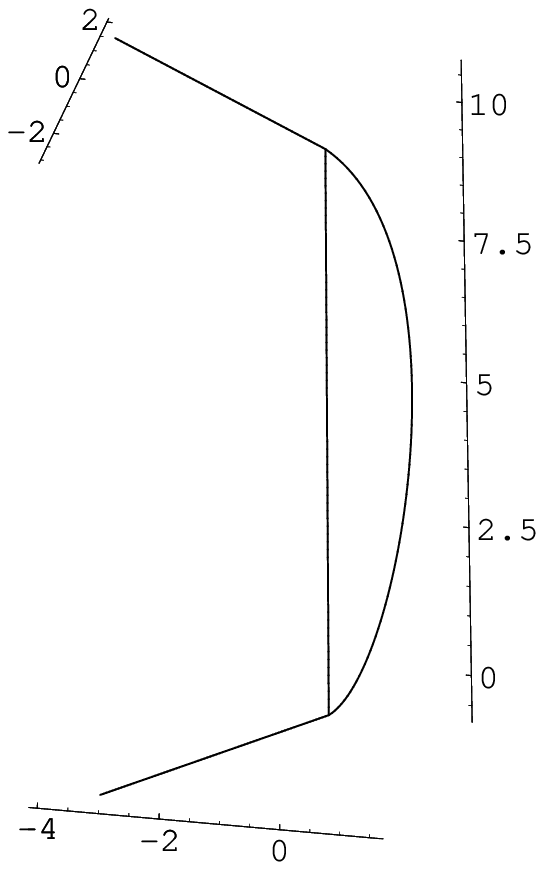}
}
\hspace{1cm}
    \subfigure[${\bf V}_{2}=(-17.458, 1.863, 23.506)$]{\label{newview4}
\includegraphics{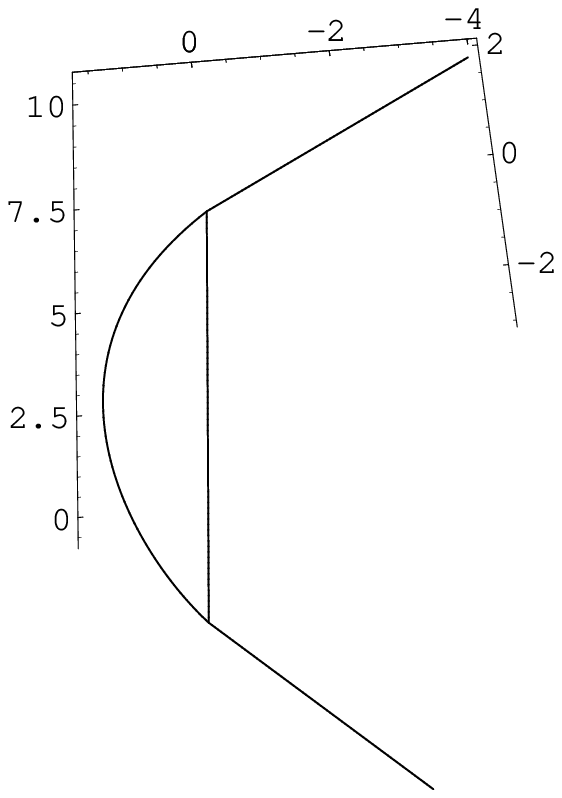}
}
\end{figure}
\begin{figure}
    \subfigure[Viewpoint ${\bf V}_{3}=(-7.458, 6.863, 33.506)$]{\label{newview5}
\includegraphics[scale=0.8]{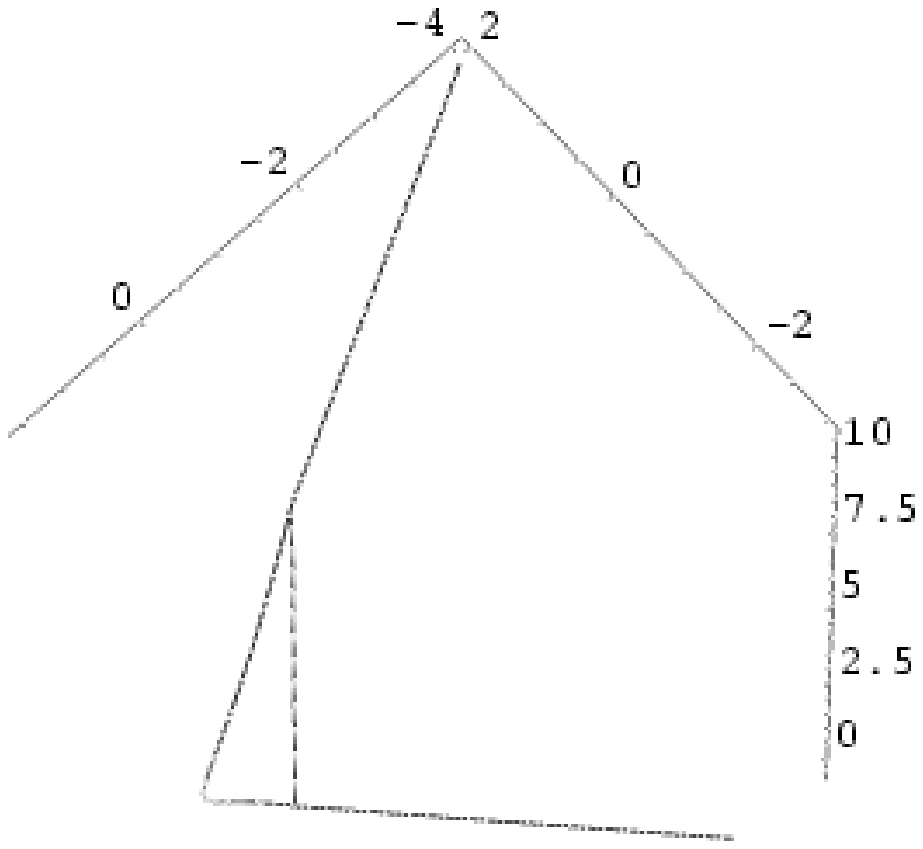}
}
\hspace{1cm}
    \subfigure[Viewpoint ${\bf V}_{4}=(-7.458, 30.863, 33.506)$]{\label{newview6}
\includegraphics[scale=0.8]{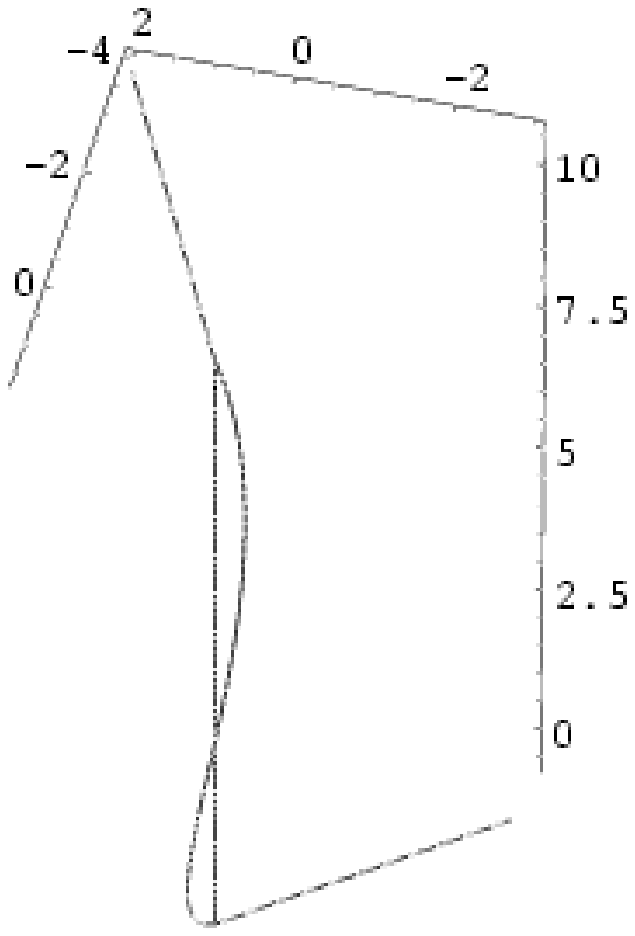}
}
\caption{Data polygonal arc with curve between ${\bf x}_{i-1}$ 
and ${\bf x}_{i}$ along different viewpoints}\label{viewpointex2}
\end{figure}
%
%
%
\begin{figure}[h]
\centerline{ 
\includegraphics{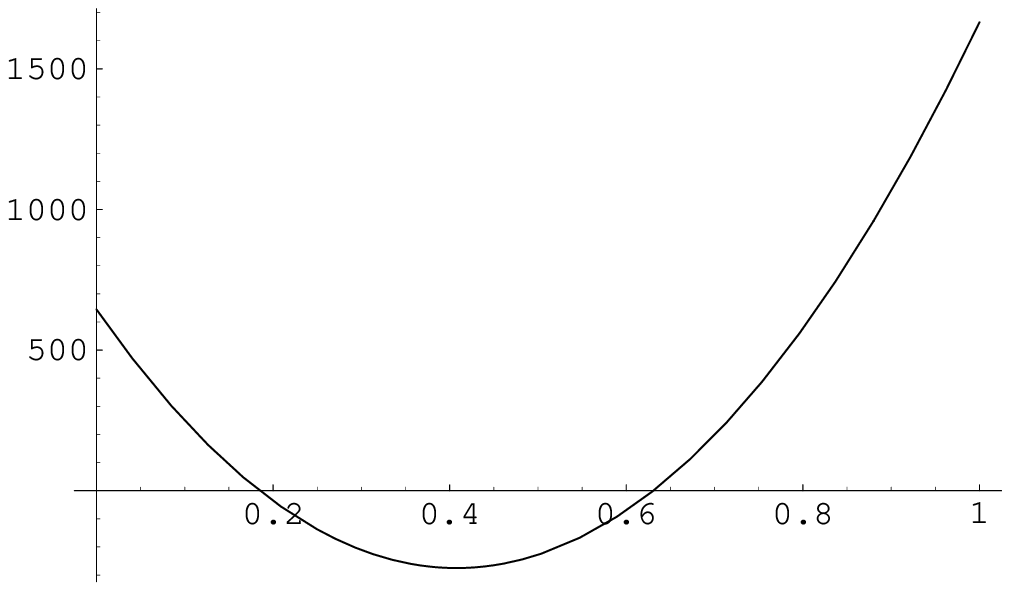} 
}
\caption{Curvature of the projection of the curve along the viewpoint 
${\bf V}_{3}$}\label{curvature5}
\end{figure}

\noindent
The mathematica code for the generation of figure \ref{curvature5}
for the curvature of the projection of the curve along the 
viewpoint ${\bf V}_{3}$ is as below:\\
{\rm (***********************************************************)\\
m1=\{1,1,0.5\};
m2=\{2,-3,0.5\};\\
A=\{0,0,0\};
B=\{0,0,10\};
Am=A+m1; 
Bm=B-m2;\\
P=((1-t)\^{ } 3)*A + (3*(1-t)\^{ } 2*t)*Am + (3*(1-t)*t\^{ } 2)*Bm + (t\^{ } 3)*B\\
DR1=D[P,\{t,1\}]
DR2=D[P,\{t,2\}]
CTR=Cross[DR1,DR2]\\
CTRPerp=CTR.\{-7.458, 6.863, 33.506\}\\
Infl5=Plot[CTRPerp,\{t,0,1\}]\\
Display["d:$\backslash$Gautam\_Viewpoint$\backslash$infl5.png", Infl5, "PNG"];\\
(**************************************************************)
}
\end{example}

\begin{remark}
Note in that in the above examples as we change the viewpoints the curve and
the data polygonal arc approximately same nature of deviation from convexity.
In figures 
the curve as well as data polygonal arc shows approximately same inflections.
This due to the fact that the curve satisfies torsion preservation criteria
described in section 
along with convexity preservation criteria.
\end{remark}

\section{Collinearity preservation criteria for interpolating splines}\label{seccollin}
\setcounter{equation}{0}
\begin{definition}\cite[Karavelas and Kaklis, 2000]{karavelas}\label{defcollincrit}
The collinearity preservation criteria is defined by the condition that if $|N_{i}|=0$ and
$L_{i-1} \cdot L_{i} > 0$, then
\begin{eqnarray}
\frac{|\gamma' (t) \times L_{j}|}{|\gamma'(t)||L_{j}|} < \epsilon_{0} \mbox{, }
t \in \eta_{i} \mbox{, } j= i-1,i, \label{defcollincriteqn} 
\end{eqnarray}
where $\epsilon_{0}$ is a user-specified small positive number in $(0,1]$,
and $\eta_{i}$ a user specified closed subinterval of $(t_{i-1}, t_{i+1})$
that includes $t_{i}$ as an interior point.
\end{definition}

We note that $ |N_{i}|=0 \implies  L_{i-1} = \alpha L_{i} $, $\alpha \in R$ 
and this condition with $L_{i-1} \cdot L_{i} >0$ implies that $L_{i-1}= \alpha L_{i}$,
$\alpha \in R^{+}$.
Equation (\ref{defcollincriteqn}) states that the (sine of the) angle between 
tangent vector at each point on the spline $\gamma (t)$ and $L$ is less than 
$\epsilon > 0$ in the user specified closed interval in $(t_{i-1} , t_{i+1})$.  
Thus collinearity preservation criteria requires that if two consecutive 
polygon 
segments are collinear and is having the same direction then the curve segments
of the corresponding indexes should be approximately collinear and parallel to the 
corresponding polygon segments.

We now investigate collinearity preservation criteria a bit more closely. 
One natural question to ask is: What if one considers ${\bf x}_{i}$ to be 
redundant? 
Well in that case, if the curve $\gamma (t)$ is not collinear to the line 
segment \{${\bf x}_{i-1}$, ${\bf x}_{i+1}$\}, then the curve between 
${\bf x}_{i-1}$ and ${\bf x}_{i+1}$ 
is required to satisfy other shape preservation criteria.
For convenience of understanding the situation let us suppose 
${\bf x}_{i-2}$, ${\bf x}_{i-1}$, ${\bf x}_{i}$, ${\bf x}_{i+1}$, 
${\bf x}_{i+2}$ are coplanar. Now consider the following cases
\begin{description}
\item[case i] $N_{i-1} \cdot N_{i+1} \geq 0$ to be refered as convex neighbourhood data
\item[case ii] $N_{i-1} \cdot N_{i+1} < 0$ to be refrered as inflection neighbourhood data
\end{description}
For {\bf case i} we propose that if the curve does not coincide with line
segments \{${\bf x}_{i-1}$, ${\bf x}_{i}$\}, \{${\bf x}_{i}$, ${\bf x}_{i+1}$\}
one must ensure that 
\begin{enumerate}
\item $\gamma (t)$ does not interpolate ${\bf x}_{i}$,
\item $\omega (t_{j}) \cdot N_{j} \geq 0$, $j=i-1,i+1$, 
\item $(\gamma'(t_{i+1}) \times L_{i+1}) \cdot (\gamma'(t_{i+1}) \times L_{i+2}) <0$
and $(\gamma'(t_{i-1}) \times L_{i}) \cdot (\gamma'(t_{i-1}) \times L_{i-1}) <0$
\item $\gamma (t)$ is globally convex between ${\bf x}_{i-1}$ and 
${\bf x}_{i+1}$,
\item $\gamma' (t)= \alpha L_{i}$, $\alpha \in R^{+}$ for 
$t \in (t_{i}- \eta, t_{i} + \delta) \subset [t_{i-1}, t_{i+1}]$ 
(suitable choice of $\eta$ and $\delta$ provides necessary tilt to the 
curve $\gamma (t)$).
\end{enumerate}

For a suitable choice of $\epsilon_{0}$, condition~\ref{defcollincriteqn} 
along with conditions 1-5 the curve $\gamma (t)$ have following properties
(see Figure~\ref{collinconvrt1} and \ref{collinconvrt2}):
\begin{itemize}
\item satisfies convexity preservation criteria between ${\bf x}_{i-1}$ 
and ${\bf x}_{i+1}$ and
\item convexity and inflection criteria preservation criteria for
data arc segments \{${\bf x}_{i-1}$, ${\bf x}_{i-2}$\} and 
\{${\bf x}_{i+1}$, ${\bf x}_{i+2}$\} achievable.
\end{itemize}

Violation of any of these conditions leads to the violation of the above
properties as illustrated by Figure~\ref{collinconvwr1} and {collinconvwr2}.\\ 
\begin{minipage}[h]{450pt}
\begin{myFirstSubFigure}{\label{collinconvrt1}Collinearity preservation 
criteria for convex neighbourhood data makes conditions for convexity 
preservation criteria between ${\bf x}_{i-1}$ and ${\bf x}_{i-2}$ 
achievable.}
\centering 
\epsfig{figure=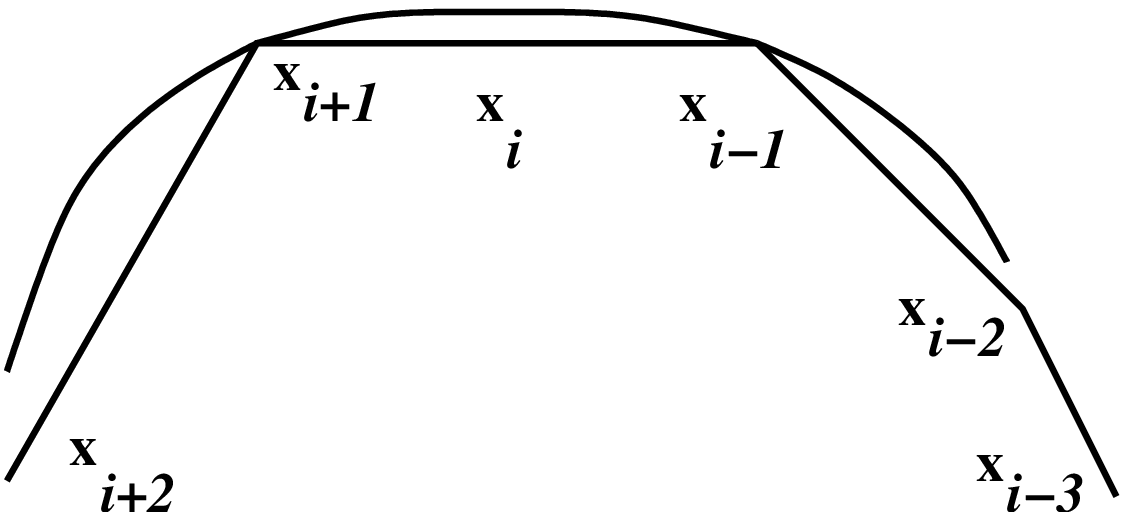}
\end{myFirstSubFigure}
\end{minipage}\hfill\\
\begin{minipage}[h]{450pt}
\begin{mySubFigure}{\label{collinconvrt2} Collinearity preservation 
criteria for convex neighbourhood data makes conditions for 
inflection preservation criteria between 
${\bf x}_{i-1}$ and ${\bf x}_{i-2}$ achievable.}
\centering
\epsfig{figure=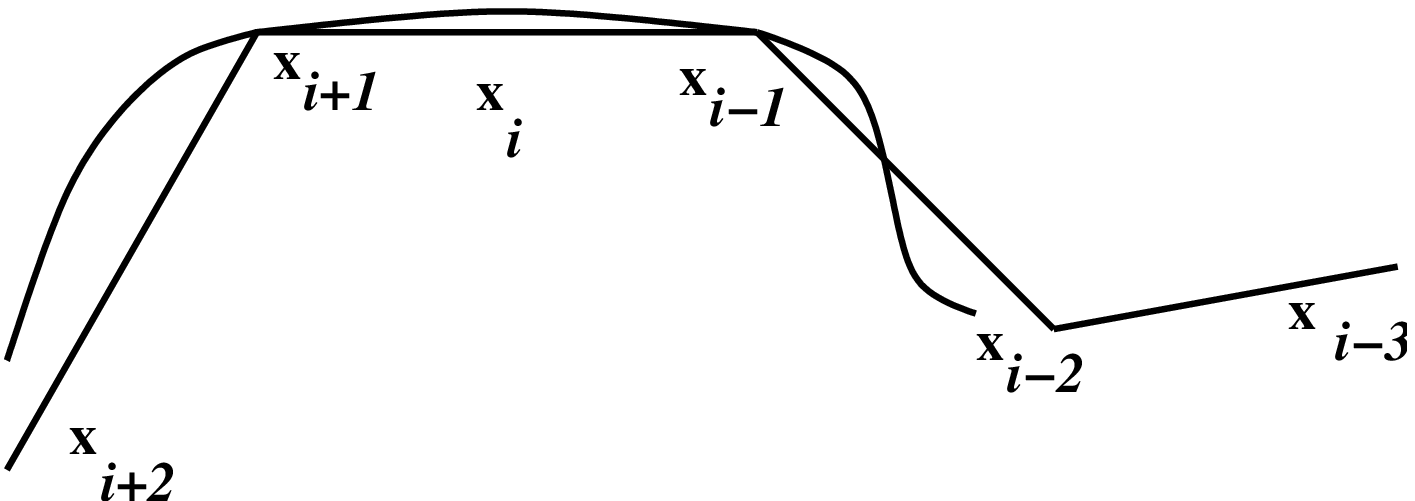}
\end{mySubFigure}
\end{minipage}\hfill\\
\begin{minipage}[h]{450pt}
\begin{mySubFigure}{\label{collinconvwr1}Violation of collinearity preservation criteria 
for convex neighbourhood data 
leads to violation of convexity preservation criteria between 
${\bf x}_{i-1}$ and ${\bf x}_{i+1}$}
\centering
\epsfig{figure=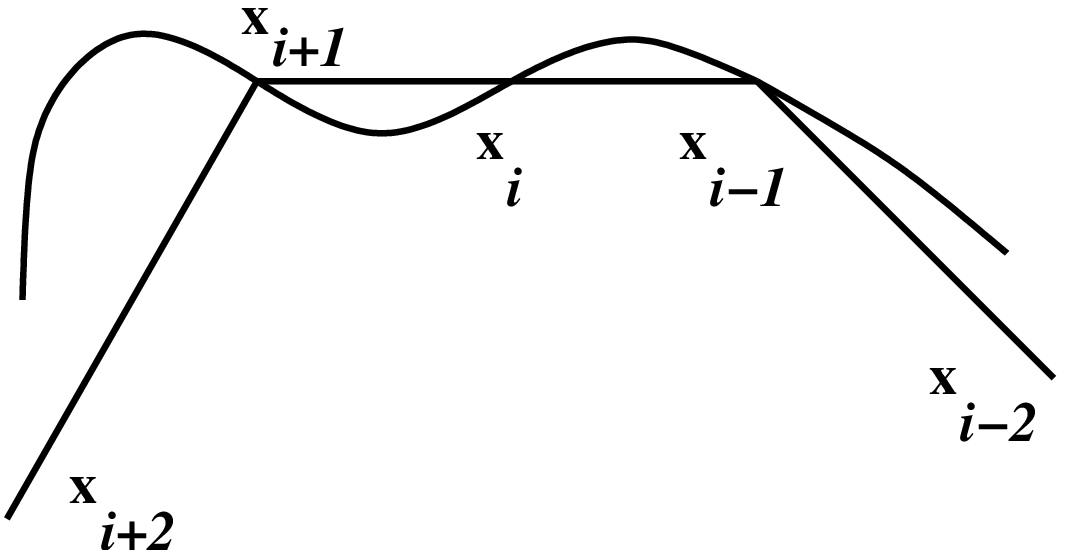}
\end{mySubFigure}
\end{minipage}\hfill\\
\begin{minipage}[h]{450pt}
\begin{myLastSubFigure}{\label{collinconvwr2}Violation of collinearity preservation criteria 
for convex neighbourhood data 
leads to violation of convexity preservation criteria between 
${\bf x}_{i-1}$ and ${\bf x}_{i+1}$}
\centering
\epsfig{figure=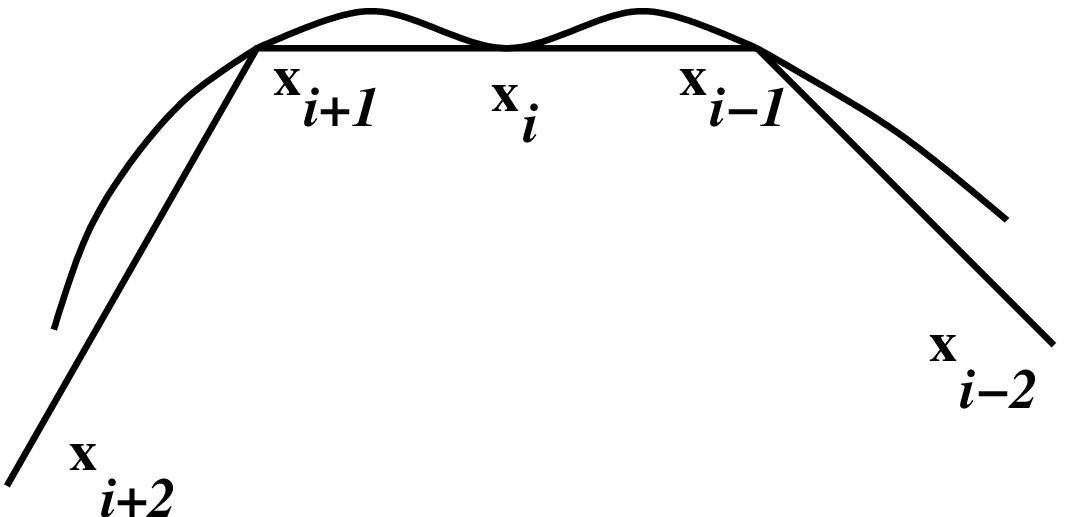}
\end{myLastSubFigure}
\end{minipage}\hfill\\
\\
For {\bf case ii} we propose that if the curve does not coincide with line
segments \{${\bf x}_{i-1}$, ${\bf x}_{i}$\}, \{${\bf x}_{i}$, ${\bf x}_{i+1}$\}
one must ensure that 
\begin{enumerate}
\item $\gamma (t)$ interpolate ${\bf x}_{i}$, 
\item $\omega (t_{j}) \cdot N_{j} \geq 0$, $j=i-1,i+1$, 
\item $(\gamma'(t_{i+1}) \times L_{i+1}) \cdot (\gamma'(t_{i+1}) \times L_{i+2}) <0$
and $(\gamma'(t_{i-1}) \times L_{i}) \cdot (\gamma'(t_{i-1}) \times L_{i-1}) <0$
\item $\omega (t)$ changes sign only once between ${\bf x}_{i-1}$ and 
${\bf x}_{i+1}$
\item $\omega (t)$ changes sign at $t=t_{i}$. 
\end{enumerate}

For a suitable choice of $\epsilon_{0}$, condition~\ref{defcollincriteqn} 
along with conditions 1-5 the curve $\gamma (t)$ have following properties
(see Figure~\ref{collininflecrt1} and \ref{collininflecrt2}):
\begin{itemize}
\item satisfies convexity preservation criteria between ${\bf x}_{i-1}$ 
and ${\bf x}_{i+1}$ and
\item convexity and inflection criteria preservation criteria for
data arc segments \{${\bf x}_{i-1}$, ${\bf x}_{i-2}$\} and 
\{${\bf x}_{i+1}$, ${\bf x}_{i+2}$\} achievable.
\end{itemize}

Violation of any of these conditions leads to the violation of the above
properties as illustrated by \ref{collininflecwr1} and \ref{collininflecwr2} .\\ 
\begin{minipage}[h]{450pt}
\begin{myFirstSubFigure}{\label{collininflecrt1}Collinearity preservation 
criteria for inflection neighbourhood data makes conditions for convexity 
preservation criteria between ${\bf x}_{i-1}$ and ${\bf x}_{i-2}$ 
achievable.}
\centering 
\epsfig{figure=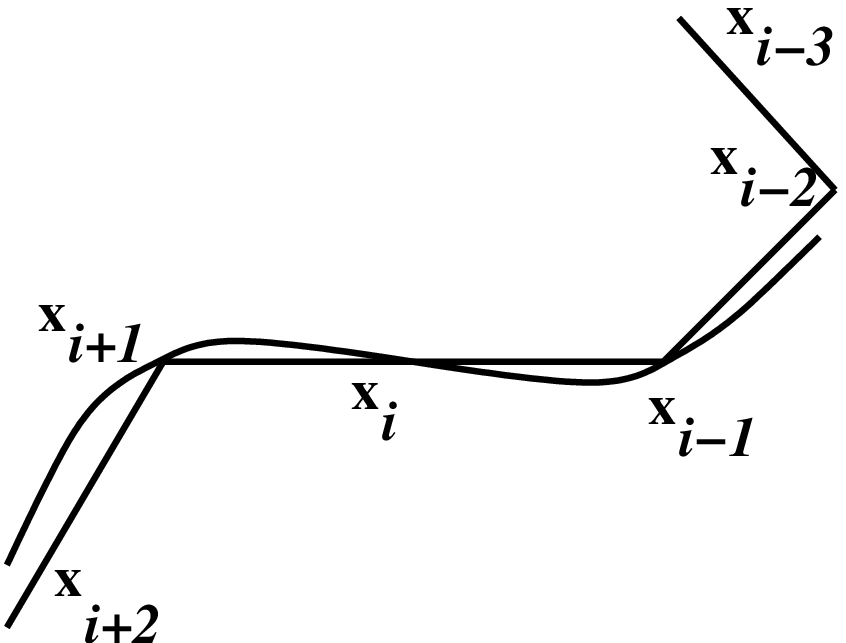}
\end{myFirstSubFigure}
\end{minipage}\hfill\\
\begin{minipage}[h]{450pt}
\begin{mySubFigure}{\label{collininflecrt2} Collinearity preservation 
criteria for inflection neighbourhood data makes conditions for 
inflection preservation criteria between 
${\bf x}_{i-1}$ and ${\bf x}_{i-2}$ achievable.}
\centering
\epsfig{figure=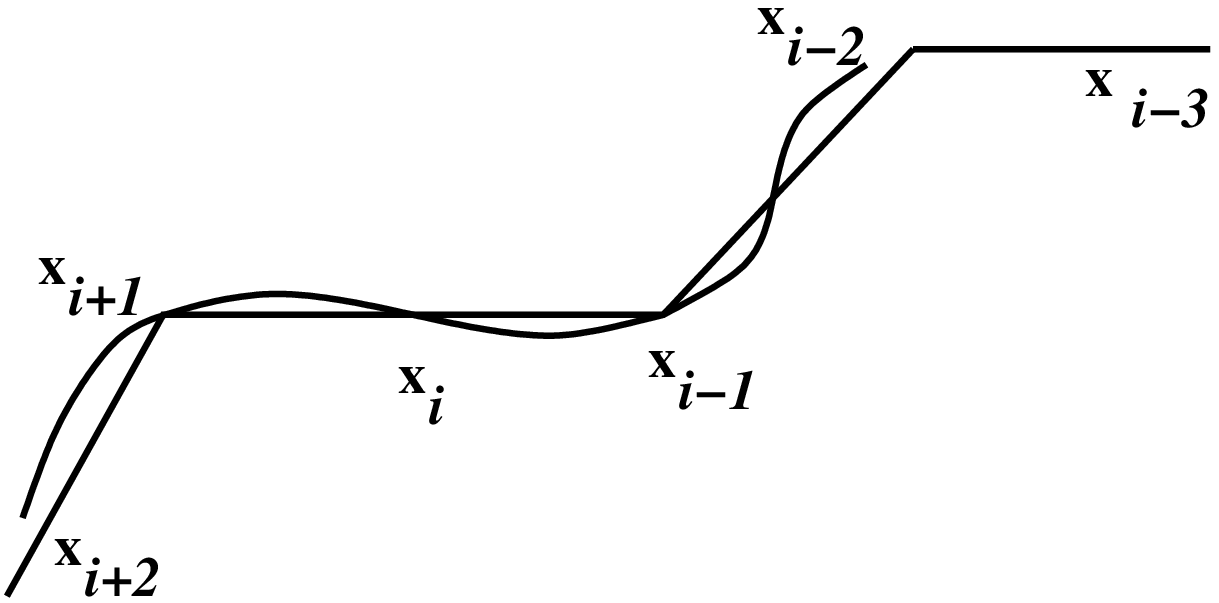}
\end{mySubFigure}
\end{minipage}\hfill\\
\begin{minipage}[h]{450pt}
\begin{mySubFigure}{\label{collininflecwr1}Violation of collinearity preservation criteria 
for inflection neighbourhood data 
leads to violation of inflection preservation criteria between 
${\bf x}_{i-1}$ and ${\bf x}_{i+1}$}
\centering
\epsfig{figure=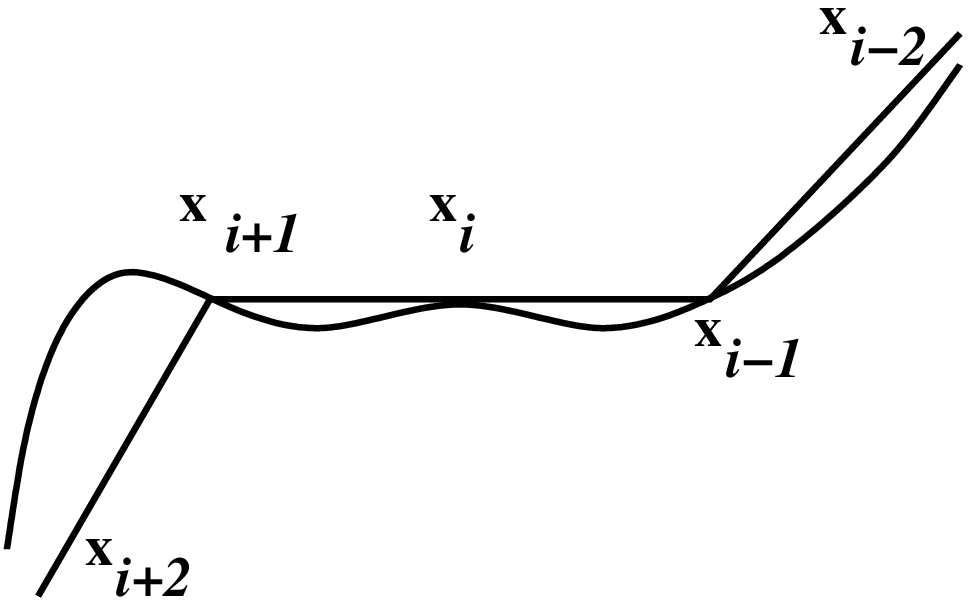}
\end{mySubFigure}
\end{minipage}\hfill\\
\begin{minipage}[h]{450pt}
\begin{myLastSubFigure}{\label{collininflecwr2}Violation of collinearity preservation criteria 
for inflection neighbourhood data 
leads to violation of inflection preservation criteria between 
${\bf x}_{i-1}$ and ${\bf x}_{i+1}$}
\centering
\epsfig{figure=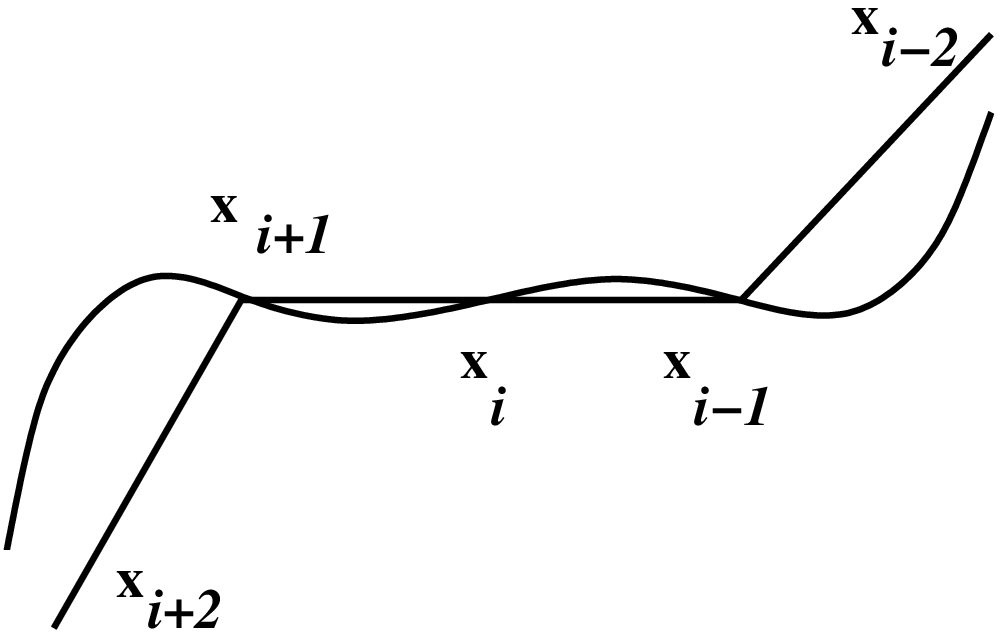}
\end{myLastSubFigure}
\end{minipage}\hfill\\

Thus we see that we need to modify the definition of collinearity 
preservation criteria for the general data, that is, when ${\bf x}_{i-2}$
${\bf x}_{i-1}$, ${\bf x}_{i}$, ${\bf x}_{i+1}$ and ${\bf x}_{i+2}$ are 
nonplanar. The modification is to be done by adding conditions according to 
the following cases:
\begin{description}
\item[case i] $N_{i-1} \cdot N_{i+1} \geq 0$ to be refered as convex neighbourhood data
\item[case ii] $N_{i-1} \cdot N_{i+1} < 0$ to be refrered as inflection neighbourhood data
\end{description}

We state our modified definition as follows:
\begin{definition}\label{moddefcollincrit}
The collinearity preservation criteria is defined by the condition that if $|N_{i}|=0$ and
$L_{i-1} \cdot L_{i} > 0$, then
\begin{eqnarray}
\frac{|\gamma' (t) \times L_{j}|}{|\gamma'(t)||L_{j}|} < \epsilon_{0} \mbox{, }
t \in \eta_{i} \mbox{, } j= i-1,i, \label{moddefcollincriteqn} 
\end{eqnarray}
where $\epsilon_{0}$ is a user-specified small positive number in $(0,1]$,
and $\eta_{i}$ a user specified closed subinterval of $(t_{i-1}, t_{i+1})$
that includes $t_{i}$ as an interior point and additionally for the case
of convex neighbourhood data
\begin{enumerate}
\item $\gamma (t)$ does not interpolate ${\bf x}_{i}$,
\item $\gamma (t)$ should satisfy convexity preservation criteria between
${\bf x}_{i-1}$ and ${\bf x}_{i+1}$, considering ${\bf x}_{i-2}$, 
${\bf x}_{i-1}$,  ${\bf x}_{i+1}$ and ${\bf x}_{i+2}$ as consecutive data 
points,
\item $|\gamma'(t_{i+1}) \times L_{i+1}| |\gamma'(t_{i+1}) \times L_{i+2}| <0$
and $|\gamma'(t_{i-1}) \times L_{i}| |\gamma'(t_{i-1}) \times L_{i-1}| <0$
\item $\gamma' (t)= \alpha L_{i}$, $\alpha \in R^{+}$ for 
$t \in (t_{i}- \eta, t_{i} + \delta) \subset [t_{i-1}, t_{i+1}]$ 
(suitable choice of $\eta$ and $\delta$ provides necessary tilt to the 
curve $\gamma (t)$).
\end{enumerate}
for the case of inflection neighbourhood data
\begin{enumerate}
\item $\gamma (t)$ interpolate ${\bf x}_{i}$, 
\item $\gamma (t)$ should satisfy inflection preservation criteria between
${\bf x}_{i-1}$ and ${\bf x}_{i+1}$, considering ${\bf x}_{i-2}$, 
${\bf x}_{i-1}$,  ${\bf x}_{i+1}$ and ${\bf x}_{i+2}$ as consecutive data 
points,
\item $|\gamma'(t_{i+1}) \times L_{i+1}| |\gamma'(t_{i+1}) \times L_{i+2}| <0$
and $|\gamma'(t_{i-1}) \times L_{i}| |\gamma'(t_{i-1}) \times L_{i-1}| <0$
\end{enumerate}
Considering ${\bf x}_{i-2}$, ${\bf x}_{i-1}$, ${\bf x}_{i+1}$ and 
${\bf x}_{i+2}$ as consecutive data points the curve $\gamma (t)$ between 
${\bf x}_{i-1}$ and ${\bf x}_{i+1}$ also satisfy torsion preservation criteria 
(to be stated in section~\ref{sectorsion}) or coplanarity preservation 
criteria (to be stated in section~\ref{seccoplan}) according to the condition
$[L_{i-1} \: L_{i} \: L_{i+1}] \neq 0$ or $[L_{i-1} \: L_{i} \: L_{i+1}] = 0$
respectively.
\end{definition}

We observe that the last two conditions guides the spatial behavior of the
curve $\gamma (t)$ between ${\bf x}_{i-1}$ and ${\bf x}_{i+1}$, with respect to 
the planes $\Pi_{i-1}$ (containing ${\bf x}_{i-2}$, ${\bf x}_{i-1}$, and 
${\bf x}_{i}$) and $\Pi_{i+1}$ (containing ${\bf x}_{i}$, ${\bf x}_{i+1}$ and  
${\bf x}_{i+2}$).
We also observe that in case ${\bf x}_{i-2}$, ${\bf x}_{i-1}$, ${\bf x}_{i}$, 
${\bf x}_{i+1}$ and ${\bf x}_{i+2}$ are nonplanar, 
then conditions of the definition~\ref{moddefcollincrit}
makes conditions of coplanarity preservation criteria (to be stated in 
section~\ref{seccoplan}) for curve $\gamma (t)$
between ${\bf x}_{i+2}$ and ${\bf x}_{i+3}$ and  
between ${\bf x}_{i-2}$ and ${\bf x}_{i-1}$, achievable.   
Thus we see that for a person, who tends to ignore ${\bf x}_{i}$ as a data 
point and considers ${\bf x}_{i-1}$ and ${\bf x}_{i+1}$ as adjacent data 
points, definition~\ref{moddefcollincrit} makes all the shape preservation 
criteria by $\gamma (t)$ between ${\bf x}_{i-1}$ and ${\bf x}_{i+1}$ achievable,
without conflict with the shape preservation criteria for curve segment
between ${\bf x}_{i-2}$ and ${\bf x}_{i-1}$ and 
between ${\bf x}_{i+1}$ and ${\bf x}_{i+2}$.  


\section{Torsion preservation criteria for interpolating splines}\label{sectorsion}
\setcounter{equation}{0}

\begin{definition}
Discrete torsion for the polygonal arc $ {\bf x}_{0} {\bf x}_{1} \cdots {\bf x}_{n}$ is defined as
\begin{eqnarray}
\triangle_{i} = [L_{i-1} \: L_{i} \: L_{i+1}]\mbox{, } i = 3,\hdots, n-1
\end{eqnarray}
where $L_{i}=  {\bf x}_{i} -  {\bf x}_{i-1}$, $i=1,..., n$
\end{definition}

\begin{definition}\cite[Costantini and Manni, 2003]{constmanni}\label{olddeftorsioncrit} 
Torsion preservation criteria consists of following conditions
\begin{enumerate}
\item $\tau_{i} (t) \triangle_{i} >0 $ in a chosen closed subinterval of 
$(t_{i-1},t_{i})$, whenever $\triangle_{i} \neq 0$.
\item $ \tau_{i} (t_{i-1}) \triangle_{j} > 0$, $j=i-1,i$,
whenever $\triangle_{i-1} \triangle_{i} > 0$  
\end{enumerate}
where $\tau_{i}(t)=\displaystyle{ 
\frac{\vert \gamma'_{i} (t) \: \gamma''_{i} (t) \: \gamma'''_{i} (t) \vert}
{\Vert \gamma'_{i} (t) \times \gamma''_{i} (t) \Vert^{2}}}$, if 
$\gamma'_{i} (t) \times \gamma''_{i} (t) \neq 0$.
\end{definition}

First condition of torsion preservation criteria states that 
$i^{th}$ curve segment should appear to twist away from its osculating 
plane in the same way as $L_{i+1}$ moves away from the plane of 
\{${\bf x}_{i-2}$, ${\bf x}_{i-1}$, ${\bf x}_{i}$\}.
 
\begin{figure}[htb]
\centerline{\epsfig{figure=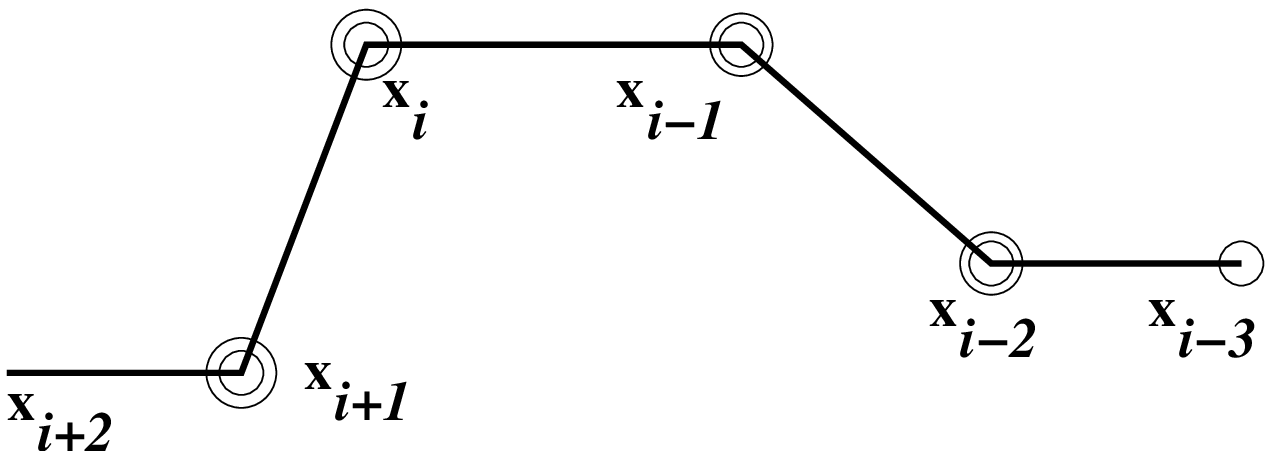}}
\caption{Data polygon arc involved in second condition of definition \ref{olddeftorsioncrit}} \label{torsiondef}
\end{figure}

Most of the author in their papers do not consider second condition
in their definition for torsion preservation criteria. Following definition
is followed by them
\begin{definition}\cite[Kong and Ong, 2002]{kong}\label{deftorsioncrit} 
Torsion preservation criteria is defined by the condition that
if $\triangle_{i} \neq 0$ then $\tau_{i} (t) \triangle_{i} \geq 0$, 
$t \in [t_{i-1},t_{i}]$, where $\tau_{i}(t)=\displaystyle{ 
\frac{\vert \gamma'_{i} (t) \: \gamma''_{i} (t) \: \gamma'''_{i} (t) \vert}
{\Vert \gamma'_{i} (t) \times \gamma''_{i} (t) \Vert^{2}}}$, if 
$\gamma'_{i} (t) \times \gamma''_{i} (t) \neq 0$.
\end{definition}
(We discuss the situation in which $\tau_{i} (t) \triangle_{i} = 0$ in the 
theorem \ref{trsntrsncmptbl}.)\\
We now discuss the second condition of definition of \ref{olddeftorsioncrit}.
In figure \ref{torsiondef}
set of points with bigger circle, \{${\bf x}_{i-2}$, ${\bf x}_{i-1}$,
${\bf x}_{i}$, ${\bf x}_{i+1}$\}, correspond to $\triangle_{i}$ and 
set of points with smaller circle, \{${\bf x}_{i-3}$, ${\bf x}_{i-2}$,
${\bf x}_{i-1}$, ${\bf x}_{i}$\}, correspond to $\triangle_{i-1}$. 
We know that the sign of $\triangle_{i}$ and $\triangle_{i-1}$ depends on 
the cosine of the angle that $L_{i+1}$ and $L_{i-2}$, respectively, makes with 
the normal of the plane $\Pi_{i-1}$ containing $L_{i-1}$ and $L_{i}$. 
The condition $\triangle_{i} \triangle_{i-1} >0$ states that $L_{i-2}$ moves
into the plane $\Pi_{i-1}$ in the same way as $L_{i+1}$ moves out of the plane, 
that is, the line segments \{${\bf x}_{i-3}$, ${\bf x}_{i-2}$\} and 
\{${\bf x}_{i}$, ${\bf x}_{i+1}$\} lies on the opposite sides of the plane 
$\Pi_{i-1}$. The condition $\tau_{i} (t_{i-1}) \triangle_{i} >0$, states that
the curve at $t=t_{i-1}$ moves out of its osculating plane in the same way 
as the vector $L_{i+1}$ moves out of the plane $\Pi_{i-1}$.  
Similarly, the condition $\tau_{i} (t_{i-1}) \triangle_{i-1} >0$, states that
the curve at $t=t_{i-1}$ moves out of its osculating plane in the same way 
as the vector $L_{i-2}$ moves into the plane $\Pi_{i-1}$ or $L_{i}$ moves out 
of the plane $\Pi_{i-2}$. 

We first observe that 
$\mbox{sign}(\tau_{i} (t_{i-1}) \triangle_{i-1})
=\mbox{sign}(\tau_{i} (t_{i-1}) \triangle_{i})$, 
whenever $\triangle_{i-1} \triangle_{i} > 0$ and
$\mbox{sign}(\tau_{i} (t_{i-1}) \triangle_{i-1})
=-\mbox{sign}(\tau_{i} (t_{i-1}) \triangle_{i})$, 
whenever $\triangle_{i-1} \triangle_{i} < 0$  
(since
$\mbox{sign}(\tau_{i} (t_{i-1}) \triangle_{i-1}) =$ 
$\mbox{sign}(\tau_{i} (t_{i-1}) \triangle_{i-1} (\triangle_{i})^{2}) =$
$\mbox{sign}(\tau_{i} (t_{i-1}) \triangle_{i} (\triangle_{i-1} \triangle_{i}))$
).\\
Given $\triangle_{i-1} \triangle_{i} >0$, 
the condition $\tau_{i}(t_{i-1}) \triangle_{i-1} >0$  
implies that 
$\tau_{i} (t_{i-1}) \triangle_{i} >0$ and vice-versa.
The conditions $\tau_{i}(t_{i-1}) \triangle_{i} >0$ and 
$\triangle_{i-1} \triangle_{i} <0$ implies that 
$\tau_{i} (t_{i-1}) \triangle_{i} <0$. For the case $\triangle_{i-1} \triangle_{i} <0$,
that is, the line segments \{${\bf x}_{i-3}$, ${\bf x}_{i-2}$\} and
\{${\bf x}_{i}$, ${\bf x}_{i+1}$\} lying on the same side of the plane
$\Pi_{i-1}$, the curve satisfying the condition $\tau_{i}(t_{i-1}) \triangle_{i} >0$,
have the property that\\ 
$\bullet$ (since condition $\tau_{i} (t_{i-1}) \triangle_{i} >0$) 
the curve at $t=t_{i-1}$ moves out of its osculating plane in the same way 
as the vector $L_{i+1}$ moves out of the plane $\Pi_{i-1}$. \\ 
$\bullet$ (since $\tau_{i} (t_{i-1}) \triangle_{i-1} <0$) 
the curve at $t=t_{i-1}$ moves out of its osculating plane opposite to the way 
as the vector $L_{i-2}$ moves into the plane $\Pi_{i-1}$ or $L_{i}$ moves out 
of the plane $\Pi_{i-2}$. 

Since in the definition \ref{deftorsioncrit}, we have $\tau_{i} (t_{i}) \triangle_{i}>0$,  
above analysis holds true when the curve is traversed in the reverse 
direction by concentrating the view on $x_{i}$ instead of ${\bf x}_{i-1}$ (with set
of points involved being \{${\bf x}_{i-2}$, ${\bf x}_{i-1}$,
${\bf x}_{i}$, ${\bf x}_{i+1}$, ${\bf x}_{i+2}$\})

Thus from the above analysis, we see that the definition \ref{deftorsioncrit} has 
better influence on
the shape of the curve than that of the definition \ref{olddeftorsioncrit}.
However, the definition \ref{olddeftorsioncrit} helped us to bring a very
important property (discussed above) of curve satisfying condition of 
definition \ref{deftorsioncrit}. 


\section{Coplanarity preservation criteria for interpolating spline}\label{seccoplan}
\setcounter{equation}{0}
\begin{definition}\cite[Karavelas and Kaklis, 2000]{karavelas}\label{defcoplancrit}
Coplanarity preservation criteria is defined by the condition that if $\triangle_{i} = 0$ and 
$|N_{i-1}||N_{i}| \neq 0$, then
\begin{eqnarray}
\frac{|\omega(t) \times N_{j}|}{|\omega(t)||N_{j}|} < \epsilon_{1} \mbox{, } 
|\omega (t)| \neq 0 \mbox{, } t \in I_{i} \mbox{, } j=i-1,i \label{defcoplancriteqn}
\end{eqnarray}
where $\epsilon_{1}$ is a user specified small positive number in $(0,1]$,
and $I_{i}$ is user-specified closed interval such that 
$[t_{i-1},t_{i}] \subseteq I_{i} \subseteq (t_{i-2}, t_{i+1})$.
\end{definition}
Coplanarity preservation criteria states that if data points ${\bf x}_{i-2}$, 
${\bf x}_{i-1}$, ${\bf x}_{i}$ and ${\bf x}_{i+1}$ are coplanar 
to a plane $\Pi$, then the interpolating curve between ${\bf x}_{i-1}$
and ${\bf x}_{i}$ and in the vicinity of ${\bf x}_{i-1}$ and 
${\bf x}_{i}$ has its binormal close to $N_{j}$, that is, its osculating 
plane should remain close to a plane parallel to $\Pi_{j}$.

We observe that in addtion to the condition (\ref{defcoplancriteqn}), the curve
segment between ${\bf x}_{i-1}$ and ${\bf x}_{i}$ should be constrained
such that 
its oscillations about the plane $\Pi_{j}$ is minimum. In fact,\\ 
$\bullet$ if $|\omega(t_{i-1}) \times N_{i-1}|=0$ and 
$|\omega(t_{i}) \times N_{i}|=0$, then the curve segment can be constrained
to be coplanar with the plane $\Pi_{i}$.\\
$\bullet$ if $|\omega(t_{i-1}) \times N_{i-1}| \neq 0$ and 
$|\omega(t_{i}) \times N_{i}|=0$ (or $|\omega(t_{i-1}) \times N_{i-1}| = 0$ 
and $|\omega(t_{i}) \times N_{i}| \neq 0$) then the curve segment can be 
constrained such that it  
oscillation about the plane $\Pi_{i}$ only once.\\
and if $|\omega(t_{i-1}) \times N_{i-1}| \neq 0$ and
$|\omega(t_{i}) \times N_{i}| \neq 0$ \\
$\bullet$ and addtionally $N_{i-1}$ and $N_{i}$ lie on the same side of the 
plane $\Pi_{j}$, then the curve segment can be constrained such that it 
does not oscillate about a (fixed) plane parallel 
to the plane $\Pi_{i}$ and\\
$\bullet$ and addtionally when $N_{i-1}$ and $N_{i}$ lie on the opposite side 
of the plane $\Pi_{j}$, the curve segment  can be constrained such that it 
oscillates about a (fixed) plane parallel to the plane $\Pi_{i}$ only once.

\section{Different shape preservation criteria on a curve segment}\label{conflictsame} 
\setcounter{equation}{0}
We now observe that the data points \{${\bf x}_{i-2}$, ${\bf x}_{i-1}$,
${\bf x}_{i}$, ${\bf x}_{i+1}$\} must satisfy one of the two conditions
\begin{description}
\item[c1] $N_{i-1} \cdot N_{i} < 0$, qualifying 
condition for inflection preservation criteria,
\item[c2] $N_{i-1} \cdot N_{i} > 0$, qualifying
condition for convexity preservation criteria
\end{description}
with one of the two conditions
\begin{description}
\item[t1] $\triangle_{i} = 0$, qualifying 
condition for torsion preservation criteria,
\item[t2] $\triangle_{i} \neq 0$, qualifying
condition for coplanarity preservation criteria.
\end{description}
We observe that there is no conflict between the conditions that the curve
needs to satisfy for one among convexity preservation criteria and inflection 
preservation criteria simultaneously with one among torsion preservation 
criteria and coplanarity preservation criteria. 

We also observe that there is no conflict between condition that curve need
to satisfy for collinearity preservation criteria simultaneously with 
torsion preservation criteria or coplanarity preservation criteria.

\section{Avoiding conflict between shape preservation behaviour of 
adjacent curve segments}\label{conflictadjacent}
\setcounter{equation}{0}
We observed in previous section that every curve segment need to satisfy
1) condition for either convexity preservation criteria or inflection 
preservation criteria
2) condition for either torsion preservation criteria or coplanarity 
preservation criteria. 
In this section we investigate the
compatibility between the shape preservation behaviour of adjacent curve
segments.
In Figure~19a-19d we
observe that if the curve $\gamma(t)$ is required to be $C^{1}$ smooth then
there is possibility that convexity as well as inflection preservation of 
a curve segment may lead to the violation of convexity and inflection 
preservation of adjacent curve segment. 
\\
\begin{minipage}[h]{450pt}
\begin{myFirstSubFigure}{\label{convconvwr}Convexity preserving curve segment
between ${\bf x}_{i-1}$ and ${\bf x}_{i}$ making violation of inflection
preservation criteria between ${\bf x}_{i-1}$ and ${\bf x}_{i-2}$ imminent.}
\centering
\epsfig{figure=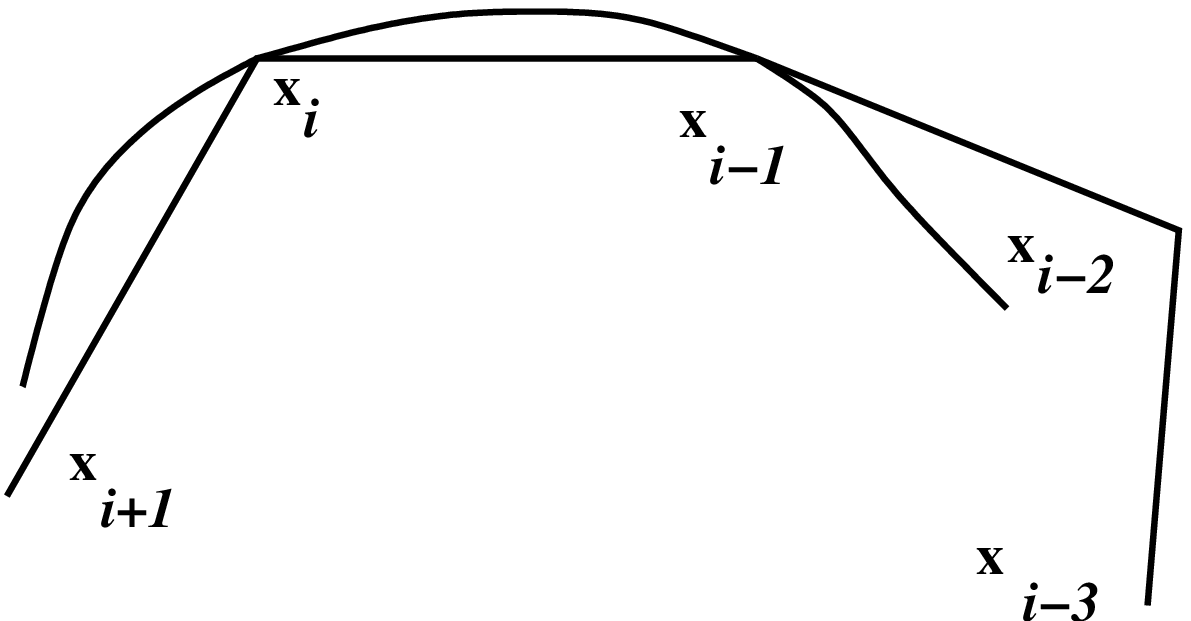}
\end{myFirstSubFigure}
\end{minipage}\hfill\\
\begin{minipage}[h]{450pt}
\begin{mySubFigure}{\label{convinflecwr}Convexity preserving curve segment
between ${\bf x}_{i-1}$ and ${\bf x}_{i}$ making violation of inflection
preservation criteria between ${\bf x}_{i-1}$ and ${\bf x}_{i-2}$ imminent.}
\centering
\epsfig{figure=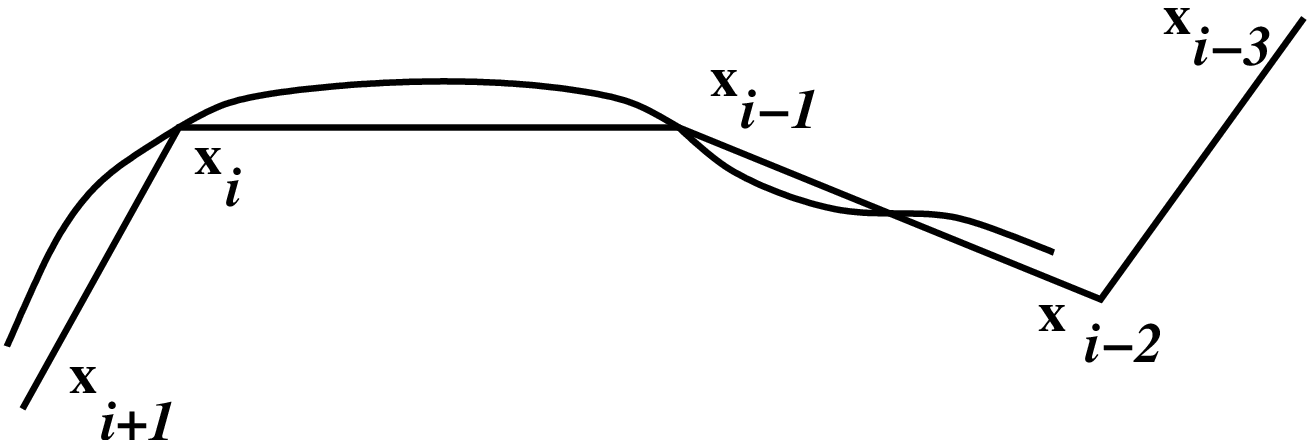}
\end{mySubFigure}
\end{minipage}\hfill\\
\begin{minipage}[h]{450pt}
\begin{mySubFigure}{\label{inflecinflecwr}Inflection preserving curve segment
between ${\bf x}_{i-1}$ and ${\bf x}_{i}$ making violation of inflection
preservation criteria between ${\bf x}_{i}$ and ${\bf x}_{i+1}$ imminent.}
\centering
\epsfig{figure=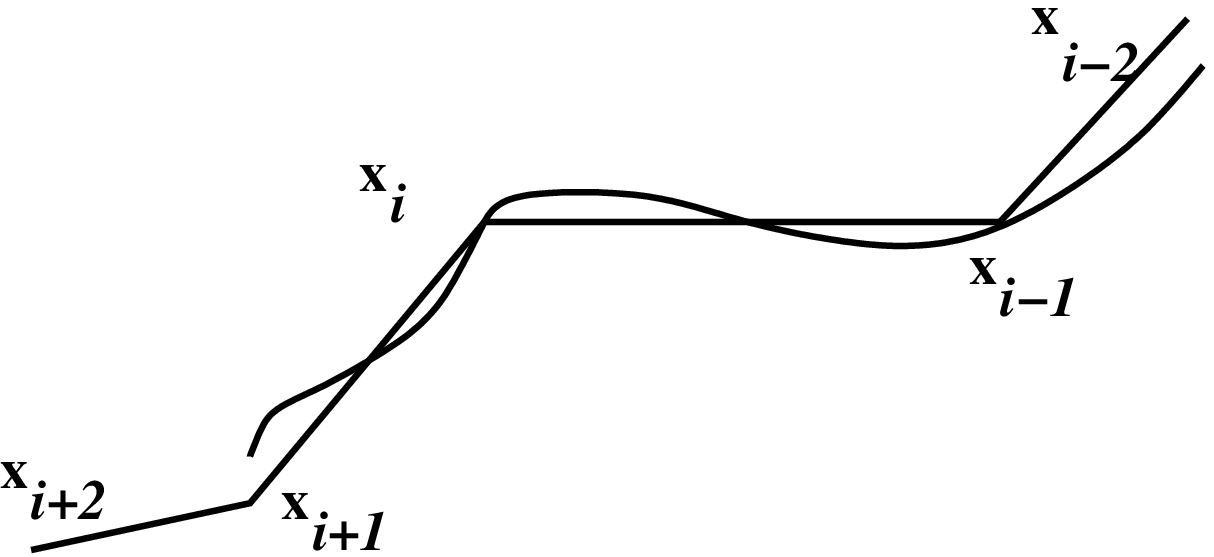}
\end{mySubFigure}
\end{minipage}\hfill\\
\begin{minipage}[h]{450pt}
\begin{myLastSubFigure}{\label{inflecconvwr}Inflection preserving curve segment
between ${\bf x}_{i-1}$ and ${\bf x}_{i}$ making violation of convexity 
preservation criteria between ${\bf x}_{i}$ and ${\bf x}_{i+1}$ imminent.}
\centering
\epsfig{figure=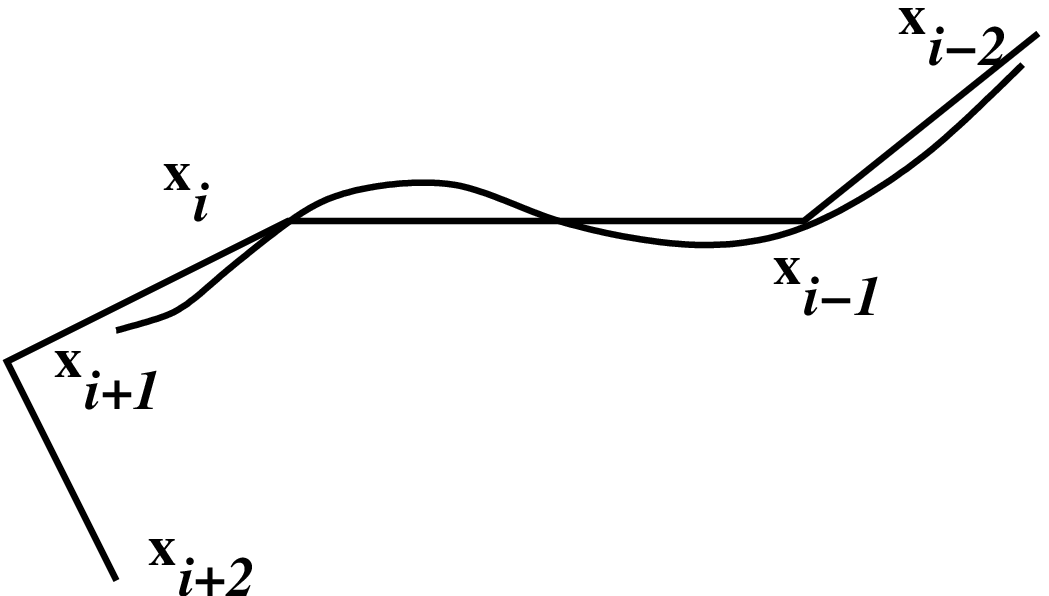}
\end{myLastSubFigure}
\end{minipage}\hfill\\
%
%
\\
We observe from Figure~19a-19d that if 
${\bf x}_{i-2}$, ${\bf x}_{i-1}$, ${\bf x}_{i}$, ${\bf x}_{i+1}$ are 
coplanar, then the conflict of convexity preservation criteria or 
inflection preservation of a curve segment with that of adjacent curve
segment (of $C^{1}$ smooth spline curve $\gamma(t)$) is resolved if and 
only if
$(\gamma'(t_{i-1}) \times L_{i}) \cdot (\gamma'(t_{i-1}) \times L_{i-1}) <0$.

For the general case, we observe from the definition of convexity and 
inflection preservation criteria that compatibility between convexity
and inflection preservation behaviour of adjacent curve segments of 
$C^{1}$ smooth spline curve $\gamma(t)$ can be guaranteed if and only if
\begin{eqnarray}
(\gamma_{N^{\perp}_{i-1}}'(t_{i-1}) \times L_{i}) \cdot 
(\gamma_{N^{\perp}_{i-1}}'(t_{i-1}) \times L_{i-1}) <0 \label{curvecompatible}
\end{eqnarray}
where 
$\gamma_{N^{\perp}_{i}} (t)$ is the orthogonal projection the curve 
$\gamma (t)$ on a plane having normal vector as $N_{i}$.
We note that the condition (\ref{curvecompatible}) does not interfere with
the conditions of convexity and inflection preservation criteria.\\

\begin{minipage}[h]{450pt}
\begin{myFirstSubFigure}{\label{convconvrt}Convexity preserving curve segment
between ${\bf x}_{i-1}$ and ${\bf x}_{i}$ satsifying (\ref{curvecompatible})
facilitates inflection
preservation between ${\bf x}_{i-1}$ and ${\bf x}_{i-2}$.}
\centering
\epsfig{figure=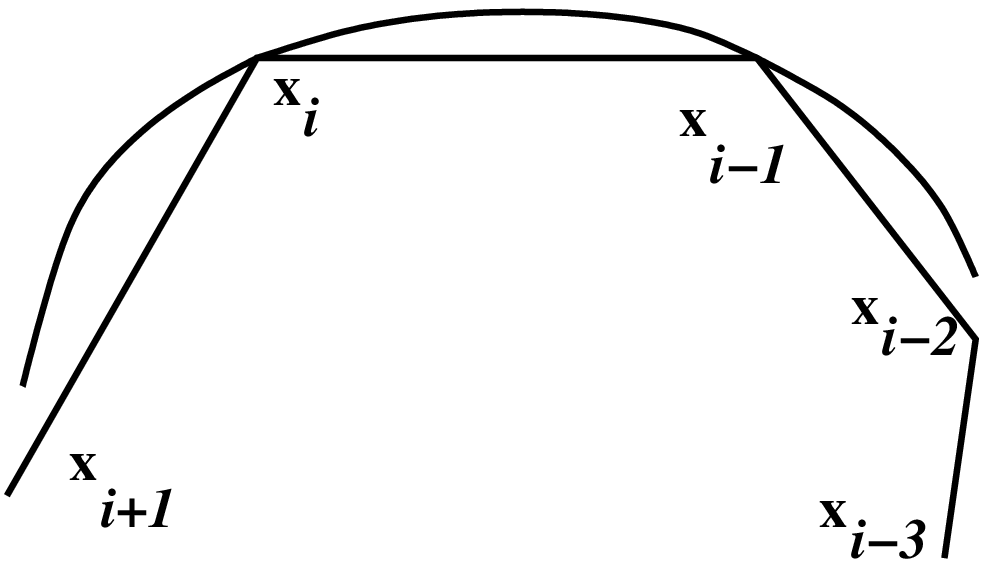}
\end{myFirstSubFigure}
\end{minipage}\hfill\\
\begin{minipage}[h]{450pt}
\begin{mySubFigure}{\label{convinflecrt}Convexity preserving curve segment
between ${\bf x}_{i-1}$ and ${\bf x}_{i}$ satsifying (\ref{curvecompatible})
facilitates inflection
preservation between ${\bf x}_{i-1}$ and ${\bf x}_{i-2}$.}
\centering
\epsfig{figure=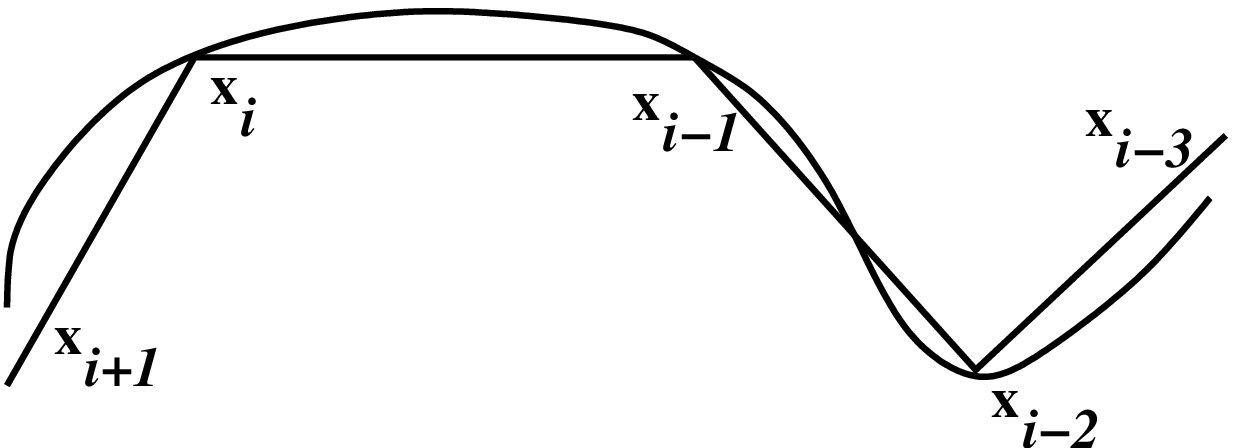}
\end{mySubFigure}
\end{minipage}\hfill\\
\begin{minipage}[h]{450pt}
\begin{mySubFigure}{\label{inflecinflecrt}Inflection preserving curve segment
between ${\bf x}_{i-1}$ and ${\bf x}_{i}$ satsifying (\ref{curvecompatible})
facilitates inflection
preservation between ${\bf x}_{i}$ and ${\bf x}_{i+1}$.}
\centering
\epsfig{figure=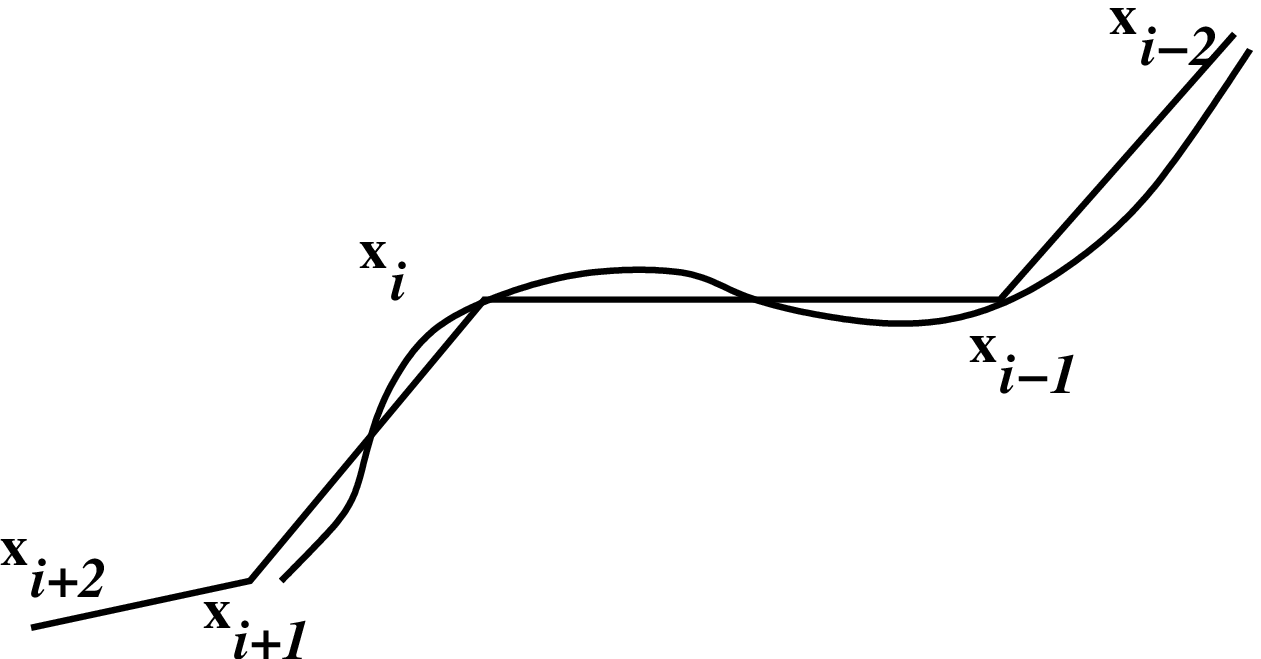}
\end{mySubFigure}
\end{minipage}\hfill\\
\begin{minipage}[h]{450pt}
\begin{myLastSubFigure}{\label{inflecconvrt}Inflection preserving curve segment
between ${\bf x}_{i-1}$ and ${\bf x}_{i}$ satsifying (\ref{curvecompatible})
facilitates convexity 
preservation between ${\bf x}_{i}$ and ${\bf x}_{i+1}$.}
\centering
\epsfig{figure=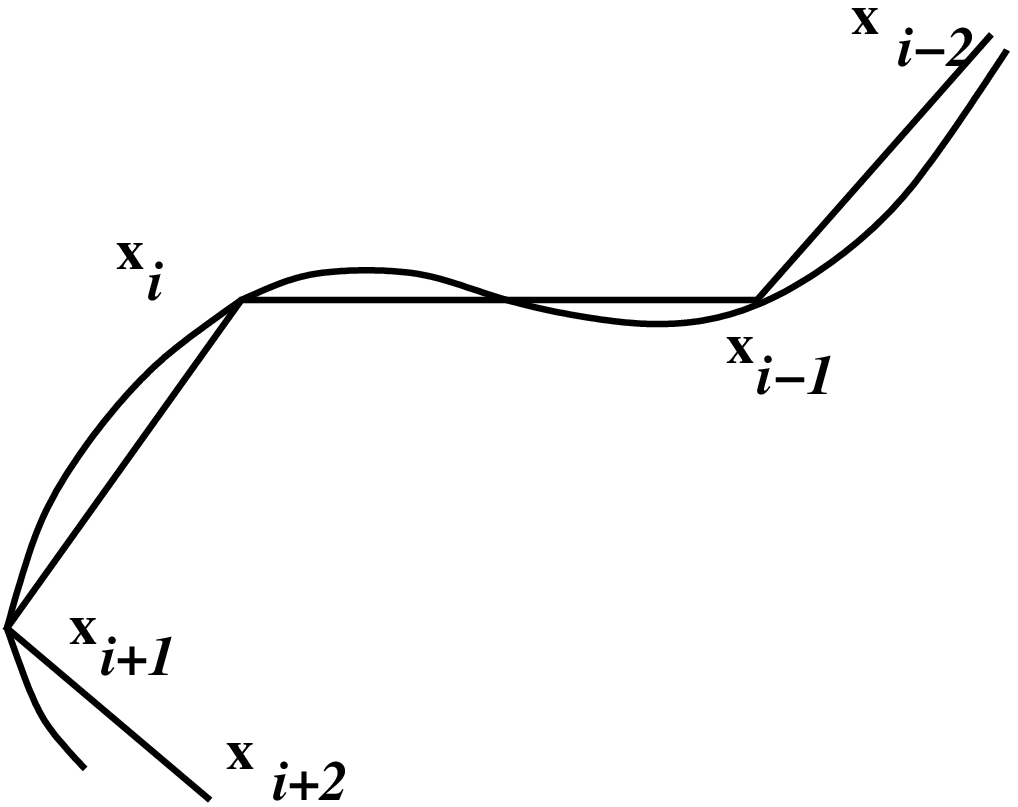}
\end{myLastSubFigure}
\end{minipage}\hfill\\
\\
%

We also observe that the condition (\ref{curvecompatible}) can be conveniently
imposed on a curve along with the conditions of torsion and coplanarity 
preservation criteria. Therefore, compatibility of torsion and coplanarity
preservation of a curve segment with the convexity and inflection preservation
of adjacent curve segment is feasible. 
From the above analysis and conditions of collinearity preservation criteria, 
we see that compatibility of convexity, inflection, torsion and coplanarity
preservation of a curve segment with the collinearity preservation of adjacent 
curve segment is also feasible. 

We now analyze the compatibility between the adjacent curve segments satisfying 
coplanarity and torsion preservation criteria. Let us consider figure 
\ref{torsiondef}. The points ${\bf x}_{i-3}$, ${\bf x}_{i-2}$, ${\bf x}_{i-1}$, 
${\bf x}_{i}$ and ${\bf x}_{i+1}$ may be such that 
1) $\triangle_{i} \triangle_{i-1} < 0$, 
2) $\triangle_{i} \triangle_{i-1} > 0$ and 
3) $\triangle_{i} \neq 0$, $\triangle_{i-1} = 0$.

\begin{theorem}\label{trsntrsncmptbl}
If $\triangle_{i} \triangle_{i-1} < 0$, then the curve segments 
$\gamma_{i-1} (t)$ (between ${\bf x}_{i-2}$ and ${\bf x}_{i-1}$) and 
$\gamma_{i} (t)$ (between ${\bf x}_{i-1}$ and ${\bf x}_{i}$), of spline
curve $\gamma (t)$, satisfies conditions for torsion preservation criteria,
if and only if either $\gamma (t)$ is torsion discontinuous or 
$\tau (t_{i-1})=0$ 
(and therefore $\tau_{i} (t_{i-1}) \triangle_{i} = 0$, 
$\tau_{i-1} (t_{i-1}) \triangle_{i-1} = 0$). 
\end{theorem}

\begin{proof}
As discussed in section \ref{sectorsion} we have 
$\mbox{sign}(\tau_{i} (t_{i-1}) \triangle_{i}) = 
\mbox{sign} (\tau_{i} (t_{i-1}) \triangle_{i-1}
\triangle_{i} \triangle_{i-1})$. Therefore, from
the definition \ref{deftorsioncrit} we see that torsion 
preservation by $\gamma_{i} (t)$ requires
$\tau_{i} (t_{i-1}) \triangle_{i-1} \leq 0$ 
and torsion preservation by $\gamma_{i-1} (t)$ requires 
$\tau_{i-1} (t_{i-1}) \triangle_{i-1} \geq 0$. Hence the 
theorem. 
\end{proof}

From the above proof it is evident that if
$\triangle_{i} \triangle_{i-1} > 0$, then torsion
preservation by $\gamma_{i-1}$ and $\gamma_{i}$ is
compatible.

If $\triangle_{i} \neq 0 $ and $\triangle_{i-1} = 0$ (requiring coplanarity
preservation by $\gamma_{i-1} (t)$), then from the definition 
\ref{defcoplancrit} we see that $\gamma_{i} (t)$ need to satisfy the 
condition of coplanarity condition \ref{defcoplancriteqn}, that is, binormal
of the curve should be close to $N_{i-1}$ in addition to satisfying
torsion preservation criteria, for $t \in [t_{i-1},t_{i}] \cap I_{i-1}$, 
where 
$[t_{i-2},t_{i-1}] \subseteq I_{i-1} \subseteq (t_{i-3}, t_{i})$.

\section{Shape preservation by cubic interpolating splines}\label{chapshapehermite}
\setcounter{equation}{0}

\subsection{Cubic B\'{e}zier segments}\label{secdervbezcubic}
\setcounter{equation}{0}
Let the control polygon for B\'{e}zier representation of $i^{th}$ cubic curve segment 
$\gamma_{i} (t)$ of cubic spline be 
$\{ {\bf P}_{i,0} \mbox{, } {\bf P}_{i,1} \mbox{, } {\bf P}_{i,2} \mbox{, } {\bf P}_{i,3} \}$,
with ${\bf P}_{i,0}={\bf x}_{i}$ and ${\bf P}_{i,3}={\bf x}_{i+1}$. 
For the $i^{th}$ cubic curve segment $\gamma_{i} (t)$, of the cubic spline $\gamma (t)$ 
we have
\begin{eqnarray}
\gamma_{i} (t) &=& {\bf P}_{i,0} B^{3}_{0} (u(t)) + {\bf P}_{i,1} B^{3}_{1} (u(t))+
                   {\bf P}_{i,2} B^{3}_{2} (u(t)) + {\bf P}_{i,3} B^{3}_{3} (u(t))
\end{eqnarray}
where $B^{n}_{i}(t)$ $t \in [0,1]$ is $i^{th}$ Bernstein's polynomial of order $n$,
$u(t) = \displaystyle{\frac{t-t_{i-1}}{t_{i}-t_{i-1}}}$. Conditions for shape preservation criteria
consists of first, second and third order derivatives of B\'{e}zier curves. 
Therefore we get their expressions in terms of the end point and slopes at
end points.
\begin{eqnarray}
\gamma'_{i} (t) &=& \frac{3}{h_{i}}(({\bf P}_{i,1}-{\bf P}_{i,0})B^{2}_{0} (u(t)) + 
                      ({\bf P}_{i,2}-{\bf P}_{i,1})B^{2}_{1} (u(t))+
                      ({\bf P}_{i,3}-{\bf P}_{i,2}) B^{2}_{2} (u(t))) \nonumber\\
&&\\
\gamma''_{i} (t)&=& \frac{6}{h^{2}_{i}}(({\bf P}_{i,2}-2{\bf P}_{i,1}+{\bf P}_{i,0}) (1-u(t))+
                      ({\bf P}_{i,3}-2{\bf P}_{i,2}+{\bf P}_{i,1}) u(t)) \\
\gamma'''_{i} (t)&=& \frac{6}{h^{3}_{i}}({\bf P}_{i,3}-3{\bf P}_{i,2}+3{\bf P}_{i,1}-{\bf P}_{i,0})
\end{eqnarray}
We have ${\bf m}_{i-1} = \displaystyle{\frac{3({\bf P}_{i,1}-{\bf P}_{i,0})}{h_{i}}}$, 
${\bf m}_{i} = \displaystyle{\frac{3({\bf P}_{i,3}-{\bf P}_{i,2})}{h_{i}}}$ 
and $L_{i}= {\bf P}_{i,3}-{\bf P}_{i,0}$. 
Therefore we can rewrite the expression for curve and its derivatives as follows
\begin{eqnarray}
\gamma_{i} (t) &=& {\bf x}_{i-1} B^{3}_{0} (u(t)) + 
                   ({\bf x}_{i-1} + \frac{h_{i}}{3} {\bf m}_{i-1}) B^{3}_{1} (u(t))+ \nonumber \\
                && ({\bf x}_{i} - \frac{h_{i}}{3}{\bf m}_{i}) B^{3}_{2} (u(t))+
                   {\bf x}_{i} B^{3}_{3} (u(t)) \mbox{,}\\
\gamma'_{i} (t) &=& {\bf m}_{i-1} B^{2}_{0} (u(t)) + 
 (\frac{3}{h_{i}}L_{i}- {\bf m}_{i-1} - {\bf m}_{i}) B^{2}_{1} (u(t))+
                    {\bf m}_{i} B^{2}_{2} (u(t)) \mbox{,}\\
\gamma''_{i} (t)&=& \frac{2}{h_{i}}(( \frac{3}{h_{i}}L_{i} - 2{\bf m}_{i-1} - {\bf m}_{i})) (1-u(t))+
                    (-\frac{3}{h_{i}}L_{i}+ {\bf m}_{i-1} + 2{\bf m}_{i}) u(t)) \nonumber\\
&&\\
\gamma'''_{i} (t)&=& \frac{6}{h^{3}_{i}}(h_{i}({\bf m}_{i-1} + {\bf m}_{i}) - 2 L_{i}) \mbox{.}
\end{eqnarray}

\subsection{Convexity preservation criteria for cubic interpolating splines}\label{secconvexcubic}
\setcounter{equation}{0}
Recall from theorem \ref{thmconvexcubic} that cubic spline $\gamma (t)$ interpolating 
data points ${\bf x}_{i}$, $0\leq i \leq n$, satisfies convexity criteria if and only 
if the control polygons of the projection of $\gamma (t)$, $t \in [t_{i-1}, t_{i}]$ on 
planes with normal vectors $N_{j}$, $j=i-1, i$ are globally convex, 
whenever $N_{i-1} \cdot N_{i} \geq 0$.

\noindent We now find simplification of the condition of global convexity control 
polygon of $P_{N^{\perp}} (\gamma (t))$, $t \in [t_{i-1}, t_{i}]$. Using this 
simplified condition we find the modified convexity preservation criteria for cubic 
splines in theorem \ref{thmconvexcubicmod}.

\begin{lemma}\cite[Goldman, 1990]{goldman}\label{linelineintrsct} Let the 
points ${\bf P}_{0}$, ${\bf P}_{1}$ ${\bf P}_{2}$, ${\bf P}_{3}$ in $R^{3}$
lie on plane with normal vector ${\bf n}$.
Then a line, through the points 
${\bf P}_{0}$ and ${\bf P}_{1}$, intersects with a line, through the
points ${\bf P}_{2}$ and ${\bf P}_{3}$, at the point 
\begin{eqnarray*}
{\bf P}=& 
{\bf P}_{0} + ({\bf P}_{1}-{\bf P}_{0}) s = &
{\bf P}_{3} + ({\bf P}_{2}-{\bf P}_{3}) t \\ 
=& {\bf P}_{1} + ({\bf P}_{0}-{\bf P}_{1}) \overline{s} = &
{\bf P}_{2} + ({\bf P}_{3}-{\bf P}_{2}) \overline{t}
\end{eqnarray*}

\begin{center}
$s = \displaystyle{
\frac{({\bf P}_{3}-{\bf P}_{0}) \times ({\bf P}_{2}-{\bf P}_{3}) \cdot {\bf n}}
{({\bf P}_{1}-{\bf P}_{0}) \times ({\bf P}_{2}-{\bf P}_{3}) \cdot {\bf n}} 
}$, 
$t = -\displaystyle{ 
\frac{({\bf P}_{1}-{\bf P}_{0}) \times ({\bf P}_{3}-{\bf P}_{0}) \cdot {\bf n}}
{(({\bf P}_{1}-{\bf P}_{0}) \times ({\bf P}_{2}-{\bf P}_{3})) \cdot {\bf n}} 
}$
\end{center}

\begin{center}
$\overline{s} = \displaystyle{
\frac{({\bf P}_{2}-{\bf P}_{1}) \times ({\bf P}_{3}-{\bf P}_{2}) \cdot {\bf n}}
{({\bf P}_{0}-{\bf P}_{1}) \times ({\bf P}_{3}-{\bf P}_{2}) \cdot {\bf n}} 
}$, 
$\overline{t} = -\displaystyle{ 
\frac{({\bf P}_{0}-{\bf P}_{1}) \times ({\bf P}_{2}-{\bf P}_{1}) \cdot {\bf n}}
{({\bf P}_{0}-{\bf P}_{1}) \times ({\bf P}_{3}-{\bf P}_{2}) \cdot {\bf n}} 
}$
\end{center}
\end{lemma}

\begin{proof} To find the point of intersection the given two lines we need
to solve the equations
\begin{eqnarray}
{\bf P} &=& {\bf P}_{0} + ({\bf P}_{1}-{\bf P}_{0})s \label{linelineeqn1}\\
{\bf P} &=& {\bf P}_{3} + ({\bf P}_{2}-{\bf P}_{3})t \label{linelineeqn2}
\end{eqnarray}
for $s$ and $t$ with the condition for the coplanarity of the four points
\begin{eqnarray*}
({\bf P}_{3}-{\bf P}_{0}) \cdot (({\bf P}_{1}-{\bf P}_{0}) \times 
({\bf P}_{2}-{\bf P}_{3})) & = & 0
\end{eqnarray*}
By subtracting equation (\ref{linelineeqn1}) from equation (\ref{linelineeqn2})
we get
\begin{eqnarray}
({\bf P}_{3} - {\bf P}_{0}) + ({\bf P}_{2}-{\bf P}_{3})t 
-({\bf P}_{1}-{\bf P}_{0})s &=& 0 \label{linelineeqn3}
\end{eqnarray}
Taking the cross product on both sides of equation (\ref{linelineeqn3}) by
$({\bf P}_{2}-{\bf P}_{3})$ we get
\begin{eqnarray}
({\bf P}_{3} - {\bf P}_{0}) \times ({\bf P}_{2}-{\bf P}_{3}) & = & 
(({\bf P}_{1}-{\bf P}_{0}) \times ({\bf P}_{2}-{\bf P}_{3})) s \label{modlinelineeqn3}
\end{eqnarray}
Now taking the scalar product on both sides of equation (\ref{modlinelineeqn3}) with {\bf n} 
we get
\begin{eqnarray}
s & = & 
\frac{(({\bf P}_{3} - {\bf P}_{0}) \times ({\bf P}_{2}-{\bf P}_{3})) \cdot {\bf n}}
{(({\bf P}_{1}-{\bf P}_{0}) \times ({\bf P}_{2}-{\bf P}_{3})) \cdot {\bf n}} 
\end{eqnarray}
Similarly we get the value for $t$. Now by interchanging ${\bf P}_{0}$ with
${\bf P}_{1}$ and ${\bf P}_{3}$ with ${\bf P}_{2}$ we get the values for 
$\overline{s}$ and $\overline{t}$.   
Hence proved.
\end{proof}


\begin{lemma}\label{lemconvexcubicpoly}
A planar polygonal arc $\{ {\bf P}_{0} \mbox{, } {\bf P}_{1} \mbox{, } {\bf P}_{2} \mbox{, } 
{\bf P}_{3} \}$ lying on a plane with normal vector $N$ is globally convex according to 
orientation induced by $N$ if and only if either
\begin{enumerate}
\item $({\bf P}_{1}-{\bf P}_{0}) \times ({\bf P}_{2}-{\bf P}_{3}) \cdot N >0$ 
with
\begin{enumerate}
\item 
$({\bf P}_{1}-{\bf P}_{0}) \times ({\bf P}_{2}-{\bf P}_{1}) \cdot N <0$ and 
$({\bf P}_{2}-{\bf P}_{1}) \times ({\bf P}_{3}-{\bf P}_{2}) \cdot N <0$
or
\item 
$ ({\bf P}_{0}-{\bf P}_{1}) \times ({\bf P}_{3}-{\bf P}_{0}) \cdot N <0 $ and
$ ({\bf P}_{3}-{\bf P}_{0}) \times ({\bf P}_{2}-{\bf P}_{3}) \cdot N <0 $ 
\end{enumerate} 
or
\item $({\bf P}_{1}-{\bf P}_{0}) \times ({\bf P}_{2}-{\bf P}_{3}) \cdot N <0$ 
with
\begin{enumerate}
\item 
$({\bf P}_{1}-{\bf P}_{0}) \times ({\bf P}_{2}-{\bf P}_{1}) \cdot N >0$ and 
$({\bf P}_{2}-{\bf P}_{1}) \times ({\bf P}_{3}-{\bf P}_{2}) \cdot N >0$
or
\item 
$ ({\bf P}_{0}-{\bf P}_{1}) \times ({\bf P}_{3}-{\bf P}_{0}) \cdot N >0 $ and
$ ({\bf P}_{3}-{\bf P}_{0}) \times ({\bf P}_{2}-{\bf P}_{3}) \cdot N >0 $ 
\end{enumerate} 
\end{enumerate} 
holds.
\end{lemma}

\begin{proof}
From the definition \ref{d3} we know that planar polygonal arc $\{ {\bf P}_{0} 
\mbox{, } {\bf P}_{1} \mbox{, } {\bf P}_{2} \mbox{, } {\bf P}_{3} \}$
is globally convex according to the orientation induced by normal vector $N$ 
if and only if
\begin{description}
\item[condition i] polygonal arc starting from ${\bf P}_{0}$ always turn towards the 
right side and 
\item[condition ii] it always lies entirely to its right side of any of its edges.
\end{description}
For the given polygonal arc, {\bf condition i} holds if and only if
\begin{eqnarray}
(({\bf P}_{1}-{\bf P}_{0}) \times ({\bf P}_{2}-{\bf P}_{1})) \cdot N 
(({\bf P}_{2}-{\bf P}_{1}) \times ({\bf P}_{3}-{\bf P}_{2})) \cdot N > 0
\end{eqnarray}
%

Now for the given polygonal arc, {\bf condition ii} holds if and only if
\begin{enumerate}
\item line through ${\bf P}_{0}$ and ${\bf P}_{1}$ does not intersect the line segment between 
${\bf P}_{2}$ and ${\bf P}_{3}$ and
\item line through ${\bf P}_{2}$ and ${\bf P}_{3}$ does not intersect the line segment between 
${\bf P}_{1}$ and ${\bf P}_{0}$.
\end{enumerate}
The above two condition holds if and only if the point of intersection 
${\bf P}$, between the line $l_{1}$
through ${\bf P}_{0}$ and ${\bf P}_{1}$ and line $l_{2}$ through ${\bf P}_{2}$ and ${\bf P}_{3}$,
does not lie in the segment between ${\bf P}_{0}$ and ${\bf P}_{1}$ or the segment between
${\bf P}_{2}$ and ${\bf P}_{3}$. 
From lemma \ref{linelineintrsct} we know that point of intersection 
of lines $l_{1}$ and $l_{2}$ is given by
\begin{eqnarray*}
{\bf P}=& 
{\bf P}_{0} + ({\bf P}_{1}-{\bf P}_{0}) s = &
{\bf P}_{3} + ({\bf P}_{2}-{\bf P}_{3}) t \\ 
=& {\bf P}_{1} + ({\bf P}_{0}-{\bf P}_{1}) \overline{s} = &
{\bf P}_{2} + ({\bf P}_{3}-{\bf P}_{2}) \overline{t}
\end{eqnarray*}

\begin{center}
$s = \displaystyle{
\frac{(({\bf P}_{3}-{\bf P}_{0}) \times ({\bf P}_{2}-{\bf P}_{3})) \cdot {\bf N}}
{(({\bf P}_{1}-{\bf P}_{0}) \times ({\bf P}_{2}-{\bf P}_{3})) \cdot {\bf N}} 
}$, 
$t = -\displaystyle{ 
\frac{(({\bf P}_{1}-{\bf P}_{0}) \times ({\bf P}_{3}-{\bf P}_{0})) \cdot {\bf N}}
{(({\bf P}_{1}-{\bf P}_{0}) \times ({\bf P}_{2}-{\bf P}_{3})) \cdot {\bf N}} 
}$
\end{center}

\begin{center}
$\overline{s} = \displaystyle{
\frac{(({\bf P}_{2}-{\bf P}_{1}) \times ({\bf P}_{3}-{\bf P}_{2})) \cdot {\bf N}}
{(({\bf P}_{0}-{\bf P}_{1}) \times ({\bf P}_{3}-{\bf P}_{2})) \cdot {\bf N}} 
}$, 
$\overline{t} = -\displaystyle{ 
\frac{(({\bf P}_{0}-{\bf P}_{1}) \times ({\bf P}_{2}-{\bf P}_{1})) \cdot {\bf N}}
{(({\bf P}_{0}-{\bf P}_{1}) \times ({\bf P}_{3}-{\bf P}_{2})) \cdot {\bf N}} 
}$
\end{center}

Now with {\bf condition i} ensured, {\bf condition ii} is satisfied if and only
if either $s<0$ with $t<0$ or $\overline{s}<0$ with $\overline{t}<0$ holds.
Hence proved.
\end{proof}

\begin{figure}[h]
  \centering
    \subfigure[case 1 (a) $\overline{s}<0$, $\overline{t}<0$]{\label{convexpoly4}
\includegraphics[scale=0.93]{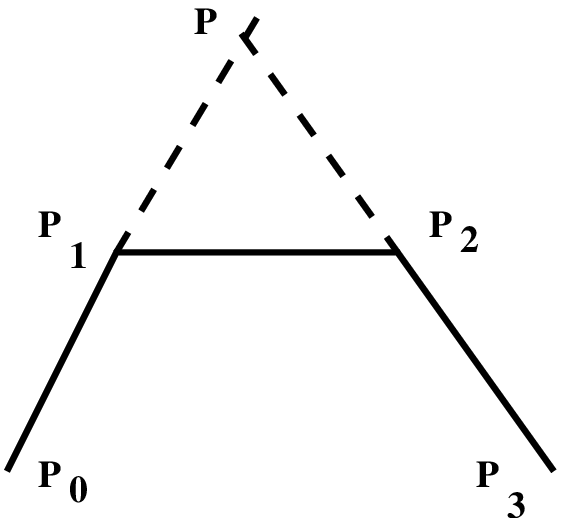}
}
\hspace{1cm}
    \subfigure[case 1 (b) $s<0$, $t<0$]{\label{convexpoly2}
\includegraphics[scale=0.93]{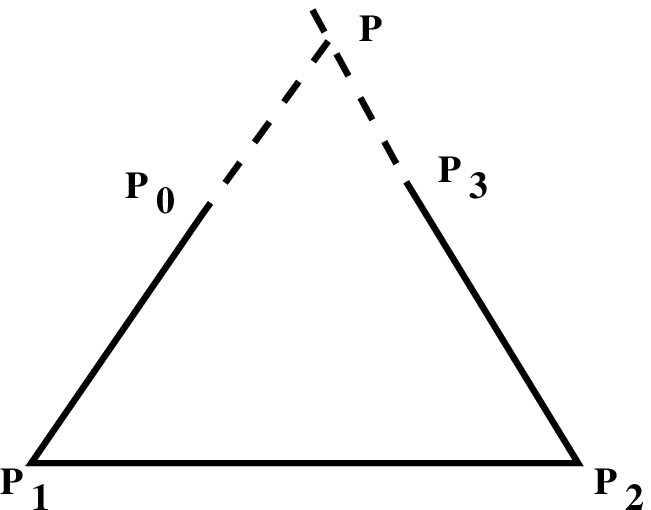}
}
\end{figure}
\begin{figure}[h]
\centering
    \subfigure[case 2 (a) $\overline{s}<0$, $\overline{t}<0$]{\label{convexpoly3}
\includegraphics[scale=0.93]{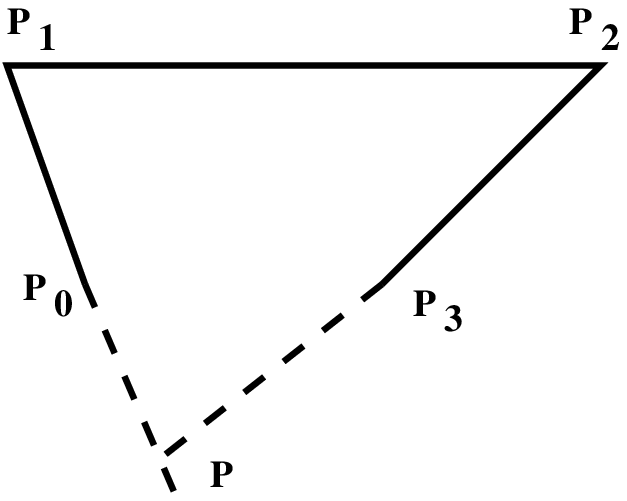}
}
\hspace{1cm}
    \subfigure[case 2 (b) $s<0$, $t<0$]{\label{convexpoly1}
\includegraphics[scale=0.93]{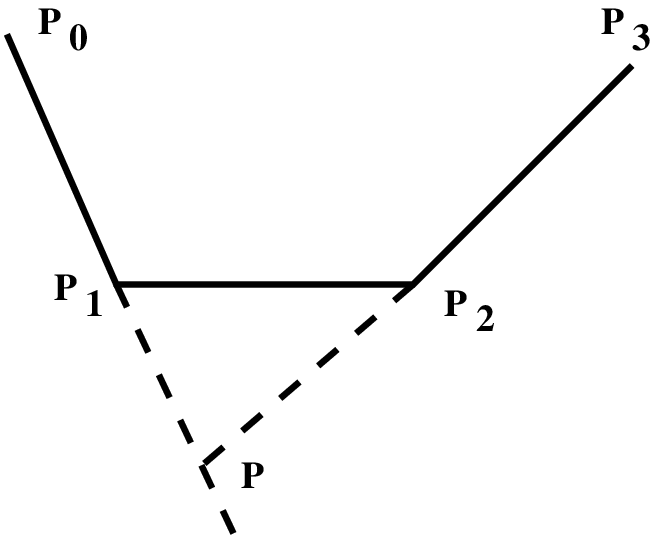}
}
\caption{Examples of different planar convex control polygons}\label{convexpoly}
\end{figure}

\begin{lemma}\label{lemconvexcubicpolyprojection} 
Let $\overline{{\bf P}}_{i} = P_{N^{\perp}} ({\bf P}_{i})$.
Then
\begin{eqnarray}
(\overline{{\bf P}}_{a}-\overline{{\bf P}}_{b}) \times (\overline{{\bf P}}_{c}-\overline{{\bf P}}_{d}) \cdot N
&=&
({\bf P}_{a}-{\bf P}_{b}) \times ({\bf P}_{c}-{\bf P}_{d}) \cdot N
\end{eqnarray}
\end{lemma}

\begin{proof} We know that projection of ${\bf P}_{i}$ on a plane 
$\displaystyle{\frac{(x,y,z) \cdot N + d}{\Vert N \Vert}}=0$ with normal vector $N$,  
$P_{N^{\perp}} ({\bf P}_{i}) = \overline{{\bf P}}_{i}$ 
is given by
\begin{eqnarray}
\overline{{\bf P}}_{i} &=& {\bf P}_{i} + \frac{{\bf P}_{i} \cdot N + d}{\Vert N \Vert^{2}} N \label{defproject} \mbox{.}
\end{eqnarray}
One can get the proof using the idea in the proof of lemma \ref{lemsuper}. 
\end{proof}

Using the definition \ref{impconvex} and following theorem from 
\cite[Liu, 2001]{liu} we get

\begin{theorem}\label{thmconvexcubicmod}
A cubic spline curve $\gamma (t)$ satisfies the convexity preservation criteria 
if and only if either
\begin{enumerate}
\item ${\bf m}_{i-1} \times {\bf m}_{i} \cdot N_{j}<0$, with
\begin{enumerate}
\item ${\bf m}_{i-1} \times L_{i} \cdot N_{j}<
 \displaystyle{\frac{h_{i}}{3}} 
{\bf m}_{i-1} \times {\bf m}_{i} \cdot N_{j}$
and $ L_{i} \times {\bf m}_{i} \cdot N_{j} <
 \displaystyle{\frac{h_{i}}{3}} 
{\bf m}_{i-1} \times {\bf m}_{i} \cdot N_{j}$
or
\item
${\bf m}_{i-1} \times L_{i} \cdot N_{j}>0$ and
$ L_{i} \times {\bf m}_{i} \cdot N_{j}>0$
\end{enumerate}
or
\item ${\bf m}_{i-1} \times {\bf m}_{i} \cdot N_{j}>0$, with
\begin{enumerate}
\item ${\bf m}_{i-1} \times L_{i} \cdot N_{j}>
 \displaystyle{\frac{h_{i}}{3}} 
{\bf m}_{i-1} \times {\bf m}_{i} \cdot N_{j}$
and $ L_{i} \times {\bf m}_{i} \cdot N_{j} >
 \displaystyle{\frac{h_{i}}{3}} 
{\bf m}_{i-1} \times {\bf m}_{i} \cdot N_{j}$ or
\item
${\bf m}_{i-1} \times L_{i} \cdot N_{j}<0$ and
$ L_{i} \times {\bf m}_{i} \cdot N_{j}<0$
\end{enumerate}
\end{enumerate} 
$j=i-1,i$, whenever $N_{i-1} \cdot N_{i} > 0$.
\end{theorem}
\begin{proof} 
We note that  
\begin{eqnarray*}
({\bf P}_{1}-{\bf P}_{0}) \times ({\bf P}_{2}-{\bf P}_{1}) \cdot N =
({\bf P}_{1}-{\bf P}_{0}) \times ({\bf P}_{3}-{\bf P}_{0}) \cdot N + 
({\bf P}_{1}-{\bf P}_{0}) \times ({\bf P}_{2}-{\bf P}_{3}) \cdot N\\ 
({\bf P}_{2}-{\bf P}_{1}) \times ({\bf P}_{3}-{\bf P}_{2}) \cdot N =
({\bf P}_{3}-{\bf P}_{0}) \times ({\bf P}_{3}-{\bf P}_{2}) \cdot N -
({\bf P}_{1}-{\bf P}_{0}) \times ({\bf P}_{3}-{\bf P}_{2}) \cdot N 
\end{eqnarray*} 
since ${\bf P}_{2}-{\bf P}_{1}=  
({\bf P}_{2}-{\bf P}_{3})-({\bf P}_{1}-{\bf P}_{0})+({\bf P}_{3}-{\bf P}_{0})$ 

Since control polygon of (orthogonal) projection of B\'{e}zier curve on a 
plane is same as the orthogonal projection of control polygon of the curve on 
the plane, using theorem \ref{thmconvexcubic}, lemma \ref{lemconvexcubicpoly} 
(by replacing ${\bf P}_{i}$ by ${\bf P}_{i,j}$) and 
lemma \ref{lemconvexcubicpolyprojection} we get the theorem.
\end{proof}

\subsection{Inflection preservation criteria for cubic interpolating splines}\label{secinflexcubic}
\setcounter{equation}{0}
The definition \ref{defninflexcrit} of inflection preservation criteria 
involves the curvature term $\omega_{i} (t)$. So, in order to get a simplified 
characterization for inflection criteria for cubic case we first get 
simplified expression for $\omega_{i} (t)$ as follows.
\begin{lemma}
Let ${\bf c} (t) = {\bf P}_{0} (1-t)^{2} + {\bf P}_{1} (2~t~(1-t)) + 
{\bf P}_{2} t^{2}$ be a quadratic B\'{e}zier curve. Then 
\begin{eqnarray}
{\bf c} (t) \times {\bf c}' (t) = 
2({\bf P}_{0} \times {\bf P}_{1}) (1-t)^{2} + 
({\bf P}_{0} \times {\bf P}_{2}) (2~t~(1-t)) + 
2({\bf P}_{1} \times {\bf P}_{2}) t^{2} \mbox{.}\label{eqnlemmacture}
\end{eqnarray}
\end{lemma}

\begin{proof} In the statement of the lemma we observe that though
the ${\bf c} (t)$ and ${\bf c}' (t)$ are B\'{e}zier curves of degree
$2$ and $1$ respectively, their cross product is a B\'{e}zier curve
of degree $2$ instead of $3$. In order to understand this we express 
the curve ${\bf c} (t)$ in power basis form as 
${\bf c} (t) = {\bf p}_{0} + {\bf p}_{1} t + {\bf p}_{2} t^{2}$, where
${\bf p}_{0} = {\bf P}_{0}$, ${\bf p}_{1} = 2({\bf P}_{1}-{\bf P}_{0})$,
${\bf p}_{2} = {\bf P}_{0} - 2 {\bf P}_{1} + {\bf P}_{2}$.
Now the $z-$coordinate of $({\bf c} (t)\times {\bf c}' (t))$ is

\begin{eqnarray*}
\left|
\begin{array}{cc}
p_{0,x} + p_{1,x} t + p_{2,x} t^{2}&
p_{0,y} + p_{1,y} t + p_{2,y} t^{2}\\
p_{1,x}  + 2 p_{2,x} t&
p_{1,y}  + 2 p_{2,y} t
\end{array}
\right| &=& 
\left| 
\begin{array}{cc}
p_{1,x} & p_{1,y} \\
p_{2,x} & p_{2,y} 
\end{array}
\right| t^{2}
+
2 \left| 
\begin{array}{cc}
p_{0,x} & p_{0,y} \\
p_{2,x} & p_{2,y} 
\end{array}
\right| t
+\\
&&
\qquad \qquad \left| 
\begin{array}{cc}
p_{0,x} & p_{0,y}\\ 
p_{1,x} & p_{1,y} 
\end{array}
\right| 
\end{eqnarray*}
The coefficient of $t^{3}$ is $0$ due to special relation between the curve and
its derivative. (We study this phenomenon, in detail, in chapter~\ref{chapGC1}.)
Thus we see that 
${\bf c} (t) \times {\bf c}' (t)= 
({\bf p}_{1} \times {\bf p}_{2}) t^{2}+
({\bf p}_{0} \times {\bf p}_{2}) t+
({\bf p}_{0} \times {\bf p}_{1}) $.
On substituting the values of ${\bf p}_{i}$ we get the 
relation~\ref{eqnlemmacture}.
\end{proof}

Now by substituting the expression for $\gamma'_{i} (t)$ from 
section~\ref{secdervbezcubic} we get the expression for curvature of cubic 
curve as
\begin{eqnarray}
\omega_{i} (t) &=& \gamma'_{i} (t) \times \gamma''_{i} (t) \nonumber\\
          &=& {\bf g}_{0,i} (1-u(t))^{2} + {\bf g}_{1,i} u(t)(1-u(t)) + {\bf g}_{2,i} (u(t))^{2} \label{cubiccture}
\end{eqnarray}
where
\begin{eqnarray}
{\bf g}_{0,i} &=& \frac{6}{h^{2}_{i}} ({\bf m}_{i-1} \times L_{i}) - \frac{2}{h_{i}}({\bf m}_{i-1} \times {\bf m}_{i})\\
{\bf g}_{1,i} &=& \frac{2}{h_{i}} ({\bf m}_{i-1} \times {\bf m}_{i})\\
{\bf g}_{2,i} &=& \frac{6}{h^{2}_{i}} (L_{i} \times {\bf m}_{i}) - \frac{2}{h_{i}}({\bf m}_{i-1} \times {\bf m}_{i})
\end{eqnarray}

\begin{theorem}\label{cubicinflecredun1}
If the $i^{th}$ curve segment of interpolating spline $\gamma_{i} (t)$ is a 
cubic curve, $\omega_{i} (t_{j}) \cdot N_{j} >0$ and
$\omega_{i} (t) \cdot N_{j} >0$ changes sign only once for 
$t \in [t_{i-1}, t_{i}]$, $j=i-1,i$ then for all 
$N = \lambda N_{i-1} + \mu N_{i}$, where $\lambda \mu < 0$, 
$\omega_{i} (t) \cdot N$ has precisely one sign 
change for $t \in [t_{i-1}, t_{i}]$.
\end{theorem}

\begin{proof} We prove the theorem for the case $\lambda >0$. 
We first note that 
$\lambda \omega_{i} (t_{i-1}) \cdot N_{i-1} >0$
$\mu \omega_{i} (t_{i-1}) \cdot N_{i} >0$. 
Since $\omega_{i} (t) \cdot N_{j} >0$ changes sign only once for
$t \in [t_{i-1}, t_{i}]$, $j=i-1,i$ we also have
$\lambda \omega_{i} (t_{i}) \cdot N_{i-1} <0$
$\mu \omega_{i} (t_{i}) \cdot N_{i} <0$.
Thus
$ \lambda \omega_{i} (t_{i-1}) \cdot N_{i-1} +
\mu \omega_{i} (t_{i-1}) \cdot N_{i}=\omega_{i} (t_{i-1}) \cdot N >0 $
and
$ \lambda \omega_{i} (t_{i}) \cdot N_{i-1} +
\mu \omega_{i} (t_{i}) \cdot N_{i} = \omega_{i} (t_{i}) \cdot N<0 $. 

From formula~(\ref{cubiccture}) we see that $\omega_{i} (t)$ is a 
quadratic curve and hence $\omega_{i} (t) \cdot N$ is quadratic polynomial 
for $j=i-1, i$. That is, $\omega_{i} (t) \cdot N$ can change sign only twice. 
And if $\omega_{i} (t) \cdot N_{i-1}$ changes sign twice for 
$t \in [t_{i-1}, t_{i}]$, then we must have 
$\omega_{i} (t_{i-1}) \cdot N >0$ and $\omega_{i} (t_{i}) \cdot N >0$. 
Therefore  $\omega_{i} (t) \cdot N_{j}$ changes sign only once for 
$t \in [t_{i-1},t_{i}]$.

The proof for the case $\lambda <0$ is similar.
\end{proof}

\begin{remark}
In \cite[Goodman and Ong, CAGD 15, 1-17, 1997]{goodmanong}, theorem 
\ref{cubicinflecredun1} is proved for rational cubic B\'{e}zier curve
for a special case in which the tangent vector ${\bf m}_{j}$ lies
on the plane with normal vector as $N_{j}$, for $j=i-1, i$.
But this is a heavy restriction for modeling curves and surfaces.
\end{remark}

\begin{theorem}\label{cubicinflecredun2}
If the $i^{th}$ curve segment of interpolating spline $\gamma_{i} (t)$ is a 
cubic curve and $\omega_{i} (t_{i-1}) \cdot N_{i-1} >0$, 
and $\omega_{i} (t_{i}) \cdot N_{i-1} <0$
then $\omega_{i} (t) \cdot N_{i-1}$ changes sign only once for
$t \in [t_{i-1}, t_{i}]$.
\end{theorem}

\begin{proof} From formula~(\ref{cubiccture}) we see that $\omega_{i} (t)$ is a 
quadratic curve and hence $\omega_{i} (t) \cdot N_{j}$ is quadratic polynomial 
for $j=i-1, i$. That is, $\omega_{i} (t) \cdot N_{i-1}$ (and 
$\omega_{i} (t) \cdot N_{i}$) can change sign only twice. And if 
$\omega_{i} (t) \cdot N_{i-1}$ changes sign twice for $t \in [t_{i-1}, t_{i}]$,
then $\omega_{i} (t_{i-1}) \cdot N_{i-1} $ and $\omega_{i} (t_{i}) \cdot N_{i-1}$ 
will have same sign. 
But we have $\omega_{i} (t_{i-1}) \cdot N_{i-1} >0$ and 
$\omega_{i} (t_{i}) \cdot N_{i-1} <0$.
Therefore  $\omega_{i} (t) \cdot N_{j}$ changes sign only once for 
$t \in [t_{i-1},t_{i}]$.
\end{proof}

\begin{theorem}\label{cubicinflecredun3}
If the $i^{th}$ curve segment of interpolating spline $\gamma_{i} (t)$ is a 
cubic curve and $\omega_{i} (t_{i-1}) \cdot N_{i} <0$, 
and $\omega_{i} (t_{i}) \cdot N_{i-1} >0$
then $\omega_{i} (t) \cdot N_{i}$ changes sign only once for
$t \in [t_{i-1}, t_{i}]$.
\end{theorem}

\begin{proof}
The proof is similar to that of theorem~\ref{cubicinflecredun2}.
\end{proof}


\begin{remark}
The ease of using cubic curve, justified by 
theorems~\ref{cubicinflecredun1},~\ref{cubicinflecredun2}~and~\ref{cubicinflecredun3}, 
is one of the reasons for using cubic curve instead of higher order curves in 
splines. 
\end{remark}

\begin{theorem}
$\gamma_{i} (t)$ satisfies inflection criteria if and only if
\begin{enumerate}
\item ${\bf g}_{0,i} \cdot N_{i-1} >0$, ${\bf g}_{0,i} \cdot N_{i} < 0$ and
\item ${\bf g}_{2,i} \cdot N_{i-1} <0$  ${\bf g}_{2,i} \cdot N_{i} > 0$
\end{enumerate}
whenever $N_{i-1} \cdot N_{i} <0$.
\end{theorem} 

\begin{proof}
Proof follows from the conditions in the definition \ref{defninflexcrit} for 
inflection criteria of splines, equation~(\ref{cubiccture}) and 
theorem~\ref{cubicinflecredun1},~\ref{cubicinflecredun2}~and~\ref{cubicinflecredun3}. 
\end{proof}

\begin{remark}\label{quadraticplanar}
It is known that quadratic curves cannot be used in splines interpolating 
non-planar set of data points as it does not exhibit torsion. But now we can 
easily see that the quadratic curves cannot even be used in splines 
interpolating planar set of data points because it cannot satisfy the 
conditions for inflection preserving criteria. 
\end{remark}

\subsection{Torsion preservation criteria for cubic interpolating splines}\label{sectorsioncubic}
\setcounter{equation}{0}
We first state our result for torsion of a cubic B\'{e}zier curve.
Using the identities in section \ref{secinflexcubic} we get
\begin{theorem}\label{torsionformula}Let 
$\overline{\tau}_{i} (t) = 
|\gamma'_{i}(t) \gamma''_{i}(t) \gamma'''_{i}(t)| \mbox{, }t \in [t_{i-1},t_{i}]
$
(numerator of ${\tau}_{i} (t)$). 
Then
$\overline{\tau}_{i} (t) = 
\displaystyle{\frac{12}{h^{4}_{i}}}({\bf m}_{i-1} \times {\bf m}_{i} \cdot L_{i})
$
\end{theorem}

\begin{proof}
We can write numerator of $\tau_{i} (t)$ as
\begin{eqnarray*}
\overline{\tau}_{i} (t) 
&=& |\gamma'_{i}(t) \gamma''_{i}(t) \gamma'''_{i}(t)| \\
&=& \omega_{i} (t) \cdot \gamma'''_{i}(t) \\
&=& h_{0,i} (1-u(t))^{2} + h_{1,i} u(t)(1-u(t)) + h_{2,i} (u(t))^{2}
\end{eqnarray*}
where
\begin{eqnarray*}
h_{0,i} &=& (\frac{6}{h^{2}_{i}} ({\bf m}_{i-1} \times L_{i}) - \frac{2}{h_{i}}({\bf m}_{i-1} \times {\bf m}_{i}))
           \cdot \frac{6}{h^{3}_{i}} (h_{i}({\bf m}_{i-1} +{\bf m}_{i}) - 2L_{i})\\ 
  &=& \frac{6}{h^{4}_{i}}(6 ({\bf m}_{i-1} \times L_{i}) \cdot {\bf m}_{i} + 4({\bf m}_{i-1} \times {\bf m}_{i}) \cdot L_{i})
=\frac{12}{h^{4}_{i}}({\bf m}_{i-1} \times L_{i} \cdot {\bf m}_{i})  \\
\\
h_{1,i} &=& \frac{2}{h_{i}} ({\bf m}_{i-1} \times {\bf m}_{i}) \cdot \frac{6}{h^{3}_{i}}(h_{i}({\bf m}_{i-1} +{\bf m}_{i}) - 2L_{i})
= \frac{24}{h^{4}_{i}} ({\bf m}_{i-1} \times L_{i} \cdot {\bf m}_{i}) \\
\\
h_{2,i} &=& (\frac{6}{h^{2}_{i}} (L_{i} \times {\bf m}_{i}) - \frac{2}{h_{i}}({\bf m}_{i-1} \times {\bf m}_{i}))
            \cdot \frac{6}{h^{3}_{i}}(h_{i}({\bf m}_{i-1} +{\bf m}_{i}) - 2L_{i})\\
      &=& \frac{6}{h^{4}_{i}}(6 (L_{i} \times {\bf m}_{i}) \cdot {\bf m}_{i-1} + 4({\bf m}_{i-1} \times {\bf m}_{i}) \cdot L_{i})
= \frac{12}{h^{4}_{i}}({\bf m}_{i-1} \times L_{i} \cdot {\bf m}_{i})
\end{eqnarray*}
$u(t) = \displaystyle{\frac{t-t_{i-1}}{t_{i}-t_{i-1}}}$, $t \in [t_{i-1}, t_{i}]$.
Therefore
\begin{eqnarray*}
\overline{\tau}_{i} (t) 
&=& \frac{12}{h^{4}_{i}}(({\bf m}_{i-1} \times L_{i} \cdot {\bf m}_{i})
((1-u(t))^{2} + 2 u(t)(1-u(t)) + (u(t))^{2})\\
&=&\frac{12}{h^{4}_{i}}({\bf m}_{i-1} \times L_{i} \cdot {\bf m}_{i})
\end{eqnarray*}
Hence proved.
\end{proof}

\noindent
From definition \ref{deftorsioncrit} of torsion preservation criteria 
and theorem \ref{torsionformula} we get the following
\begin{theorem}\label{torsioncritcubic}
A cubic spline curve $\gamma (t)$ satisfies torsion preservation criteria, 
for $t \in [t_{i-1}, t_{i}]$, if and only if
$ [{\bf m}_{i-1} \: L_{i} \: {\bf m}_{i}] \triangle_{i} >0 $, whenever $\triangle_{i} \neq 0$.
\end{theorem}

\subsection{Collinearity preservation criteria for cubic interpolating spline}\label{seccollincubic}
\setcounter{equation}{0}
With the interpretation given in section \ref{seccollin} we get a sufficient condition 
for collinearity preservation criteria for splines in terms of B\'{e}zier control points of 
curve segments as below. Let the control points of curve segment $\gamma (t)$, 
$t \in [t_{i},t_{i+1}]$ be ${\bf P}_{i,j}$, $j = 0,1, \hdots , m_{i}$ and the control 
polygon be represented as ${\cal P}_{i}$. The control points of the derivative of curve segment 
are ${\bf P}'_{i,j} = \displaystyle{\frac{3}{h_{i}}}({\bf P}_{i,j+1}-{\bf P}_{i,j})$, 
$j = 0, \hdots, m_{i}-1$ and the control polygon be represented as ${\cal P}'_{i}$.
Now we make a observation about B\'{e}zier curves based
on following interpretation about vector product between two vectors in $R^{3}$. 
\begin{eqnarray}
\frac{|{\bf A} \times {\bf B}|}{|{\bf A}||{\bf B}|} &=& \sin(\theta)\mbox{, } {\bf A} \mbox{, }
{\bf B} \in R^{3} \mbox{, } \theta \mbox{ is the angle between } {\bf A} \mbox{ and } {\bf B} 
\label{sinetheta}
\end{eqnarray}

For any vector $L \in R^{3}$ equation \ref{sinetheta} implies the following
\begin{lemma}\label{linecrossbnd}If 
$\theta_{i} = \angle {\bf p}_{i} L \leq 90^{o}$, $i = 1,2$, 
${\bf p}_{1} \mbox{, } {\bf p}_{2} \in R^{3}$,
then 
\begin{eqnarray*}
\sup \left\{\frac{|{\bf p} \times L|}{|{\bf p}||L|} : 
{\bf p} = {\bf p}_{1} (1-t) + {\bf p}_{2} t \mbox{, } t \in [0,1]\right\}=
\sup \left\{\frac{|{\bf p}_{1} \times L|}{|{\bf p}_{1}||L|} \mbox{, } 
\frac{|{\bf p}_{2} \times L|}{|{\bf p}_{2}||L|} \right\}
\mbox{.}
\end{eqnarray*}
\end{lemma}
Following lemma follows from lemma \ref{linecrossbnd}
\begin{lemma}\label{trianglecrossbnd}If
$\theta_{i} = \angle {\bf p}_{i} L \leq 90^{o}$, $i = 1,2,3$, then 
\begin{eqnarray*}
\sup \bigg\{\frac{|{\bf p} \times L|}{|{\bf p}||L|} : 
{\bf p} \mbox{ is a point inside the planar triangle formed by the points } \\
{\bf p}_{1} \mbox{, } {\bf p}_{2} \mbox{ and } {\bf p}_{3} \in R^{3} \bigg\}
= 
\sup \left\{\frac{|{\bf p}_{i} \times L|}{|{\bf p}_{i}||L|} \mbox{ : } 
i=1,2,3 \right\}&
\end{eqnarray*}
\end{lemma}

\noindent We observe that 
\begin{theorem}\label{thmnewbezproperty} For a B\'{e}zier curve ${\bf c} (t)$ with control points ${\bf p}_{i}$, 
$\theta_{i} = \angle {\bf p}_{i} L \leq 90^{o}$,  
$i = 0,1, \hdots, m$. 
\begin{equation*}
 \sup \left\{\frac{|{\bf c} (t) \times L|}{|{\bf c}(t)||L|} : t \in [0, 1] \right\}
<  
\end{equation*}
\begin{equation}
\sup \left\{\frac{|{\bf p} \times L|}{|{\bf p}_{i}||L|} : {\bf p} \mbox{ belongs 
to set of vertices of convex hull formed by } {\bf p}_{i} i= 0,1, \hdots, m \right\} 
\end{equation}
\end{theorem}

\begin{proof} Proof follows from equation (\ref{sinetheta}) convex hull property of B\'{e}zier
curves (which states that B\'{e}zier curves lie inside the convex hull, that is, smallest convex 
polyhedra, with triangular sides, formed by its control points) and lemma \ref{trianglecrossbnd}.
\end{proof}



\begin{theorem}\label{collincritcubic}
$\gamma (t)$ satisfies collinearity criteria if 
\begin{eqnarray}
\sup \left\{\frac{|{\bf P}'_{i,k} \times L_{j}|}{|{\bf P}'_{i,k}||L_{j}|} \mbox{ : } k=0,1,2 \right\}
< \epsilon 
 \mbox{, } j= i-1,i, \label{defcollincritbnd}
\end{eqnarray}
whenever $|N_{i}|=0$ and $L_{i-1} \cdot L_{i} > 0$,
$0 < \theta_{k} = \angle {\bf P}'_{i,k} L_{j} \leq 90^{o}$, $k=0,1,2$, $j=i-1,i$
(a reasonable assumption to make).  
\end{theorem}

\begin{proof}
Since for a quadratic B\'{e}zier curve convex hull of its control points is the triangle formed by
the control points, therefore
\begin{eqnarray}
\sup \left\{\frac{|\gamma' (t) \times L_{j}|}{|\gamma'(t)||L_{j}|} : t \in [t_{i-1}, t_{i}] \right\}
< \nonumber \\ 
\sup \left\{\frac{|{\bf P}'_{i,k} \times L_{j}|}{|{\bf P}'_{i,k}||L_{j}|} \mbox{ : } k=0,1,2 \right\}
 \mbox{, } j= i-1,i, 
\end{eqnarray}
Hence proved.
\end{proof}

In case in the definition \ref{defcoplancrit} the condition 
\ref{defcoplancriteqn} is replaced by $\omega_{i} (t) =0$, 
$t \in \eta_{i}$ as stated in papers 
\cite[Goodman and Ong, CAGD 15, 1997]{goodmanong} etc., 
the collinearity condition for cubic spline
would have required ${\bf m}_{i} = \alpha L_{i}$, 
$\alpha >0$ in place of \ref{defcollincritbnd}. 


\subsection{Coplanarity preservation criteria for interpolating cubic splines}\label{seccoplancubic}
\setcounter{equation}{0}
From the definition of coplanarity criteria \ref{defcoplancrit} and the analysis in the previous
section we have following theorem stating the sufficient condition for cubic spline to satisfy
co-planarity preservation criterion.
\begin{theorem}\label{coplancritcubic}
$\gamma_{i} (t)$ satisfies the co-planarity preservation criteria if
\begin{eqnarray}
\sup \left\{\frac{|{\bf g}_{k,i} \times N_{j}|}{|{\bf g}_{k,i}||N_{j}|} \mbox{ : } k=0,1,2 \right\}
< \epsilon 
 \mbox{, } j= i-1,i, \label{defcoplancritbnd}
\end{eqnarray}
whenever $\triangle_{i} =0$ and $|N_{i-1}||N_{i}| \neq 0$ 
$\theta_{k} = \angle {\bf g}_{i,k} N_{j} \leq 90^{o}$, $k=0,1,2$, $j=i-1 \mbox{ or }i$
(a reasonable assumption to make).  
\end{theorem}

\begin{proof}
Proof is same as the proof of theorem \ref{collincritcubic}
\end{proof}

In case in the definition \ref{defcoplancrit} the condition 
\ref{defcoplancriteqn} is replaced by $\tau_{i} (t) =0$, 
$t \in [t_{i-1}, t_{i}]$ as stated in papers 
\cite[Goodman and Ong, CAGD 15, 1997]{goodmanong} etc., 
the coplanarity condition for cubic spline
would have required ${\bf m}_{i-1} \times L_{i} \cdot {\bf m}_{i} =0$,
that is, ${\bf m}_{i} = \alpha L_{i} + \beta {\bf m}_{i-1}$, 
$\alpha$, $\beta \in R$ in place of \ref{defcoplancritbnd}. 

Alternatively (actually more precisely) the condition 
\ref{defcoplancritbnd} can be replaced by the pair of conditions
$m_{i-1}= \alpha_{1} L_{i} + \beta_{1} L_{i-1}$, 
$m_{i}= \alpha_{2} L_{i} + \beta_{2} L_{i+1}$, with 
$\alpha_{1}$, $\alpha_{2}$, $\beta_{1}$, $\beta_{2} >0$. 

\section{Shape preserving properties of cubic Catmull-Rom splines}\label{secshapenetwork}
\setcounter{equation}{0}
Let denote the vector ${\bf x}_{i+1}-{\bf x}_{i-1}$ as
${\bf t}_{i}$, $i=2,...,n-1$, the plane containing the data points 
${\bf x}_{i-1}$, ${\bf x}_{i}$, ${\bf x}_{i+1}$ as $\Pi_{i}$, $i=2,...,n-1$
and for a curve $\gamma(t)=[x(t),y(t),z(t)]$,
$t \in [0,1]$ in $R^{3}$ let $\omega(t)= \gamma'(t) \times \gamma''(t)$. 
We now prove that cubic Catmull-Rom splines, with magnitude of tangent vectors 
calculated according to our algorithm, preserve convexity, 
inflection, torsion, collinearity and coplanarity behavior of the data
polygon, as follows. 

The tangent vector of the Catmull-Rom spline at a data point ${\bf x}_{i}$, 
$i=2,...,n-1$ is parallel to ${\bf t}_{i}$ so that the tangent vector is 
coplanar with the plane $\Pi_{i}$. Due to this cubic segment of the spline,
between the data points ${\bf x}_{i}$ and ${\bf x}_{i+1}$, at ${\bf x}_{i}$ 
lies on $\Pi_{i}$ and at ${\bf x}_{i+1}$ lies on the plane $\Pi_{i+1}$. Thus 
from the interpretations torsion preservation and coplanarity preservation
criteria in sections \ref{sectorsion} and \ref{seccoplan} we observe that the 
Catmull-Rom splines (having tangent vectors with our magnitudes) preserve the 
torsion and coplanarity behavior of the data polygon. 

\subsection{Torsion and coplanarity preservation}
Torsion preservation
by cubic Catmull-Rom spline is also assured by the theorem 
\ref{torsioncritcubic} as follows.
According to the theorem \ref{torsioncritcubic}, the spline curve $\gamma (t)$
must satisfy the condition 
$[{\bf m}_{i-1} \: L_{i} \: {\bf m}_{i}] \triangle_{i} >0$, 
where $\triangle_{i} = [L_{i-1} \: L_{i} \: L_{i+1}]$. But for Catmull-Rom 
spline we have ${\bf m}_{i}= {\bf x}_{i+1} - {\bf x}_{i-1} = L_{i} + L_{i+1}$, 
so that
$[{\bf m}_{i-1} \: L_{i} \: {\bf m}_{i}] = [L_{i-1} \: L_{i} \: L_{i+1}] = 
\triangle_{i}$ 
and therefore the torsion preservation condition 
$[{\bf m}_{i-1} \: L_{i} \: {\bf m}_{i}] \triangle_{i} >0$ holds $\forall i$.
Also if ${\bf x}_{i-2}$, ${\bf x}_{i-1}$, ${\bf x}_{i}$, ${\bf x}_{i+1}$ are
coplanar then 
$\tau_{i} (t) = [{\bf m}_{i-1} \: L_{i} \: {\bf m}_{i}] = 
[L_{i-1} \: L_{i} \: L_{i+1}] = 0$. Thus coplanarity condition is also 
satisfied.

\subsection{Convexity and inflection preservation}
It satisfies convexity and inflection preserving criteria with suitably chosen 
tangent length. This is mainly due to the reason that apart from the tangent 
vector at a data 
point ${\bf x}_{i}$ being coplanar with the plane $\Pi_{i}$ the two consecutive 
sides $ L_{i}$ and $ L_{i+1}$ lies on one side of it.

\subsection{Collinearity preservation}
Also if the data points ${\bf x}_{i-1}$, ${\bf x}_{i}$ and ${\bf x}_{i+1}$ are 
collinear and $L_{i-1} \cdot L_{i} > 0$ then the tangent vector of the spline
at ${\bf x}_{i}$ is collinear with ${\bf x}_{i-1}$, ${\bf x}_{i}$ and 
${\bf x}_{i+1}$. Thus the Catmull-Rom splines having tangent vectors with our 
magnitudes satisfies collinearity preservation criteria.

We also observe that for collinear data arc, the shape of Catmull-Rom splines
may not be aesthetically pleasing. We need to deviate from Catmull-Rom splines
in accrodance with the conditions of modified definition of collinearity 
preservation criteria stated in section \ref{seccollin}.

\begin{figure}[h]
\centering{
\includegraphics*{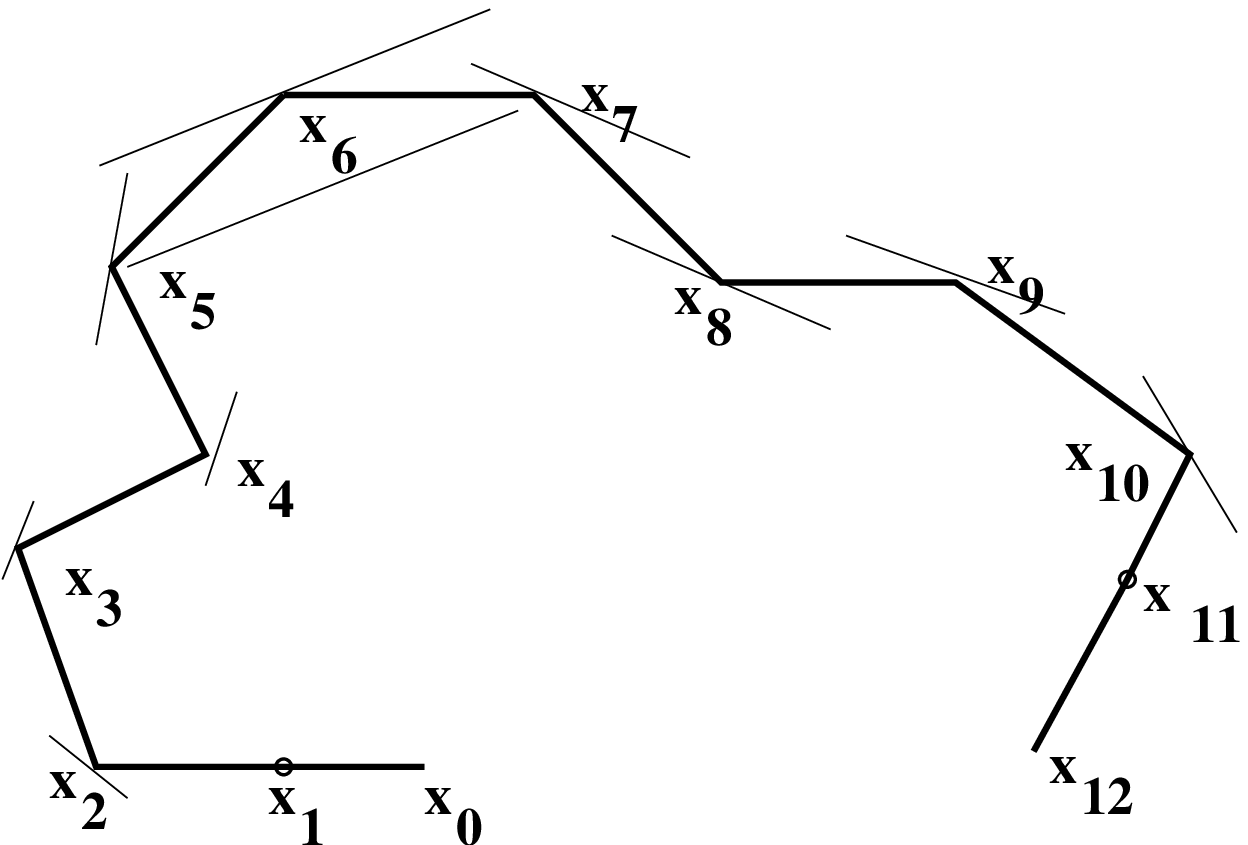}
\caption{By suitable choice of magnitudes of tangent vectors of 
Catmull-Rom splines can preserve the shape of data polygon}\label{catmullshape}
}
\end{figure}

\section{Conclusion}\label{conclusion}
\setcounter{equation}{0}
We have analyzed the characterization for shape preservation criteria for splines.
We have improved upon the definitive criteria for convexity preservation for splines.
We have studied in detail the inflection criteria and various results concerning it.
We have also stated the results from the literature which in conjunction with our 
analysis are observed to give negative results regarding convexity and inflection 
preservation criteria for splines. Such negative results would have been difficult 
to perceive intuitively. We have also discussed the analysis for collinearity, torsion
and coplanarity preservation criteria.

We obtained a very important theorem \ref{trsntrsncmptbl}, which states 
that there is a possibility that torsion preserving spline may need to have 
torsion set to zero at some nodepoints in order to be torsion continuous spline 
curve.
From the literature we find that
\begin{itemize}
\item Shape preservation criteria also gives better way of segmentation
of curves.
\item Such curves can be very usefull tool for data reduction, which is 
very important for data transmission.
\item Such interpolation can be used for robot path determination.
\end{itemize}
We are currently working towards getting profound theoretical and 
experimental results in the above directions.

We have found the characterization of all the shape preservation criteria for 
splines for the cubic case in terms of data points
and slope vectors on them. 


\section*{Acknowledgments}
I am grateful to my guide Prof. Sachin Balkrishna Patkar for
encouraging me understand and solve problems related to industrial applications.
I am also grateful to Prof. Mohan C. Joshi and Prof. Amiya K. Pani for giving
me an opportunity to work in the IMG team of IIT Bombay.

\end{document}